\documentclass[aip,jcp,reprint]{revtex4-1}

\usepackage[T1]{fontenc} % Use modern font encodings
\usepackage{graphicx}% Include figure files
\usepackage{dcolumn}% Align table columns on decimal point
\usepackage{bm}% bold math
\usepackage{amsmath}
\usepackage{amssymb}
\usepackage{amsfonts}
\usepackage{siunitx}
\DeclareSIUnit\molar{\mole\per\liter}
\DeclareSIUnit\Molar{M}
\DeclareSIUnit{\Litre}{L}
\DeclareSIUnit\calorie{cal}
\DeclareSIUnit\atmosphere{atm}
\DeclareSIUnit\angstrom{\text{\AA}}
\usepackage[normalem]{ulem} % strikethrough text
\usepackage{acro}
\usepackage[version=4]{mhchem} % chemical formulae
\usepackage{xr}
\usepackage{filecontents}

\newcommand{\rmd}{\mathrm{d}}
\newcommand{\kB}{k_\mathrm{B}}
\newcommand{\dbulk}{d_\mathrm{b}}
\newcommand{\lbulk}{l_\mathrm{b}}
\newcommand{\fKA}{f^\mathrm{KA}}
\newcommand{\fid}{f^\mathrm{ideal}}
\newcommand{\tauc}{\tau_\mathrm{c}}
\newcommand{\fu}{f_\mathrm{u}}
\newcommand{\fv}{f_\mathrm{v}}
\newcommand{\Nu}{N_\mathrm{u}}
\newcommand{\Nv}{N_\mathrm{v}}

\newcommand{\Nudot}{\dot{N}_\mathrm{u}}
\newcommand{\Nvdot}{\dot{N}_\mathrm{v}}
\newcommand{\siguw}{\sigma_\mathrm{uw}}
\newcommand{\epsuw}{\epsilon_\mathrm{uw}}

\newcommand{\Qv}{Q_\mathrm{v}}
\newcommand{\Ju}{J_\mathrm{u}}
\newcommand{\Jv}{J_\mathrm{v}}
\newcommand{\cu}{c_\mathrm{u}}

\newcommand{\citarg}{c_{i0}}
\newcommand{\cutarg}{c_\mathrm{u0}}
\newcommand{\curatio}{\cutarg^+/\cutarg^-}
\newcommand{\cubar}{\bar{c}_\mathrm{u}}
\newcommand{\cvbar}{\bar{c}_\mathrm{v}}

\newcommand{\cubulk}{c_{\mathrm{u}\infty}}

\newcommand{\rhobulk}{\rho_\infty}
\newcommand{\Ptarg}{P_0}

\newcommand{\ah}{a_\mathrm{h}}

\newcommand{\kappaDO}{\kappa_\mathrm{DO}}
\newcommand{\Ps}{\mathcal{P}_\mathrm{s}}
\newcommand{\Psdiff}{\mathcal{P}_\mathrm{s,diff}}

\newcommand{\deletion}[1]{}                 % Don't show deletions

\graphicspath{{./images/}}

\makeatletter
\def\@email#1#2{%
 \endgroup
 \patchcmd{\titleblock@produce}
  {\frontmatter@RRAPformat}
  {\frontmatter@RRAPformat{\produce@RRAP{*#1\href{mailto:#2}{#2}}}\frontmatter@RRAPformat}
  {}{}
}%
\makeatother

\begin{document}

%\preprint{}

\title{Non-equilibrium molecular dynamics of steady-state fluid transport through a 2D membrane driven by a concentration gradient} %Title of paper



\author{Daniel J. Rankin}
%\thanks{}
\affiliation{Department of Chemistry, School of Physics, Chemistry and Earth Sciences, The University of Adelaide, SA 5005, Australia}

\author{David M. Huang}
\email{david.huang@adelaide.edu.au}
%\thanks{}
\affiliation{Department of Chemistry, School of Physics, Chemistry and Earth Sciences, The University of Adelaide, SA 5005, Australia}


% \date{\today}

\begin{abstract}
	We use a novel non-equilibrium algorithm to simulate steady-state fluid transport through a two-dimensional (2D) membrane due to a concentration gradient by molecular dynamics (MD) for the first time. We confirm that, as required by the Onsager reciprocal relations in the linear-response regime, the solution flux obtained using this algorithm agrees with the excess solute flux obtained from an established non-equilibrium MD algorithm for pressure-driven flow. In addition, we show that the concentration-gradient solution flux in this regime is quantified far more efficiently by explicitly applying a transmembrane concentration difference using our algorithm than by applying Onsager reciprocity to pressure-driven flow. The simulated fluid fluxes are captured with reasonable quantitative accuracy by our previously derived continuum theory of concentration-gradient-driven fluid transport through a 2D membrane [\textit{J. Chem. Phys.} \textbf{151}, 044705 (2019)] for a wide range of solution and membrane parameters even though the simulated pore sizes are only several times the size of the fluid particles. The simulations deviate from the theory especially for strong solute--membrane interactions relative to the thermal energy, for which the theoretical approximations break down. Our findings will be beneficial for molecular-level understanding of fluid transport driven by concentration gradients through membranes made from 2D materials, which have diverse applications in energy harvesting, molecular separations, and biosensing.
\end{abstract}

%\pacs{}% insert suggested PACS numbers in braces on next line

\maketitle %\maketitle must follow title, authors, abstract and \pacs


\section{Introduction}\label{sec:intro}

Transport of liquid mixtures and solutions through porous membranes is pivotal to many applications, including water desalination and purification,\cite{werber2016} chemical separations, energy generation \cite{logan2012} and storage \cite{pomerantseva2019}, and biological and chemical sensing.\cite{venkatesan2011} But inadequate performance of current membranes limits widespread adoption of these technologies.\cite{siria2017} Membranes made from two-dimensional (2D) materials, such as graphene, molybdenum disulfide (\ce{MoS_2}), or hexagonal boron nitride (hBN), hold great promise for tackling these challenges.\cite{sahu2019,wang2017h,macha2019,heerema2016} But gaps in knowledge of fundamental aspects of transport processes in 2D membranes, particularly those driven by the concentration gradients \cite{marbach2019} that are central to many applications, hinders predictive design of 2D membranes.

The atomic-scale thickness of 2D membranes confers fundamentally different properties compared with conventional membranes that are highly beneficial. For example, 2D membranes can circumvent the permeability--selectivity trade-off that often plagues conventional desalination or filtration membranes,\cite{sahu2019} whereby selectivity against transport of an unwanted component of a mixture is generally associated with reduced permeability to a desired component, with computer simulations of 2D graphene membranes achieving almost complete salt rejection along with water fluxes orders of magnitude larger than those of current desalination membranes.\cite{cohen-tanugi2012} The extreme thinness of 2D membranes also means that transmembrane gradients of e.g. pressure, concentration, or electric potential can be enormous, resulting in huge driving forces for fluid transport. The orders-of-magnitude higher power densities for osmotic power generation of single-layer \ce{MoS_2} membranes compared with conventional membranes  was attributed partly to the massive salt gradient.\cite{feng2016c} In nanopore-based sensing or sequencing of macromolecules, which provides a cheap, fast, and portable method of chemical or biological sensing, the single-atom thickness of 2D membranes offers the possibility of discriminating single monomers in biopolymers due to the membrane thickness being comparable to the inter-monomer spacing. \cite{heerema2016,sahu2019} Achieving this goal, however, requires precise quantification of ion fluxes both in the absence and presence of the biomolecule.  

Although numerical simulations of fluid transport across 2D membranes have been carried out in the presence of concentration gradients using both continuum \cite{ji2019,graf2019} and molecular \cite{cohen-tanugi2012,heiranian2015,feng2016c,li2016i,cao2021,wang2021f,cheng2023} models, until our recent work,\cite{rankin2019} no analytical theory existed to quantify the relationship between the solution and solute flow through a 2D membrane driven by a solute concentration gradient and basic parameters such as the membrane pore size and strength and range of solute--membrane interactions. This theory was derived from a continuum hydrodynamic model of the fluid and its quantitative accuracy was verified by comparison with computational fluid dynamics simulations.\cite{rankin2019} However, the continuum fluid approximation can break down when the dimensions of the confining pores becomes comparable to the size of the fluid molecules \cite{bocquet2010}, which could be an issue for 2D membranes even with large pores, due to their atomic-scale thickness. Furthermore, continuum models cannot easily account for the effects of mechanical flexibility of the membrane, which can be significant for 2D membranes \cite{heerema2015,li2016i,noh2022a}. They also do not readily predict from first principles some phenomena that can strongly impact nano-scale fluid transport, such as finite solid--fluid friction, inhomogeneous fluid properties, non-electrostatic ion-specific interactions, and non-ideal solutions. 

But modelling fluid transport due to a concentration difference across a porous membrane using molecular simulations is not straightforward. This is because the periodic boundary conditions that are generally used in such simulations would create an artificial concentration jump across the periodic boundary when a concentration difference is applied across the membrane. Without constraints, mixing of the solution would occur across the periodic boundary instead of across the membrane. Until recently, a generally suitable molecular algorithm for simulating steady-state transport of liquid mixtures driven by concentration gradients has not existed. To the best of our knowledge, all molecular simulations of concentration-gradient-driven transport through 2D membrane until now have used a non-equilibrium molecular dynamics (NEMD) method in which the solute concentrations on either side of the membrane are set to be initially different and the resulting transient fluxes measured over time.\cite{cohen-tanugi2012,heiranian2015,feng2016c,li2016i,cao2021,wang2021f,cheng2023} Such simulations in general do not enable the solution and solute fluxes for a given transmembrane concentration difference to be accurately quantified, since the concentrations change measurably over time in the relatively small systems that can be practically simulated. This can be especially problematic for electrolyte solutions, since electrostatic screening depends on concentration, which could have a significant impact for the large concentration gradients that occur across 2D membranes.

Methods for calculating concentration-gradient-driven transport from equilibrium simulations in the absence of a concentration gradient \cite{liu2016f} also have deficiencies, particularly in the case of electrolytes, since they do not account for the effects of spatial variations of concentration-dependent properties such as electrostatic screening. On the other hand, stochastic algorithms that combine molecular dynamics (MD) with a grand canonical Monte Carlo (GCMC) method for maintaining constant reservoir concentrations by inserting, deleting, or exchanging particles in the reservoirs \cite{heffelfinger1994,cracknell1995a,kutzner2011,hato2016} suffer from low acceptance rates of stochastic particle moves at the high densities in liquids and unphysical particle insertions can perturb the flow.\cite{arya2001} Deterministic MD algorithms have also been developed to simulate steady-state flow in a periodic system due to a concentration gradient by applying a supplementary force to solute and/or solvent molecules to mimic a chemical potential gradient \cite{frentrup2012,khalili-araghi2013,ozcan2017,yoshida2017,monet2023}, either throughout the system \cite{yoshida2017} or in small transition regions far from the membrane.\cite{frentrup2012,khalili-araghi2013,ozcan2017,monet2023} Significant advantages of these methods over stochastic algorithms are that they are efficient even at liquid densities and they do not involve unphysical particle insertion or deletion steps. However, most of these algorithms have significant deficiencies that limit their utility, such as not being applicable to complex geometries such as 2D membranes,\cite{yoshida2017} not explicitly modeling the concenentration difference (and thus not accounting for spatial variations in concentration-dependent properties),\cite{yoshida2017} or not enabling independent control of differences in solute concentration and pressure across the membrane ,\cite{frentrup2012,khalili-araghi2013,ozcan2017} which can lead to spurious effects, particularly at high solute concentrations, as discussed later on. One recent algorithm\cite{monet2023} has been developed that does not suffer from these issues, but it controls the solute chemical potential difference across the membrane, whereas the solute concentration difference is a more useful control parameter for direct comparison with experiments.

In this work, we have modified a previous deterministic NEMD algorithm\cite{khalili-araghi2013} for constraining the solute concentration difference across a porous membrane in a periodic system to enable independent control of the transmembrane pressure difference. As a proof-of-principle of its ability to simulate and accurately quantify steady-state concentration-gradient-driven fluid transport through 2D membranes, we have applied this algorithm to a model system comprising a binary Lennard--Jones (LJ) liquid mixture and 2D LJ membrane. Using this model system, we have systematically investigated the effects of key parameters such as the membrane pore size, average solute mole fraction, and the strength and range of the solute--membrane interactions on the solution and solute fluxes. We have used these molecular simulation results to evaluate the accuracy of our previously developed continuum theory for predicting concentration-gradient-driven fluid transport through these atomically thin membranes.

\section{Computational methods}\label{sec:methods}

\subsection{System details}\label{sec:system}

\begin{figure}
	\includegraphics{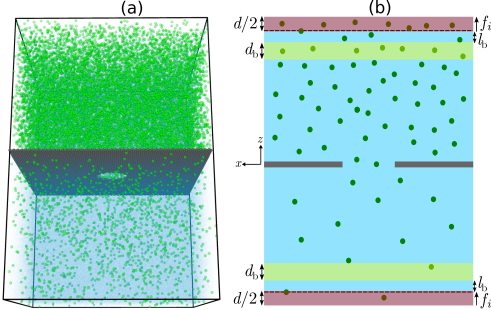}
	\caption{(a) Snapshot of a typical MD simulation system, consisting of a binary LJ liquid mixture (solvent particles are translucent) and a single-layer planar fcc (111) lattice membrane with a circular pore. (b) Schematic of constrained concentration- and pressure-difference algorithm: a force $f_i$ is applied perpendicular to the membrane to each particle of type $i$ (solute or solvent) within a transition region (red) of width $d$; the concentration ratio and pressure difference is measured between control regions (yellow) of width $\dbulk$ on either side of the membrane a distance $\lbulk$ from the transition region; the applied forces are dynamically adjusted to converge the concentration ratio and pressure difference to target values.}
	\label{fig:energy_step}
\end{figure}

All MD simulations were performed using LAMMPS (3 Mar 2020 GPU-accelerated version),\cite{plimpton1995,brown2011} with initial simulation configurations constructed using Moltemplate (version 2.16.1)\cite{jewett2021} and visualization of simulation trajectories carried out using OVITO (version 3.0.0).\cite{stukowski2009} The system comprised a binary liquid mixture (solute and solvent) and single-layer planar solid membrane parallel to the $x,y$ plane containing an approximately circular hole centered at the origin (Figure~\ref{fig:energy_step}(a)). The solid particles were placed in a single layer of the (111) surface of a close-packed face-centred cubic (fcc) lattice (lattice constant $\sqrt{2}\sigma$), and their positions were fixed throughout the simulations. The Lennard--Jones (LJ) potential, 
\begin{equation}
	u_{ij}(r_{ij}) = 4\epsilon_{ij}\left[\left(\frac{\sigma_{ij}}{r_{ij}}\right)^{12}-\left(\frac{\sigma_{ij}}{r_{ij}}\right)^6\right],
	\label{eqn:uLJ}
\end{equation} 
was used for the interactions between particles, where $r_{ij}$ is the distance between particles $i$ and $j$, and $\epsilon_{ij}$, and $\sigma_{ij}$ are parameters quantifying the strength and range, respectively of their interactions. The solute--solute, solvent--solvent, and solute--solvent interaction parameters, $\epsilon_{ij}$, and $\sigma_{ij}$, were set to the same values, $\epsilon$ and $\sigma$, respectively, in all simulations, while the parameters for the interactions between solute and solid (wall) particles, $\epsuw$ and $\siguw$, respectively, were varied for different simulations. Thus, the solute and solvent particles were identical in all respects except for their interactions with the membrane. The potential was cut-off at a distance of \num{4}$\sigma$ and  all particles had mass $m$. LJ units are used throughout this work, with masses, distances, energies, temperatures, pressures, and times in units of $m$, $\sigma$, $\epsilon$, $\epsilon/\kB$, $\epsilon/\sigma^3$, and $\tau=\sqrt{m\sigma^2/\epsilon}$, respectively, where $\kB$ is the Boltzmann constant. In all simulations, time integration was done with the velocity-Verlet integrator with a time step of $\num{0.005}\tau$ and periodic boundary conditions were applied in all dimensions. Unless otherwise stated, simulations were carried out in the canonical (NVT) ensemble at a temperature of $\epsilon/\kB$ using a Nos\'e--Hoover\cite{nose1984,hoover1985} style thermostat,\cite{shinoda2004} with only the velocity components perpendicular to the flow ($z$) direction thermostatted in NEMD simulations.

\subsection{Constrained concentration- and pressure-difference algorithm}\label{sec:algorithm}

To carry out NEMD simulations of steady-state flow due to concentration and/or pressure gradients, we adapted an algorithm by \citeauthor{khalili-araghi2013}\cite{khalili-araghi2013} designed to maintain unequal solute concentrations across a membrane in a system with periodic boundary condition. Specifically, we modified it to enable independent control of concentration and pressure differences. Here, we use the convention that the membrane lies in the $x,y$-plane at $z=0$ in the primary simulation box.

The algorithm in Ref.~\citenum{khalili-araghi2013} applies a supplementary constant force $f_i$ in the direction perpendicular to the membrane to particles of type $i$ in a small transition region of width $d$ far from the membrane, as illustrated in Fig.~\ref{fig:energy_step}(b). The algorithm was called the nonperiodic energy step method, since the addition of this force is equivalent to applying a nonperiodic energy step or ramp of size $\Delta\varepsilon_i = -f_i d$ across the transition region to the particles, which induces a concentration difference between the two sides of the membrane. The energy step $\Delta\varepsilon_i$ needed to achieve a target ratio of the concentrations $\citarg^+$ and $\citarg^-$ of species $i$ in the upper and lower fluid reservoirs, respectively, can be estimated from the relationship for a system of non-interacting (ideal) particles at infinite dilution, $\citarg^+/\citarg^- = \exp\left(\Delta\varepsilon_i/(\kB T)\right)$. The applied force required to maintain this concentration ratio in this case is
\begin{equation}
	\fid_i = -\frac{\kB T}{d}\ln \left(\frac{\citarg^+}{\citarg^-}\right),
	\label{eqn:fideal}
\end{equation}
which is used as the initial value of the applied force in the algorithm, i.e. $f_i(t=0) = \fid_i$. In general, an analytical relationship between the force and concentration difference or ratio does not exist for interacting particles. To account for the effect of particle--particle interactions, the forces are adjusted dynamically to converge to the target concentration ratio according to 
\begin{equation}
	\fKA_i(t+\Delta t) = \fKA_i(t) + \frac{\Delta t}{\alpha \tauc}\Delta \fKA_i(t)
	\label{eqn:energy_step_f}
\end{equation} 
with 
\begin{equation}
	\Delta \fKA_i(t) = \frac{\kB T}{d}\left[\left\langle \ln \left(\frac{c^+_i}{c^-_i}\right)\right\rangle - \ln \left(\frac{\citarg^+}{\citarg^-}\right)\right], 
	\label{eqn:energy_step_Df}
\end{equation} 
where $\Delta t$ is the simulation time step, $\alpha$ and $\tauc$ are tunable parameters, $c^+_i$ and $c^-_i$ are the instantaneous concentrations in the upper and lower reservoirs, respectively, and $\langle \cdots \rangle$ denotes a time average over the interval $(t - \tauc, t]$, i.e. over a duration $\tauc$ immediately preceding the current time step. (We use the superscript "KA" to distinguish the equations in  Ref.~\citenum{khalili-araghi2013} from those in our modified algorithm, which are described below.) The instantaneous concentrations are measured in control regions each of width $\lbulk$ on either side of the membrane and far from it (shown in yellow in Fig.~\ref{fig:energy_step}(b)) that is sufficiently wide to calculate the average in Eq.~\eqref{eqn:energy_step_Df} accurately. From Eq.~\eqref{eqn:energy_step_f}, the concentration ratio is expected to converge to the target ratio $\citarg^+/\citarg^-$ over a duration on the order of $\alpha\tauc$.  

To avoid calculating the average in Eq.~\eqref{eqn:energy_step_f} every time step and to reduce the chance of $\Delta f_i(t)$ becoming undefined due to $c^+_i$ or $c^-_i$  being zero, we have modified the force update algorithm from that in Ref.~\citenum{khalili-araghi2013} such that the applied force is updated every $\tauc$ time steps instead of every time step and the instantaneous concentrations are averaged over time before taking the logarithm. Thus, we replace Eqs.~\eqref{eqn:energy_step_Df} and~\eqref{eqn:energy_step_f} by
\begin{equation}
	f_i(t+\tauc) = f_i(t) + \frac{\Delta f_i(t)}{\alpha}
	\label{eqn:energy_step_mod_f}
\end{equation} 
and
\begin{equation}
	\Delta f_i(t) = \frac{\kB T}{d}\left[\ln \left(\frac{\langle c^+_i\rangle}{\langle c^-_i\rangle}\right)
	- \ln \left(\frac{\citarg^+}{\citarg^-}\right)\right], 
	\label{eqn:energy_step_mod_Df}
\end{equation} 
respectively, where the time average $\langle \cdots \rangle$ is taken over the interval $[n\tauc,(n+1)\tauc)$ for $n  \leq t/\tauc < n+1$ and $n \in \mathbb{Z}$. As in Ref.~\citenum{khalili-araghi2013}, Eq.~\eqref{eqn:fideal} is used to initialize the applied force and convergence to the target concentration ratio is expected to occur over a duration on the order of $\alpha\tauc$. 

In principle, the force update scheme specified by Eqs.~\eqref{eqn:energy_step_f} and \eqref{eqn:energy_step_Df} from Ref.~\citenum{khalili-araghi2013} or our modified version specified by Eqs.~\eqref{eqn:energy_step_mod_f} and \eqref{eqn:energy_step_mod_Df} can be used to constrain the concentration difference across of the membrane of any or all species in a multicomponent mixture. In Ref.~\citenum{khalili-araghi2013}, the external force was only applied to solute species (electrolyte ions in that case), whereas no external force was applied to the solvent (water). However, in general, applying an external force to solutes in the transition region without doing the same to the solvent will induce a hydrostatic pressure difference across the membrane that will affect the fluid fluxes, as we show below. Thus, an external force must also be applied to the solvent particles to achieve a desired pressure difference. Instead of constraining the solvent concentration using Eq.~\eqref{eqn:energy_step_mod_Df} to achieve this goal, we use a force update scheme for the solvent that directly controls the pressure difference, which more closely mimics how applied fields would be controlled experimentally. Thus, while we use Eqs.~\eqref{eqn:energy_step_mod_f} and \eqref{eqn:energy_step_mod_Df} to update the applied force on the solute particles (which we label as particles of type $i = \mathrm{u}$), for the solvent particles (type $i = \mathrm{v}$), we replace Eq.~\eqref{eqn:energy_step_mod_Df} by
\begin{equation}
	\Delta \fv(t) = \frac{A}{\Nv(t+\tauc)}\left(\left\langle \Delta P \right\rangle- \Delta \Ptarg \right),
	\label{eqn:energy_step_mod_Dfv}
\end{equation} 
where $A$ is the cross-sectional area of the simulation box, $\Delta \Ptarg$ is the target pressure difference, $\Nv(t+\tauc)$ is the instantaneous number of solvent particles in the transition region at time $t+\tauc$, $\Delta P = P^+ - P^-$ is the instantaneous pressure difference between the control regions on either side of the membrane, and the time average $\langle \cdots \rangle$ is computed identically to that in Eq~\eqref{eqn:energy_step_mod_Df}. (Note that $\left\langle \Nv \right\rangle$ may be a better choice than $\Nv(t+\tauc)$ in Eq.~\eqref{eqn:energy_step_mod_Dfv} as its use would reduce the fluctuations in the applied force.)  $P^\pm$ is calculated from the diagonal components of the per-atom stress tensor summed over atoms in the control regions using the \texttt{stress/atom} compute in LAMMPS.\cite{thompson2009} We note that this method is known to be inaccurate for computing the local pressure in inhomogeneous systems,\cite{todd1995,ikeshoji2003} for which more accurate but more computationally expensive and less widely implemented alternatives\cite{todd1995,ikeshoji2003,lion2012} exist. However, by keeping the control regions away from strong inhomogeneities in the fluid, this problem can be mitigated. For the initial applied force on the solvent particles,  $\fv(t=0) = 0$ is used.

\subsection{Simulation details}\label{sec:simulation}

Two system sizes were simulated. Unless otherwise stated, a $50 \times 50$ unit-cell membrane and a total of \num{396000} fluid particles were used. In addition, some non-equilibrium simulations were carried out with a larger $80 \times 80$ unit-cell membrane and a total of \num{1622016} fluid particles to verify that the simulations of the smaller system did not suffer from finite-size effects.

For each system size, fluid particles were placed on two cubic lattices of equal size on either side of the solid surface, which initially contained no pore, with the $z$ dimension of the box sufficiently large that the particles did not overlap. All fluid particles were initially set to be solvent particles. The system was initially equilibrated in the isothermal--isobaric (NPT) ensemble for \num{e6} time steps at a temperature of $\epsilon/\kB$ and pressure of $\epsilon/\sigma^3$ using a Nos\'e--Hoover\cite{nose1984,hoover1985} style thermostat and barostat\cite{shinoda2004} with only the $z$ dimension barostatted for the fluid particles. The average box length in the $z$ dimension measured over the last \num{e5} time steps, by which time the instantaneous box length had plateaued, was used as the box length in all subsequent constant-volume simulations, which were at a temperature of $\epsilon/\kB$. The $z$ dimension of the box was deformed at constant velocity over \num{e3} time steps to reach this value. All solid atoms within a distance $a$ of the origin were deleted to create an approximately circular pore of radius $a$ in the membrane, as illustrated in Fig.~\ref{fig:pore_image} of the supplementary material. 

Equilibrium MD and NEMD simulations were carried out for various combinations of the solute--membrane interaction parameters $\epsuw$ and $\siguw$, pore radius $a$, and average solute mole fraction $\bar{\chi}$, with $\epsuw/\epsilon = 0.5$, 0.8, 1.2 or 1.5, $\siguw/\sigma = 0.8$, 1.2 or 1.5, $a/\sigma = 0$ (equilibrium simulations only) 3, 4, 6, or 8, and $\bar{\chi} = 0.05$ or 0.2. In addition to NEMD simulations of concentration-gradient-driven flow, simulations of pressure-driven-flow without a concentration gradient were also carried out for selected systems using the algorithm in Ref.~\citenum{zhu2002}, which is similar in some respects to our constrained concentration- and pressure-difference algorithm, but in which a constant and equal force ($\fu = \fv$) is applied to solute and solvent molecules in the transition region. Most simulations with a constrained concentration difference used a target transmembrane solute concentration ratio of $\curatio = 20$, but ratios of 2, 3, and 5 were also used for selected systems to verify linear response of the fluid fluxes to the applied driving force. Unless otherwise stated, the transition region width $d$, control region width $\dbulk$, and distance $\lbulk$ between the transition and control regions in the NEMD simulations were all $2\sigma$ (see Fig.~\ref{fig:energy_step}) and the target pressure difference $\Delta \Ptarg$ was zero. Details of the simulated systems and their properties are given in Tables~\ref{tbl:eq_sys_size}--\ref{tbl:neq_Dp_sys} of the supplementary material.

\begin{figure}
	\centering
	\includegraphics{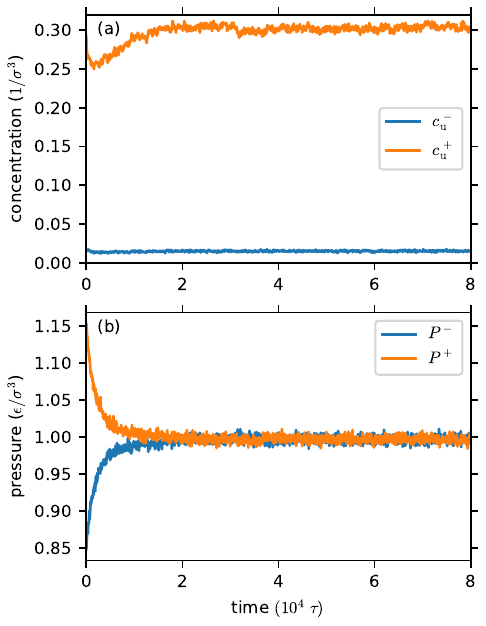}
 \caption{(a) Solute concentration $\cu$ and (b) pressure $P$ in the upper ($+$) and lower ($-$) control regions vs time when applying the constrained concentration- and pressure-difference algorithm for a target concentration ratio $\curatio = 20$ and target pressure difference $\Delta \Ptarg =0$, starting from a configuration in which solutes were uniformly distributed on either side of the membrane at the target concentration ratio ($a = 6\sigma$, $\epsuw = \epsilon$, $\siguw = \sigma$, $\bar{\chi} = 0.2$).}
 \label{fig:conc_press_vs_time}
\end{figure}

Starting from the final simulation configuration from the previous NPT equilibration step, fluid particles were randomly converted into solute particles to give the desired average solute mole fraction. Additionally, for the NEMD simulations, the fluid particles were converted to achieve the maximum target solute concentration ratio of 20. Then, for each system geometry, a NEMD simulation was carried out using the constrained concentration- and pressure-difference algorithm with $\epsuw = \epsilon $ and $\siguw = \sigma$, reducing the target solute concentration ratio at intervals of $\sim \num{8e6}$ time steps to obtain steady-state simulation configurations at each of the desired concentration ratios. Fig.~\ref{fig:conc_press_vs_time} shows the variation of the solute concentration and pressure in the control regions with time from one of these simulations, which illustrates the ability of the algorithm to converge  and maintain the concentration and pressure difference at the target values.  The final configuration at each target concentration ratio was used as the starting configuration for simulations with other values of $\epsuw$ and $\siguw$ at that target ratio. All these simulations used $\alpha = 50$ and $\tauc = 50\tau$.

For the NEMD simulations, an automated equilibration detection method\cite{chodera2016} was used to determine when each system had reached a non-equilibrium steady state and to estimate the effective number of uncorrelated samples in order to calculate steady-state averages and statistical uncertainties (at the 95\% confidence level) of fluctuating variables. Distribution functions such as solute concentration profiles, fluid density profiles, and pressure profiles were calculated only using data after the first \num{8e6} time steps of each simulation, which ensured the system was at equilibrium or in a non-equilibrium steady state.

It should be noted that changing solvent particles into solute particles with different solute-membrane interaction parameters results in deviations of the bulk pressure in the fluid reservoirs from the target pressure of $\epsilon/\sigma^3$ in the NPT equilibration simulations, particularly at high concentrations, with average pressures in the equilibrium simulations varying from 0.93 to 1.00$\epsilon/\sigma^3$ across the range of systems studied (Table~\ref{tbl:eq_sys} of the supplementary material). However, the bulk solution density remained approximately constant, varying from 0.782 and 0.787$\sigma^{-3}$ across the range of solute--membrane interactions.

\section{Results and discussion}\label{sec:results}

\subsection{Application of constrained concentration- and pressure-difference algorithm}

\begin{figure}
	\centering
	\includegraphics{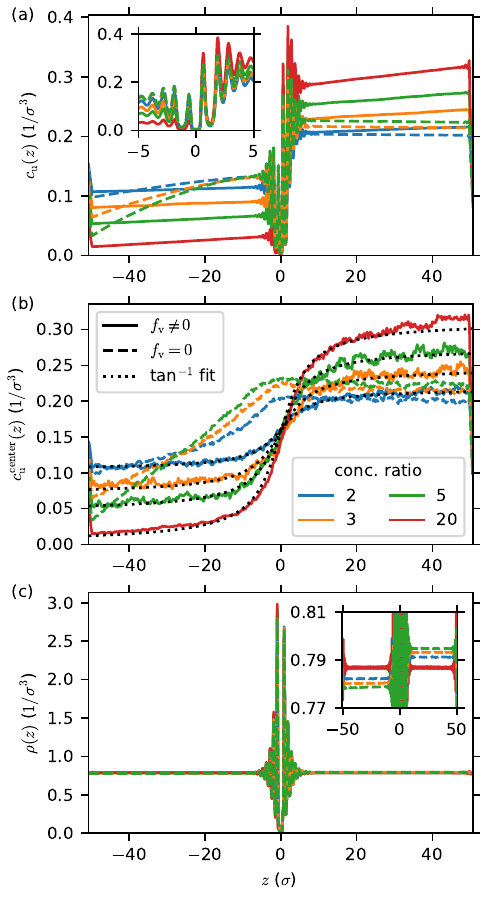}
 \caption{(a) Solute concentration, (b) centerline solute concentration, and (c) total fluid density vs $z$ coordinate in non-equilibrium simulations with ($\fv \neq 0$, solid lines) and without ($\fv = 0$, dashed lines) the transmembrane pressure difference constrained to be zero for $a = 6\sigma$ , $\epsuw = 0.5\epsilon$, $\siguw = 0.8\sigma$, $\bar{\chi} = 0.2$, and various target solute concentration ratios $\curatio$. The inset in (a) zooms in on $z$ values near the membrane, whereas the inset in (c) zooms in on density values around the bulk density. The dotted black lines in (b) are least-squares fit of the solid line to a function of the form $b_0 + b_1\tan^{-1}(z/b_2)$ with fit parameters $b_0$, $b_1$, and $b_2$.}
 \label{fig:conc_profiles_rep_highc}
\end{figure}

Fig.~\ref{fig:conc_profiles_rep_highc} depicts the solute concentration, centerline solute concentration (calculated for solute particles within a distance $\sigma$ of the axis passing through the pore center), and total (solute + solvent) fluid density profiles perpendicular to the membrane for a system to which our constrained concentration- and pressure-difference algorithm was applied, either with or without enforcing the pressure constraint $\Delta \Ptarg = 0$ using Eq.~\eqref{eqn:energy_step_mod_Dfv}. (Results for the highest concentration ratio without the pressure constraint are not shown because the concentration ratio never converged to a steady state.) The method without the pressure constraint is equivalent to the original constrained concentration-difference algorithm of  \citeauthor{khalili-araghi2013}\cite{khalili-araghi2013} The system in Fig.~\ref{fig:conc_profiles_rep_highc} had a pore radius $a = 6\sigma$, solute mole fraction $\bar{\chi}=0.2$, and overall repulsive solute--membrane interactions, as indicated by the solute depletion near the membrane in Fig.~\ref{fig:conc_profiles_rep_highc}(a). Qualitatively similar results were obtained for other systems, as illustrated in the supplementary material for a system with same solute--membrane interactions but lower ($\bar{\chi}=0.05$) solute mole fraction in Fig.~\ref{fig:conc_profiles_rep_lowc} and for a system with the same solute mole fraction but with attractive effective solute--membrane interactions in Fig.~\ref{fig:conc_profiles_attr_highc}. 

\begin{figure}
	\centering
	\includegraphics{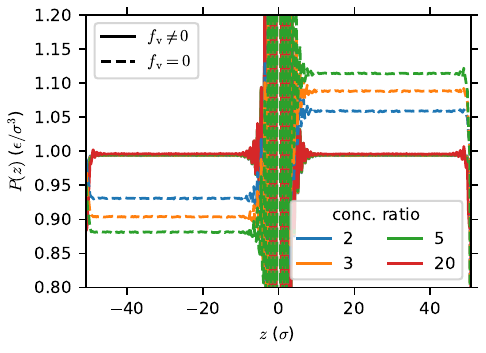}
 \caption{Pressure profile in non-equilibrium simulations with ($\fv \neq 0$, solid lines) and without ($\fv = 0$, dashed lines) the transmembrane pressure difference constrained to be zero for the same conditions in Fig.~\ref{fig:conc_profiles_rep_highc}.}
 \label{fig:press_profiles_cdrive_rep_highc}
\end{figure}

The solute concentration profiles with and without the pressure constraint in Fig.~\ref{fig:conc_profiles_rep_highc}(a) and (b) are dramatically different. At first glance, the total fluid density profiles in Fig.~\ref{fig:conc_profiles_rep_highc}(c) are similar. But closer inspection, as shown in the inset, reveals a transmembrane density difference without the pressure constraint, which is induced by the net force exerted on the solution by the applied force on the solute particles used to constrain the concentration difference. The net force is manifested in a transmembrane pressure difference, as shown in the pressure profiles for the same system in Fig.~\ref{fig:press_profiles_cdrive_rep_highc} when the pressure difference is not constrained. On the other hand, constraining the pressure difference to $\Delta \Ptarg = 0$ results in equal pressures and total fluid densities on either side of the membrane. Fluid flow driven by the pressure difference polarizes the solute concentration near the membrane, resulting in the differing solute concentration profiles in Fig.~\ref{fig:conc_profiles_rep_highc}(a) and (b) with and without the pressure constraint. As shown by the fitted curves in Fig.~\ref{fig:conc_profiles_rep_highc}(a), the centerline solute concentration when both the concentration and pressure difference are constrained is consistent with theoretical predictions under conditions in which there is a concentration difference but no pressure difference and the solute--membrane interaction range is small compared with the pore radius, in which the concentration profile is expected to be an inverse tangent function of the axial coordinate.\cite{rankin2019}

\begin{figure}
	\centering
	\includegraphics{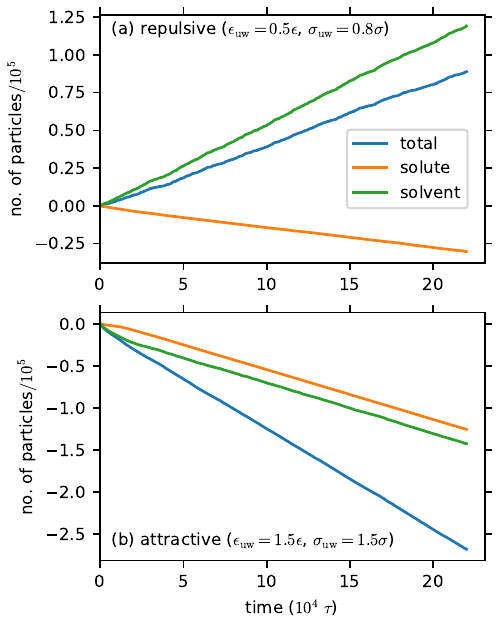}
 \caption{Cumulative number of fluid (total, solute, or solvent) particles crossing the membrane vs time in simulations with $\curatio = 20$ and $\Delta \Ptarg = 0$ for $a = 6\sigma$, $\bar{\chi} = 0.2$, and (a) repulsive ($\epsuw = 0.5\epsilon, \siguw = 0.8\sigma$) or (b) attractive ($\epsuw = 1.5\epsilon, \siguw = 1.5\sigma$) effective solute--membrane interactions. (Each point is an average over a time interval of $\tauc$.)}
 \label{fig:N_vs_time}
\end{figure}

Fluid fluxes were determined by counting the number of solute and solvent particles crossing the membrane as a function of time, as illustrated in Fig.~\ref{fig:N_vs_time} for two systems to which the constrained concentration- and pressure-difference algorithm was applied with zero pressure difference. (In practice, we measured fluxes at the boundary of the simulation box, which was in the center of the transition region (see Fig.~\ref{fig:energy_step}), which gives the same result as any plane parallel to the membrane at steady steady due to particle conservation.)  The effective solute--membrane interactions in Fig.~\ref{fig:N_vs_time}(a) and (b) are repulsive and attractive, respectively, i.e. solute is depleted and enhanced near the membrane relative to the bulk, respectively. The linearity of the curves vs time shows both systems are in the steady state with constant fluid fluxes for most of the simulation  (Similar behavior was observed for all systems studied.) Consistent with expectations for systems with a transmembrane concentration difference but no pressure difference and for which solute diffusion dominates advection (P\'eclet number $\mathrm{Pe} < 1$), the solute flux is in the direction of decreasing concentration, while the total solution flux due to concentration-gradient-driven diffusiosmosis is opposite in direction for membranes that repel and attract the solute, with flow towards increasing and decreasing concentration, respectively.\cite{marbach2019}

\begin{figure}
	\includegraphics{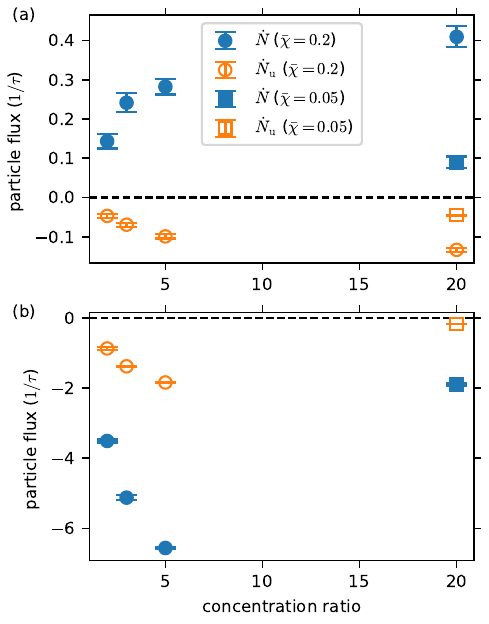}
	\caption{Total ($\dot{N}$, filled symbols) and solute ($\Nudot$, empty symbols) flux vs target concentration ratio $\curatio$ for systems with $a = 6\sigma$, $\epsuw = 0.5\epsilon, \siguw = 0.8\sigma$, and $\bar{\chi} =$ 0.2 (circles) or 0.05 (squares) in simulations with (a) $\Delta \Ptarg = 0$ pressure constraint and (b) no pressure constraint .}
	\label{fig:flux_vs_time-press_constraint_effect}
\end{figure}

The flux of particles of type $i$  was calculated as a numerical derivative, $\dot{N}_i = \Delta N_i/\Delta t$, of the cumulative particle number $N_i$ crossing the membrane vs time $t$ in the steady state. The simulation trajectory was divided into intervals of \num{2e5} timesteps = $\num{e3} \tau$ for which the flux was calculated, from which the average flux with statistical uncertainties was obtained using the method described in Sec.~\ref{sec:simulation}. (We verified that the calculated uncertainties were insensitive to halving or doubling this time interval.) Fig.~\ref{fig:flux_vs_time-press_constraint_effect} shows the total ($\dot{N}=\Nudot + \Nvdot$) and solute ($\Nudot$) fluxes vs the target concentration ratio for a system with repulsive effective solute--membrane interactions for low ($\bar{\chi} = 0.05$) and high ($\bar{\chi} = 0.2$) solute mole fractions with or without the $\Delta \Ptarg = 0$ pressure constraint. Consistent with expectations for this system when no transmembrane pressure difference is applied, the total flux is towards increasing concentration whereas the solute flux is towards decreasing concentration when the pressure constraint is enforced. On the other hand, both the total and solute flux are towards decreasing concentration without the pressure constraint, due pressure-driven flow as a result of the non-zero transmembrane pressure difference. The relative discrepancy between the total solution fluxes with and without the pressure constraint is similar for both solute mole fractions and different concentration ratios, highlighting the general importance of applying the pressure constraint to obtain accurate fluxes in constrained concentration-difference simulations. The solute flux is significantly different with and without the pressure constraint, but the relative discrepancy is much smaller at the lower solute mole fraction, due to the greater importance of solute diffusion over advection (lower P\'eclet number) at the lower mole fraction, suggesting that there may be circumstances in which the pressure constraint may not greatly affect the solute flux.

The pressure difference was not constrained in the original constrained concentration-difference algorithm in Ref.~\citenum{khalili-araghi2013}. The study for which the algorithm was developed focused on measuring the ionic current through biological membrane channels due to a concentration difference for a relatively dilute aqueous KCl electrolyte. A concentration ratio of \num{0.1}:\SI{1}{\molar}, pore radius $a \approx \SI{0.55}{\nano\metre}$, and pore length $L \approx\SI{6}{\nano\metre}$ was considered for the OmpF porin that was studied. To maintain the concentration ratio across the pore an external force $f = \SI{0.557}{\kilo\calorie\per\mol}$ was applied to the ions within a $d = \SI{2.5}{\angstrom}$ transition region. As explained in the next section, this force would create a pressure difference of magnitude $\lvert\Delta P\rvert = \Nu f/A$ across the membrane, where $\Nu$ is the number of ions in the transition region and $A$ is its cross-sectional area. Using $A=\SI{123.5}{\angstrom}\times \SI{123.5}{\angstrom}$ and the average ion concentration of $\cubar = \SI{1.1}{\molar}$ in these simulations gives $\Nu \approx 50$, and thus $\lvert\Delta P\rvert \approx \SI{2.5}{\kilo\pascal}$. Ignoring the hydraulic resistance of the pore ends for simplicity, which would reduce the flux further, the total solution flux due to this pressure difference can be estimated from the Hagen--Poiseuille equation,\cite{kirby2010} $Q =  \pi a^4\Delta P/(8\eta L)$, where $\eta$ is the solution shear viscosity, which we have taken to be that of pure water, $\eta = \SI{8.94e-4}{\pascal \second}$.\cite{huber2009} Using the parameters above gives an estimate of the convective  ion flux of $\cubar Q \approx \SI{e4}{\per\second}$.  The lowest total ionic current that was measured in Ref.~\citenum{khalili-araghi2013} was $\approx \SI{10}{\pico\ampere}$, which gives a lower bound (corresponding to a perfectly ion-selective channel) on the total ion flux of $\approx \SI{6e7}{\per\second}$. Thus, the convective ion flux due to the induced pressure difference would have been negligible compared with that due to the applied concentration difference in this study, and so the application of a pressure constraint would have made little difference to the results.

\subsection{Transport coefficients: verification of linear response and Onsager reciprocity}

From now on, we focus on NEMD simulations using our constrained concentration- and pressure-difference algorithm in which the pressure difference has been constrained to be zero. To quantify the concentration-gradient-driven fluid fluxes for all of the systems studied, we define two transport coefficients -- the diffusioosmotic  mobility,
\begin{equation}
	\kappaDO \equiv -\frac{Q}{\Delta \Pi /(\kB T)},
	\label{eqn:kappaDO}
\end{equation}
and the solute permeance,
\begin{equation}
	\Ps \equiv -\frac{\Ju}{\Delta \Pi /(\kB T)},
	\label{eqn:Ps}
\end{equation}
which characterize the total volumetric solution flux $Q$ and solute flux $\Ju = \Nudot$, respectively for a given transmembrane osmotic pressure difference $\Delta \Pi$ at temperature $T$. These definitions follow the notation in our previous work,\cite{rankin2019} in which we derived a theory of concentration-gradient-driven flow through 2D membranes for dilute solutions, but the equations above generalize them to arbitrary solute concentrations.\cite{marbach2017a,yoshida2017} Eqs.~\eqref{eqn:kappaDO} and \eqref{eqn:Ps} reduce to the corresponding equations (Eqs.~(33) and~(34)) in Ref.~\citenum{rankin2019} in the dilute solution limit, where $\Delta \Pi = \kB T\Delta\cu$.\cite{marbach2017a} The solution flux can be determined from the simulations as $Q = \frac{\dot{N}}{\bar{\rho}}$, where $\bar{\rho}$ is the bulk total fluid density, which we calculated as the average of the total fluid density in the upper and lower control regions, i.e.  $\bar{\rho} = (\rho^+ + \rho^-)/2$. 

As described recently for a similar NEMD simulation algorithm,\cite{monet2023} the transmembrane osmotic pressure difference $\Delta \Pi$ and hydrostatic pressure difference $\Delta P$ can be calculated by considering the balance of applied forces of the fluid particles in the transition region. Decomposing the applied force on each solute particle and each solvent particle as $\fu = f + \delta\fu$ and $\fv = f + \delta\fv$, respectively, such that $\Nu\delta\fu + \Nv\delta\fv = 0$, the force due to the osmotic pressure difference is $-A\Delta \Pi = \Nu \delta\fu = -\Nv \delta\fv$, while the force due to the hydrostatic pressure difference is $-A\Delta P = (\Nu + \Nv)f = Nf$, where $\Nu$ and $\Nv$ are the number of solute and solvent particles, respectively, in the transition region, $N$ is the total number of fluid particles in the transition region, and $A$ is its cross-sectional area.\cite{monet2023} (Note that, following the convention in Sec.~\ref{sec:algorithm} we have defined $\Delta \Pi$ and  $\Delta P$ as differences across the membrane rather than across the transition region as was done in Ref.~\citenum{monet2023}, so the sign in the previous equations is opposite that in the equivalent equations in Ref.~\citenum{monet2023}.) From these equations, $\Delta \Pi$ and $\Delta P$ can be calculated in terms of the number of particles and applied force on particles of each type in the transition region as
\begin{equation}
	\Delta \Pi = -\frac{1}{A}\left(\frac{\Nu\Nv}{\Nu+\Nv}\right)\left(\fv-\fu\right)
	\label{eqn:DPi_trans}
\end{equation}
and
\begin{equation}
	\Delta P = -\frac{1}{A}\left(\Nu\fu+\Nv\fv\right).
	\label{eqn:Dp_trans}
\end{equation}

Alternatively, $\Delta \Pi$ and $\Delta P$ can be calculated from the solute concentration and pressure, respectively, in the control regions. In this case, $\Delta P = P^+ - P^-$, i.e. the pressure difference evaluated in the constrained concentration- and pressure-difference algorithm. On the other hand, $\Delta \Pi$ can be estimated using the osmotic pressure of an incompressible ideal binary mixture, which can be derived from the entropy of mixing at any solute concentration (assuming the same solute and solvent molecular volume $v$) to be\cite{barrat2003,lewis1908,yoshida2017}
\begin{equation}
	\Pi = -\frac{\kB T}{v} \ln (1-\chi) = -\rho\kB T\ln(1-\chi)
	\label{eqn:Pi_id}
\end{equation}
where $\chi$ is the solute mole fraction and $\rho = 1/v$ is the total fluid density (which is assumed to be independent of $\chi$). This equation reduces to the standard van't Hoff equation, $\Pi = \cu\kB T$, for $\chi \ll 1$. We have evaluated $\Delta \Pi = \Pi^+ - \Pi^-$ from Eq.~\eqref{eqn:Pi_id} using the solute mole fraction, $\chi^+$ and $\chi^-$, and bulk fluid density, $\rho^+$ and $\rho^-$, in the upper and lower control regions, respectively. We have compared $\Delta \Pi$ and $\Delta P$ calculated from the applied force balance in the transition region and from the concentration/density or pressure in the control regions in Figs.~\ref{fig:DPi_trans_vs_DPi_control} and \ref{fig:Dp_trans_vs_Dp_control}, respectively, in the supplementary material for all our non-equilibrium simulations. This includes simulations in which the osmotic pressure difference was non-zero and the pressure difference was constrained to be zero, simulations in which the osmotic pressure difference was non-zero and the pressure difference was unconstrained and thus non-zero, and simulations in which the osmotic pressure difference was zero and pressure difference was non-zero. This comparison shows perfect agreement between $\Delta P$ calculated using either method, but $\Delta \Pi$ calculated from the control regions using Eq.~\eqref{eqn:Pi_id} overestimates (by up to $\approx$15\%) the value obtained from the transition region force balance using Eq.~\eqref{eqn:DPi_trans} for larger $\Delta \Pi$. The origin of this discrepancy is likely the absence of a well-defined "bulk" solute concentration in the upper or lower fluid reservoirs to unambiguously define the osmotic pressure in Eq.~\eqref{eqn:Pi_id} in simulations with a non-zero concentration difference, since the concentration varies throughout the system (see e.g. Fig.~\ref{fig:conc_profiles_rep_highc}). By contrast, even with a transmembrane pressure difference, the pressure profile is essentially flat in either reservoir except in the immediate vicinity of the transition region or membrane (see e.g. Fig.~\ref{fig:press_profiles_cdrive_rep_highc}), so the $\Delta P$ between the control regions is representative of the transmembrane pressure difference.

Using $\Delta\Pi$ from the force balance in the transition region, we have verified that the transport coefficients in Eqs.~\eqref{eqn:kappaDO} and~\eqref{eqn:Ps} were independent of the system size and transition region width for selected systems in Fig.~\ref{fig:kappaDO_Ps_vs_radius-res_trans_width}.  We have have also verified that the transport coefficients were independent of $\Delta\Pi$, i.e. the systems were in the regime of linear response of the fluid fluxes to the applied osmotic driving force, for selected systems as shown in Fig.~\ref{fig:linear_response} for $\kappaDO$ and in Fig.~\ref{fig:linear_response_solute_permeance} of the supplementary material for $\Ps$. The three selected systems encompassed those most likely to deviate from linear response, namely those with the three highest solution flux magnitudes (which includes those with the two highest solute flux magnitudes) at the highest target concentration ratio $\curatio = 20$ for the pore radius $a = 6\sigma$ that was used in most of simulations. As indicated by solid symbols in these figures, $\kappaDO$ and $\Ps$ are independent of $\Delta \Pi$ up to the highest $\Delta \Pi$, except for the system with the largest fluxes ($\epsuw = 1.5\epsilon$, $\siguw = 1.5\sigma$, $\chi = 0.2$), for which linear response appears to hold up to $\Delta \Pi \approx 0.15\epsilon/\sigma^3$ ($\curatio = 3$). Although a few simulations were carried out for systems with a larger pore radius of $a = 8\sigma$, for which the fluid fluxes were greater than those with $a = 6\sigma$ for the same solution and membrane properties, the fluxes were well within the range in which the system with the largest fluxes was still in the linear-response regime. Thus, we can be confident that all systems besides that one were in the linear-response regime for the conditions simulated. 

\begin{figure}
	\includegraphics{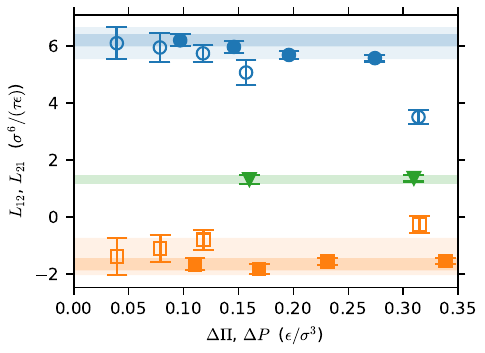}
	\caption{$L_{12} = \kappaDO/(\kB T)$ (filled symbols) and $L_{21}$ (empty symbols) transport coefficients vs transmembrane osmotic pressure difference $\Delta \Pi$ and pressure difference $\Delta P$, respectively, for flow driven exclusively by a concentration or a pressure difference for systems with $\bar{\chi} = 0.2$, $a = 6\sigma$, and various solute--membrane interaction parameters:  $\epsuw = 1.5\epsilon, \siguw = 1.5\sigma$ (circles); $\epsuw = 0.5\epsilon, \siguw = 0.8\sigma$ (squares); $\epsuw = 0.5\epsilon, \siguw = 1.5\sigma$ (triangles).  The dark and light translucent bands are 95\% confidence intervals for $L_{12}$ and $L_{21}$, respectively, for the lowest value of $\Delta \Pi$ and $\Delta P$, respectively,  for each set of solute--membrane interaction parameters. At least for the smaller values of  $\Delta \Pi$ or $\Delta P$, $L_{12}$ and $L_{21}$ are approximately constant and equal, verifying linear response and Onsager reciprocity.}
	\label{fig:linear_response}
\end{figure}

In the linear response regime, the fluid fluxes due to transmembrane differences in the hydrostatic pressure $\Delta P$ and osmotic pressure $\Delta\Pi$ (or, equivalently, chemical potential, $\Delta\mu$), are given by\cite{degroot1984} %\cite[pp.~405--426, 435--438]{degroot1984}
\begin{equation}
	\left[\begin{array}{c} Q\\ \Ju - \cubar\Qv \end{array} \right] 
	= 
	\left[\begin{array}{cc}\Lambda_{11} & \Lambda_{12} \\ 
		\Lambda_{21} & \Lambda_{22} \\  \end{array} \right]
		\left[\begin{array}{c} -\Delta P \\ -\Delta \mu \end{array}\right]\,,
	\label{eqn:onsager_mu}
\end{equation}
or using $\Delta \mu = \cubar\Delta\Pi$, \cite{monet2023} with $L_{11} = \Lambda_{11}$, $L_{12} = \Lambda_{12}/\cubar$, $L_{21} = \Lambda_{21}/\cubar$, and $L_{22} = \Lambda_{22}/\cubar^2$,
\begin{equation}
	\left[\begin{array}{c} Q\\ \Ju/\cubar - \Qv \end{array} \right] 
	= \left[\begin{array}{cc}L_{11} & L_{12} \\ 
	L_{21} & L_{22} \\  \end{array} \right]
	\left[\begin{array}{c} -\Delta P \\ -\Delta \Pi\end{array}\right]\,,
 \label{eqn:onsager_Pi}
\end{equation}
where $\Lambda_{12} = \Lambda_{21}$ and $L_{12} = L_{21}$ by the Onsager reciprocal relations,\cite{onsager1931,degroot1984} $\Qv = \Jv/\cvbar = \Nvdot/\cvbar$ is the volumetric solvent flux,  and $\cubar$ and $\cvbar$ are the average bulk solute and solvent concentrations, respectively, which we have calculated as the average of the concentrations in the upper and lower control regions, i.e. $\bar{c}_i = (c_i^+ +c_i^-)/2$. Note that a number of previous studies on concentration-gradient-driven transport\cite{gross1968,ajdari2006,huang2008c,marbach2017a,yoshida2017,monet2023,lee2017a} have not distinguished between the total volumetric solution flux $Q$ and volumetric solvent flux $\Qv$ in Eqs.~\eqref{eqn:onsager_mu} or \eqref{eqn:onsager_Pi}, although this distinction is clear in the derivation by de Groot and Mazur\cite{degroot1984} and in equations used in other studies.\cite{fair1971,kedem1958} For the dilute solutions investigated in most of these studies, this distinction is not important, since $Q \approx \Qv$ in this regime, but $Q$ and $\Qv$ can differ significantly at high solute concentrations such as studied here, especially when the solute and solvent fluxes are in opposite directions.

From the definitions of the transport coefficients in Eqs.~\eqref{eqn:kappaDO} and~\eqref{eqn:onsager_Pi}, $L_{12} = \kappaDO/(\kB T)$, and given that $T = \epsilon/\kB$ in all our simulations, this means that $L_{12}$ and $\kappaDO$ have the same numerical value in reduced LJ units in this study. For two of the systems for which $L_{12}$ was measured in Fig.~\ref{fig:linear_response}, we also measured $L_{21}$ using the definition in Eq.~\eqref{eqn:onsager_Pi} in simulations of pressure-driven flow in the absence of a transmembrane concentration difference for several pressure differences using the NEMD algorithm in Ref.~\citenum{zhu2002}. These results are presented in Fig.~\ref{fig:linear_response}, and show that the $L_{12} = L_{21}$ reciprocial relation is verified, at least for the lowest applied pressures, with the consistency between our algorithm and the previously established and widely applied method in Ref.~\citenum{zhu2002} demonstrating the validity of our method for quantifying non-equilibrium concentration-gradient-driven transport. 

Fig.~\ref{fig:linear_response} also shows that the pressure-driven flow simulations deviate from linear response at much lower values of $\Delta P$ than the $\Delta \Pi$ values at which deviations occur in the concentration-gradient-driven flow simulations. This means that much longer simulations were required to obtain roughly comparable statistical uncertainties in the linear-response regime for the pressure-driven flow simulations compared with the concentration-gradient-driven flow simulations, highlighting the computational benefits of directly measuring the diffusioosmotic mobility in NEMD simulations with an applied concentration gradient, at least for 2D membrane systems.

\subsection{Comparison of simulation vs theory}

We have previously derived a theory of fluid transport through a circular pore in an infinitesimally thin planar membrane due to a transmembrane concentration difference\cite{rankin2019} by solving the continuum hydrodynamic (Stokes, advection--diffusion, and continuity) equations for low-Reynolds-number steady-state flow of a dilute solution of an incompressible Newtonian fluid, under the assumptions that solute diffusion dominates solute advection (P\'eclet number $\mathrm{Pe} \ll 1$) and that the effective solute--membrane interaction potential $U$ is small compared with the thermal energy $\kB T$. This theory is straightforwardly generalized to arbitrary solute concentrations, by analogy with a related theory derived for concentration-gradient-driven fluid transport parallel to a planar surface at high solute concentration:\cite{marbach2017a,yoshida2017}  the equations derived in Ref.~\citenum{rankin2019} for the transport coefficients quantifying the fluid fluxes at low concentration apply at high concentration, while the distinction between low and high solute concentrations is manifested in the concentration dependence of the osmotic pressure driving force in Eqs.~\eqref{eqn:kappaDO} and \eqref{eqn:Ps}. Thus, the diffusioosmotic mobility $\kappaDO$ quantifying the total solution flux is\cite{rankin2019}
\begin{equation}
	\kappaDO = \frac{2\kB T a^3}{\pi\eta} \int_0^1 \rmd\zeta\,\zeta^2\int_0^\infty \rmd\nu\,\left(\frac{e^{-U/(\kB T)}-1}{1+\nu^2}\right),
	\label{eqn:kappaDO_th}
\end{equation}
and the solute permeance $\Ps$ quantifying the solute flux (evaluated at the pore mouth at $z = 0$) is\cite{rankin2019}
\begin{equation}
	\Ps = 2D \int_0^a \rmd r \frac{ re^{-U/(\kB T)}}{\sqrt{a^2-r^2}},
	\label{eqn:Ps_th}
\end{equation} 
where $a$ is the pore radius, $\eta$ is the solution shear viscosity, $D$ is the solute diffusivity, and the  oblate--spheroidal coordinates $\zeta$ and $\nu$ are defined in terms of the radial and axial coordinates by $r = a \sqrt{(1+\nu^2)(1-\zeta^2)}$ and $z = a\nu \zeta$, respectively. The effective solute--membrane interaction potential $U$, which includes contributions both from direct solute--membrane interactions and indirect solvent-mediated interactions, is defined by\cite{rankin2019}  
\begin{equation}
	c(\zeta,\nu) \equiv \cubulk(\zeta,\nu) e^{-U(\zeta,\nu)/(\kB T)},
	\label{eqn:cneq}
\end{equation}
where $c(\zeta,\nu)$ is the solute concentration distribution and $\cubulk(\zeta,\nu)$ is a hypothetical solute concentration distribution for the same boundary conditions but with $U =0$. 

\begin{figure}
	\includegraphics{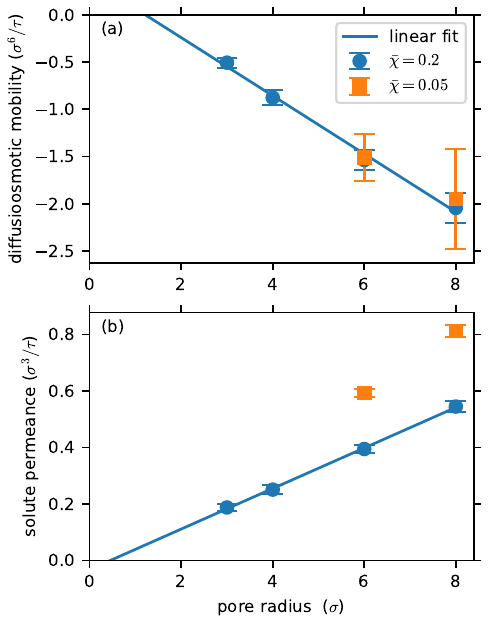}
	\caption{(a) Diffusioosmotic mobility and (b) solute permeance versus pore radius for simulations with $\curatio = 20$ and $\Delta \Ptarg = 0$ for systems with $\epsuw = 0.5\epsilon, \siguw = 0.8\sigma$, $a = 6\sigma$, and $\bar{\chi} = 0.2$ (circles) or 0.5 (squares). The solid lines are linear fits to the data for $\bar{\chi} = 0.2$.}
	\label{fig:kappaDO_Ps_vs_radius-chi}
\end{figure}

The theory yields simple scaling relationships for the transport coefficients as a function of the pore radius $a$ and strength $\epsilon$ and range $\lambda$ of $U$ in the limits of weak interactions ($\epsilon \ll \kB T $) and a small ($\lambda \ll a$) or large   ($\lambda \gg a$) interaction range relative to the pore size.\cite{rankin2019} Although none of our simulations correspond strictly to the $\lambda \ll a$ or $\lambda \gg a$ limits, $\lambda$ is significantly smaller than $a$ for all but the smallest pore studied. Our simulation results are consistent with the predicted scaling in the $\lambda \ll a$ limit, in which $\kappaDO$ and $\Ps$ are both expected to be proportional to the pore radius $a$,\cite{rankin2019} as shown in Fig.~\ref{fig:kappaDO_Ps_vs_radius-chi}. Fig.~\ref{fig:kappaDO_Ps_vs_radius-chi} also shows that $\kappaDO$ is approximately independent of the average solute mole fraction for the simulated conditions, whereas $\Ps$ depends significantly on the average solute mole fraction. The behavior of $\Ps$ appears to be due to a subtle interplay of the opposing diffusive and advective solute fluxes for the systems in Fig.~\ref{fig:kappaDO_Ps_vs_radius-chi}, which is not captured by the theory as it assumes that solute advection is negligible.  

The dependence of $\kappaDO$ and $\Ps$ on the solute--membrane interaction strength and range parameters, $\epsuw$ and $\siguw$, are given in in the supplementary material in Figs.~\ref{fig:kappaDO_Ps_vs_eps} and \ref{fig:kappaDO_Ps_vs_sig}, respectively. It should be noted that the strength and range of $U$ is not simply proportional to these parameters, due to the complex many-body contributions to the effective interactions. Interestingly, Fig.~\ref{fig:kappaDO_Ps_vs_eps} shows a linear dependence of $\kappaDO$ and $\Ps$ on $\epsuw$ for fixed $\siguw$, which is consistent with the predicted scaling with the effective solute--membrane interaction strength $\epsilon$ for $\epsilon \ll \kB T$, even though the direct solute--membrane interactions are certainly not weak in all cases. The dependence of $\kappaDO$ and $\Ps$ on $\siguw$ for fixed $\epsuw$ shown in Fig.~\ref{fig:kappaDO_Ps_vs_sig} is more complex, in part because $\siguw$ controls not only the range but also the strength of the direct solute--membrane interactions, since a given solute particle interacts with more membrane particles as $\siguw$ increases; thus, $\Ps$ varies non-monotonically while $\kappaDO$ even changes sign with increasing $\siguw$.

We have used Eqs.~\eqref{eqn:kappaDO_th}--\eqref{eqn:cneq} to predict $\kappaDO$ and $\Ps$ for all the simulated systems without using any information from the NEMD simulations. We used analytical equations fitted to equilibrium MD simulation data for the diffusivity and shear viscosity of the LJ fluid over a wide range of density and temperature for the same interaction cutoff distance in our simulations \cite{rowley1997} to obtain $D = 0.0697 \sigma^2/\tau$ and $\eta = 1.84\epsilon\tau/\sigma^3$ for the temperature $T = \epsilon/\kB$ and total bulk density $\rho \approx 0.787 \sigma^{-3}$ in all the simulations. To obtain $U$ from Eq.~\eqref{eqn:cneq}, we used the solute concentration profile from an equilibrium MD simulation with the same solute--membrane interaction parameters, pore radius, and average solute mole fraction as the NEMD simulation for which the transport coefficients were being predicted. In this case, $\cubulk(\zeta,\nu)$ in Eq.~\eqref{eqn:cneq} is a constant and equal to the bulk solute concentration in the equilibrium simulation. Distributions of the solute concentration and total fluid density in all the equilibrium simulations, in one dimension (1D) as a function of the axial ($z$) or radial ($r$) coordinate and in two dimensions (2D) as a function of both $r$ and $z$, are given in Figs.~\ref{fig:cuz_eq}--\ref{fig:crz_eq_chi0.05} of the supplementary material. The integrals in Eqs.~\eqref{eqn:kappaDO_th} and \eqref{eqn:Ps_th} were computed using the \texttt{quad} and \texttt{nquad} functions, respectively, in the SciPy Python package and 2D solute concentration distributions were interpolated using a bivariate spline with the \texttt{RectBivariateSpline} SciPy function. 

The pore radius $a$ in Eqs.~\eqref{eqn:kappaDO_th} and \eqref{eqn:Ps_th} corresponds to where the hydrodynamic boundary conditions are applied in the continuum theory and does not necessarily correspond to the definition of the pore radius used up to this point, which was the distance from the pore center within which the centers of solid atoms were absent. The Gibbs dividing surface from equilibrium MD simulations has previously been shown previously to describe the hydrodynamic boundary position in NEMD simulations of fluid flow accurately,\cite{herrero2019} and so we have used this prescription to define an effective pore radius $\ah$ to replace the actual pore radius $a$ in Eqs.~\eqref{eqn:kappaDO_th} and \eqref{eqn:Ps_th}, given by    
\begin{equation}
	\int_0^{\ah} r \left[ \rhobulk - \rho(r,z=0)\right] \, \rmd r = \int_{\ah}^\infty r \rho(r,z = 0) \, \rmd r, 
	\label{eqn:ah}
\end{equation}
where the total fluid density $\rho(r,z=0)$ in the plane of the membrane pore at $z =0$ was obtained from the 2D equilibrium distribution $\rho(r,z)$ by the same bivariate spline interpolation described above for the solute concentration distribution. (In practice the second integral was calculated up to finite value of $r$ beyond which the total fluid density $\rho(r,z = 0)$ was zero.) As shown in Fig.~\ref{fig:a_h_vs_a} of the supplementary material, $\ah$ is within a few percent of $a$, so either $a$ or $\ah$ could be used in Eqs.~\eqref{eqn:kappaDO_th} and \eqref{eqn:Ps_th} with little difference. (The dependence of $\ah$ on the solute--membrane interaction strength and range parameters, $\epsuw$ and $\siguw$, for fixed $a$ are also given in the supplementary material in Figs.~\ref{fig:a_h_vs_eps} and \ref{fig:a_h_vs_sig}.)

\begin{figure}
	\includegraphics{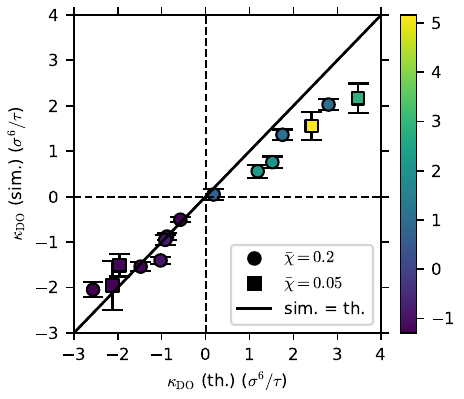}
	\caption{Diffusioosmotic mobility from simulations vs theory for all systems except for those with the strongest attractive solute--membrane interactions ($\epsuw = 1.5\epsilon$, $\siguw = 1.5\sigma$) for $\curatio=20$. Symbols are colored by the surface solute excess $\Gamma$ (units: $\sigma$) and different symbol shapes distinguish high ($\bar{\chi} = 0.2$, circles) and low (($\bar{\chi} = 0.05$, squares) average solute mole fractions. The simulation and theory values are equal along the solid line.}
	\label{fig:kappaDO_sim_vs_th_Gamma}
\end{figure}

Fig.~\ref{fig:kappaDO_sim_vs_th_Gamma} compares the diffusioosmotic mobility $\kappaDO$ from the simulations with that calculated from the theory for all the simulated systems (which includes variations in pore radii, average solute mole fraction, and the strength and range of solute--membrane interactions)  except for those with the strongest attractive solute--membrane interactions ($\epsuw = 1.5\epsilon$, $\siguw = 1.5\sigma$). The latter data are not included because they occur on a very different scale from the rest of the data, making visualization on the same figure difficult, and because the assumption in the theory of weak solute--membrane interactions certainly breaks down for these systems. A comparison of all the simulated systems in the linear-response regime is given in Fig.~\ref{fig:kappaDO_sim_vs_th_Gamma_all} of the supplementary material. The simulations and theory are also compared in Fig.~\ref{fig:kappaDO_sim_vs_th_Gamma_a} of the supplementary material for calculations using the actual pore radius $a$ instead of the effective pore radius $\ah$ in the theory, showing that this choice makes little quantitative difference. 

There is good quantitative agreement between the theory and simulations for most of the data in Fig.~\ref{fig:kappaDO_sim_vs_th_Gamma} , although deviations are evident for larger magnitudes of $\kappaDO$, especially for positive $\kappaDO$. Discrepancies between the theory and simulations are not entirely surprising, given the number of approximations made in the theory, namely weak effective solute--membrane interactions, a small P\'eclet number, and an infinitesimally thin planar membrane. The discrepancies appear to be largely due to the break down of the assumption of weak effective solute--membrane interactions, as indicated by the correlation between the deviation of the theory from the simulation results and the surface solute excess $\Gamma$ used to color the data points in Fig.~\ref{fig:kappaDO_sim_vs_th_Gamma}, which quantities the degree of adsorption ($\Gamma > 0$) or depletion ($\Gamma < 0$) of the solute at the membrane surface relative to the bulk. $\Gamma$ in Fig.~\ref{fig:kappaDO_sim_vs_th_Gamma} was calculated from  the solute concentration profile $\cu(z)$ perpendicular to a membrane containing no pore in an equilibrium MD simulation of a system with otherwise identical fluid and membrane properties to the NEMD simulation using
\begin{equation}
	\Gamma = \int_0^{\infty}  \left(\frac{\cu(z)}{\cubulk} -1 \right)\, \rmd z,
	\label{eqn:Gamma}
\end{equation}
where, in practice, the upper integration limit in Eq.~\eqref{eqn:Gamma} was taken to be the maximum value of $z$ in the simulation box (this choice was not crucial as $\cu(z) \rightarrow \cubulk$ within the simulation box). 

On the other hand, there does not appear to be a clear correlation between the discrepencies between the theory and simulation for $\kappaDO$ and other potentially relevant parameters such as the  P\'eclet number $\mathrm{Pe}$, pore radius $a$, or solute--membrane interaction strength and range parameters, $\epsuw$ and $\siguw$, as shown in Figs.~\ref{fig:kappaDO_sim_vs_th_Pe}--\ref{fig:kappaDO_sim_vs_th_sig} of the supplementary material, in which the data points have been colored by the value of these parameters. If anything, the discrepancies show the opposite trend vs $\mathrm{Pe}$ to that expected, with the deviation between theory and simulation increasing with decreasing  $\mathrm{Pe}$.  We estimated $\mathrm{Pe}$, which measures the relative magnitude of solute advection to solute diffusion, by
\begin{equation}
	\mathrm{Pe} = \left\lvert\frac{\cubar\Qv}{\Ju - \cubar\Qv}\right\rvert,
	\label{eqn:peclet}
\end{equation}
where we have used the solvent volumetric flux $\Qv$ instead of the total volumetric flux $Q$ to quantify the advective solute flux because the total flux $Q$ includes the diffusive component. The assumption of an infinitesimally thin membrane is expected to become less accurate as the aspect ratio of the pore decreases, which corresponds to decreasing pore radius. As noted above, although the effective solute--membrane interaction strength depends on $\epsuw$ and $\siguw$, the dependence on either parameter is not straightforward, and thus the lack of correlation of the discrepancies in $\kappaDO$ with either parameter is not unexpected.

\begin{figure}
	\includegraphics{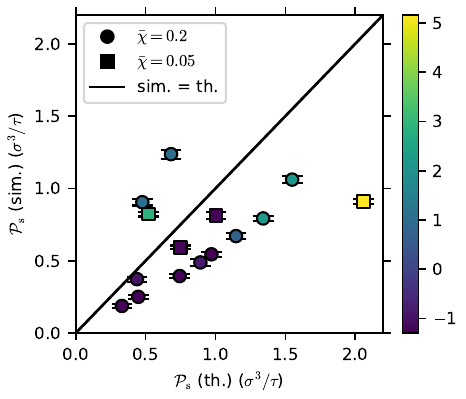}
	\caption{Solute permeance from simulations vs theory for all systems for $\curatio = 20$ except for those with the strongest attractive solute--membrane interactions ($\epsuw = 1.5\epsilon$, $\siguw = 1.5\sigma$). Symbols are colored by the surface solute excess $\Gamma$ (units: $\sigma$) and different symbol shapes distinguish high ($\bar{\chi} = 0.2$, circles) and low ($\bar{\chi} = 0.05$, squares) average solute mole fractions. The simulation and theory values are equal along the solid line.}
	\label{fig:Ps_sim_vs_th_Gamma}
\end{figure}

\begin{figure}
	\includegraphics{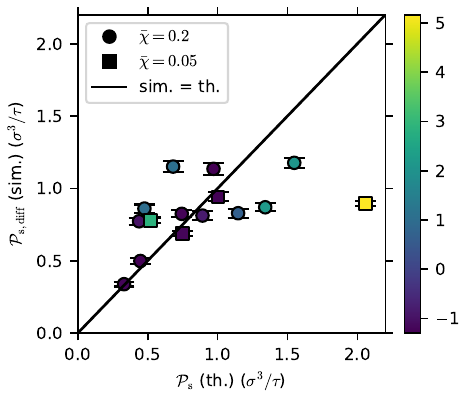}
	\caption{Solute permeance from simulations (calculated from diffusive flux only) vs theory for all systems for $\curatio = 20$ except for those with the strongest attractive solute--membrane interactions ($\epsuw = 1.5\epsilon$, $\siguw = 1.5\sigma$). Symbols are colored by the surface solute excess $\Gamma$ (units: $\sigma$)  and different symbol shapes distinguish high ($\bar{\chi} = 0.2$, circles) and low ($\bar{\chi} = 0.05$, squares) average solute mole fractions. The simulation and theory values are equal along the solid line.}
	\label{fig:Psdiff_sim_vs_th_Gamma}
\end{figure}

Fig.~\ref{fig:Ps_sim_vs_th_Gamma} shows a similar comparison between theory and simulation to Fig.~\ref{fig:kappaDO_sim_vs_th_Gamma}, but for the solute permeance $\Ps$. Since solute advection is assumed to be neglible ($\mathrm{Pe} \ll 1$) in the theory, but is clearly is significant in many of the simulations, for which $0.02 \lesssim  \mathrm{Pe} \lesssim 0.5$, in Fig.~\ref{fig:Psdiff_sim_vs_th_Gamma} we have also compared $\Ps$ from the theory with the solute permeance from the simulations calculated from the diffusive flux only, which we define as
\begin{equation}
	\Psdiff \equiv -\frac{\left(\Ju - \cubar\Qv\right)}{\Delta \Pi /(\kB T)}.
	\label{eqn:Psdiff}
\end{equation}
As with Fig.~\ref{fig:kappaDO_sim_vs_th_Gamma}, we have excluded data for the strongest attractive solute--membrane interactions ($\epsuw = 1.5\epsilon$, $\siguw = 1.5\sigma$) simulated from Figs.~\ref{fig:Ps_sim_vs_th_Gamma} and \ref{fig:Psdiff_sim_vs_th_Gamma}, but the corresponding plots containing all the simulation data can be found in Figs.~\ref{fig:Ps_sim_vs_th_Gamma_all} and \ref{fig:Psdiff_sim_vs_th_Gamma_all}, respectively, in the supplementary material. As for $\kappaDO$ the use of the actual pore radius $a$ or effective pore radius $\ah$ in the theory makes little difference, as shown in Fig.~\ref{fig:Ps_sim_vs_th_Gamma_a} in the supplementary material.  

The agreement between the theory and simulations is not as good for the solute permeance as for $\kappaDO$, but the agreement appears to improve by excluding the advective flux from the simulation definition of the solute permeance. As for $\kappaDO$, the discrepancies between the theory and simulation appear to be most strongly correlated with the strength of the effective solute--membrane interactions as quantified by the surface solute excess used to color the symbols in Figs.~\ref{fig:Ps_sim_vs_th_Gamma} and \ref{fig:Psdiff_sim_vs_th_Gamma} (similar plots for all the simulated systems colored by the P\'eclet number $\mathrm{Pe}$, pore radius $a$, or solute--membrane interaction strength and range parameters, $\epsuw$ and $\siguw$, are given in Figs.~\ref{fig:Ps_sim_vs_th_Pe_all}--\ref{fig:Psdiff_sim_vs_th_sig_all} in the supplementary material).

\section{Conclusion}

We have developed a constrained concentration- and pressure-difference algorithm for non-equilibrium molecular dynamics simulations of steady-state fluid transport driven by concentration and/or pressure differences across a porous membrane in a system with periodic boundary conditions. Our algorithm adapts a previous algorithm by \citeauthor{khalili-araghi2013},\cite{khalili-araghi2013} which controls the transmembrane concentration difference by applying an external force to solute particles in a transition region far from the membrane, by also applying an external force to solvent particles in the transition region to control the transmembrane pressure difference. Applying this algorithm to a model system comprising a binary Lennard--Jones liquid mixture and a 2D Lennard--Jones membrane containing a circular pore, we have simulated steady-state concentration-gradient-driven fluid tranport across a 2D membrane with molecular resolution for the first time, enabling accurate quantification of the solution and solute fluxes due to a given applied concentration difference. We have shown that the application of the pressure-difference constraint has a significant effect on the solute concentration distribution across the membrane and the steady-state solution flux due to a transmembrane concentration difference for both low and high average solute concentrations, although the solute flux is less affected by the pressure-difference constraint at low solute concentrations. We have also shown that the solution flux due to an applied concentration difference generated by our algorithm is consistent with Onsager reciprocity  in the linear-response regime by comparison with fluid fluxes due to an applied pressure difference. Furthermore, we have shown that directly simulating a transmembrane concentration difference is far more efficient for quantifying the concentration-gradient-driven solution flux in the 2D membrane systems studied than the indirect approach of applying the Onsager reciprocal relations to pressure-driven flow simulations. This is because only very small pressure differences can be applied before the pressure-driven fluid fluxes deviate from linear behavior.  Finally, we have shown that our recently developed theory of concentration-gradient-driven flow across a 2D membrane,\cite{rankin2019} although derived for a continuum fluid model, gives reasonably good quantitative agreement with the molecular simulations, especially for the total fluid flux, demonstrating its utility for quantifying the fluid transport even in molecular systems. Nevertheless, deviations from the simulation results are evident particularly for strong solute--membrane interactions, for which the assumptions of the theory break down.

\section*{Supplementary Material}

The supplementary material contains details of the parameters and properties of all the simulated systems, additional non-equilibrium simulation results (concentration, density, and pressure distributions; comparison of transmembrane osmotic pressure and hydrostatic pressure differences calculated by different methods; verification of linear response for the solute permeance; verification of the independence of measured transport coefficients of system size and transition region width; additional plots of transport coefficients vs system parameters; additional comparisons of transport coefficients obtained from simulation and theory), and solute concentration and total density distributions from equilibrium simulations.

\begin{acknowledgments}
This work was supported by the Australian Research Council under the Discovery Projects funding scheme (Grant No. DP210102155). This research was undertaken with the assistance of resources and services from the National Computational Infrastructure (NCI), which is supported by the Australian Government, and from the University of Adelaide’s Phoenix High-Performance Computing service.
\end{acknowledgments}

\section*{Author Declarations}

\subsection*{Conflict of Interest}

The authors have no conflicts to disclose.

\subsection*{Author Contributions}

Daniel J. Rankin: Methodology (equal); Investigation (supporting); Formal Analysis (supporting); Writing -- original draft (equal); Writing -- review \& editing (supporting).
David M. Huang: Conceptualization (lead); Methodology (equal); Investigation (lead); Data curation (lead); Formal Analysis (lead); Funding acquisition (lead); Project administration (lead); Supervision (lead); Writing -- original draft (equal); Writing -- review \& editing (lead)

\section*{Data Availability}

Moltemplate\cite{jewett2021} input scripts for creating the initial simulation configurations and sample LAMMPS\cite{plimpton1995,brown2011} input scripts for running each type of MD simulation in this study can be found at https://doi.org/10.25909/17139593. Other data that support the findings of this study are available from the corresponding author upon reasonable request.

% \section{}
%\label{}
% \subsection{}
% \subsubsection{}

% If in two-column mode, this environment will change to single-column format so that long equations can be displayed. 
%\begin{widetext}
%$$\mbox{put long equation here}$$
%\end{widetext}

% Figures should be put into the text as floats. 
% Use the graphics or graphicx packages (distributed with LaTeX2e).
% Put the label that you will use with \ref{} command in the braces of the \label{} command.
%
% \begin{figure}
% \end{figure}

% \begin{table}
% \end{tabular}
% \end{table}

\section*{References}
\bibliography{lj-2d-cgrad}

% add cross references to external document
\makeatletter\@input{lj-2d-cgrad_si-aux.tex}\makeatother
\end{document}

\begin{center}
	\Large{\textbf{Electronic Supplementary Information}} \\
	\vspace{0.5cm}
	\Large{\textbf{Non-equilibrium molecular dynamics of steady-state fluid transport through a 2D membrane driven by a concentration gradient}} \\ 
	\vspace{0.5cm}
	\large{Daniel J. Rankin and David M. Huang} \\
	\large{\textit{Department of Chemistry, School of Physics, Chemistry and Earth Sciences, The University of Adelaide, SA 5005, Australia}} \\
	\end{center}

	% \tableofcontents
	% \clearpage

\section{Systems}

\begin{table}[htb!]
	\centering
	\caption{System sizes in equilibrium simulations:   $L_{x/y} =$ system $x$ and $y$ dimensions, $L_z =$ system $z$ dimension, $a =$ pore radius, $\bar{\chi} =$ average solute mole fraction, $\Nu=$ number of solute particles, $\Nv=$ number of solvent particles, and $\Nw=$ number of solid particles.}
	\label{tbl:eq_sys_size}
	\begin{tabular}{ccccccccc}
	   \hline
		$a$ ($\sigma$) & $\epsuw$ ($\epsilon$) & 	$\siguw$ ($\sigma$) & $\bar{\chi}$ & $L_{x/y}/\sqrt{2}$ ($\sigma$) & $L_z$($\sigma$) &	$\Nu$ & $\Nv$ & $\Nw$ 	\\
	   \hline
	   0 &                 0.5 &               0.8 &   0.20 &  	50 & 101.44 & 	\num{79148} & \num{316852}	& \num{5000}  \\
	   0 &                 0.5 &               1.2 &   0.20 &   50 & 101.44 & 	\num{79006} & \num{316994}	& \num{5000}  \\
	   0 &                 0.5 &               1.5 &   0.20 &   50 & 101.44 & 	\num{79233} & \num{316767}	& \num{5000}  \\
	   0 &                 0.8 &               0.8 &   0.20 &   50 & 101.44 & 	\num{78796} & \num{317204}	& \num{5000}  \\
	   0 &                 1.2 &               0.8 &   0.20 &   50 & 101.44 & 	\num{79406} & \num{316594}	& \num{5000}  \\
	   0 &                 1.5 &               0.8 &   0.20 &   50 & 101.44 & 	\num{79236} & \num{316764}	& \num{5000}  \\
	   0 &                 1.5 &               1.5 &   0.20 &   50 & 101.44 & 	\num{78835} & \num{317165}	& \num{5000}  \\
	   0 &                 0.5 &               0.8 &   0.05 &   50 & 101.44 & 	\num{19695} & \num{376305}	& \num{5000}  \\
	   0 &                 0.5 &               1.5 &   0.05 &   50 & 101.44 & 	\num{19505} & \num{376495}	& \num{5000}  \\
	   0 &                 1.5 &               0.8 &   0.05 &   50 & 101.44 & 	\num{19799} & \num{376201}	& \num{5000}  \\
	   0 &                 1.5 &               1.5 &   0.05 &   50 & 101.44 & 	\num{19824} & \num{376176}	& \num{5000}  \\
	   3 &                 0.5 &               0.8 &   0.20 &   50 & 101.44 & 	\num{79287} & \num{316713}	& \num{4971}  \\
	   4 &                 0.5 &               0.8 &   0.20 &   50 & 101.44 & 	\num{79277} & \num{316723}	& \num{4951}  \\
	   6 &                 0.5 &               0.8 &   0.20 &   50 & 101.44 & 	\num{79121} & \num{316879}	& \num{4887}  \\
	   8 &                 0.5 &               0.8 &   0.20 &   50 & 101.44 & 	\num{79504} & \num{316496}	& \num{4803}  \\
	   6 &                 0.5 &               1.2 &   0.20 &   50 & 101.44 & 	\num{78975} & \num{317025}	& \num{4887}  \\
	   6 &                 0.5 &               1.5 &   0.20 &   50 & 101.44 & 	\num{79046} & \num{316954}	& \num{4887}  \\
	   8 &                 0.5 &               1.5 &   0.20 &   50 & 101.44 & 	\num{79245} & \num{316755}	& \num{4803}  \\
	   6 &                 0.8 &               0.8 &   0.20 &   50 & 101.44 & 	\num{79058} & \num{316942}	& \num{4887}  \\
	   6 &                 1.2 &               0.8 &   0.20 &   50 & 101.44 & 	\num{79192} & \num{316808}	& \num{4887}  \\
	   6 &                 1.5 &               0.8 &   0.20 &   50 & 101.44 & 	\num{78742} & \num{317258}	& \num{4887}  \\
	   8 &                 1.5 &               0.8 &   0.20 &   50 & 101.44 & 	\num{79405} & \num{316595}	& \num{4803}  \\
	   6 &                 1.5 &               1.5 &   0.20 &   50 & 101.44 & 	\num{79394} & \num{316606}	& \num{4887}  \\
	   6 &                 0.5 &               0.8 &   0.05 &   50 & 101.44 & 	\num{19766} & \num{376234}	& \num{4887}  \\
	   8 &                 0.5 &               0.8 &   0.05 &   50 & 101.44 & 	\num{19957} & \num{376043}	& \num{4803}  \\
	   6 &                 0.5 &               1.5 &   0.05 &   50 & 101.44 & 	\num{20082} & \num{375918}	& \num{4887}  \\
	   6 &                 1.5 &               0.8 &   0.05 &   50 & 101.44 & 	\num{19752} & \num{376248}	& \num{4887}  \\
	   6 &                 1.5 &               1.5 &   0.05 &   50 & 101.44 & 	\num{19707} & \num{376293}	& \num{4887}  \\
	   \hline
	\end{tabular}
   \end{table}

\begin{table}[htb!]
	\centering
	\caption{System sizes in non-equilibrium simulations: $a =$ pore radius, $\bar{\chi} =$ average solute mole fraction, $L_{x/y} =$ system $x$ and $y$ dimensions, $L_z =$ system $z$ dimension, $\Nu=$ number of solute particles, $\Nv=$ number of solvent particles, and $\Nw=$ number of solid particles. }
	\label{tbl:neq_sys_size}
	\begin{tabular}{ccccccc}
	   \hline
		$a$ ($\sigma$)	& $\bar{\chi}$	& $L_{x/y}/\sqrt{2}$ ($\sigma$) 	& $L_z$ ($\sigma$)	& $\Nu$ 		& $\Nv$ 		& $\Nw$ 	\\
	   \hline
		3				&	0.20	& 50 			& 101.44 		&\num{78712}	& \num{317288} 	& \num{4971}	\\
	   	4				&	0.20	& 50 			& 101.44 		& \num{79233}	& \num{316767} 	& \num{4951}	\\
	    6				&	0.20	& 50 			& 101.44 		&\num{79610}	& \num{316390} 	& \num{4887}	\\
	    8				&	0.20	& 50 			& 101.44 		&\num{79610}	& \num{316390} 	& \num{4803}	\\
	    6				&	0.05	& 50 			& 101.44 		&\num{19673}	& \num{376327} 	& \num{4887}	\\
	    8				&	0.05	& 50 			& 101.44 		&\num{19950}	& \num{376050} 	& \num{4803}	\\
	    8				& 	0.20	& 80			& 161.81		&\num{324193}	& \num{1297823}	& \num{12603}	 \\
	   \hline
	\end{tabular}
\end{table}

\begin{table}[htb!]
	\centering
 \caption{Systems properties in equilibrium simulations: $a =$ pore radius, $\epsuw =$ solute--membrane interaction strength parameter, $\siguw =$ solute--membrane interaction range parameter, $\bar{\chi} =$ average solute mole fraction, $\ttotal =$ total simulation duration, and $\ah =$ effective pore radius (calculated from Gibbs dividing surface). The bulk solute concentration $\cubulk$, bulk total fluid density $\rhobulk$, and bulk pressure $\Pbulk$ were calculated by averaging the axial profiles of the solute concentration ($\cu(z)$), total fluid density ($\rho(z)$) and pressure ($P(z)$) for distances $> 30\sigma$ from the membrane, where the profiles were constant.}
 \label{tbl:eq_sys}
 \begin{tabular}{ccccccccc}
	\hline
	$a$ ($\sigma$) & $\epsuw$~($\epsilon$) & $\siguw$ ($\sigma$) & $\bar{\chi}$ & $\ttotal$ ($\times 10^4~\tau$) & $\ah$ ($\sigma$) & $\rhobulk$ ($\sigma^{-3}$) & $\cubulk$ ($\sigma^{-3}$) & $\Pbulk$ ($\epsilon/\sigma^3$) \\
	\hline
	0 &                   0.5 &                 0.8 &   0.20 &                           13.0 &              -- &                     0.787 &                      0.160 &                          0.996 \\
	0 &                   0.5 &                 1.2 &   0.20 &                           13.0 &              -- &                     0.787 &                      0.159 &                          1.003 \\
	0 &                   0.5 &                 1.5 &   0.20 &                            9.0 &              -- &                     0.786 &                      0.153 &                          0.988 \\
	0 &                   0.8 &                 0.8 &   0.20 &                           13.0 &              -- &                     0.786 &                      0.158 &                          0.986 \\
	0 &                   1.2 &                 0.8 &   0.20 &                           13.0 &              -- &                     0.784 &                      0.154 &                          0.957 \\
	0 &                   1.5 &                 0.8 &   0.20 &                           13.0 &              -- &                     0.782 &                      0.150 &                          0.929 \\
	0 &                   1.5 &                 1.5 &   0.20 &                            9.0 &              -- &                     0.784 &                      0.139 &                          0.961 \\
	0 &                   0.5 &                 0.8 &   0.05 &                           13.0 &              -- &                     0.787 &                      0.040 &                          0.999 \\
	0 &                   0.5 &                 1.5 &   0.05 &                            4.0 &              -- &                     0.787 &                      0.036 &                          0.995 \\
	0 &                   1.5 &                 0.8 &   0.05 &                            4.0 &              -- &                     0.785 &                      0.036 &                          0.968 \\
	0 &                   1.5 &                 1.5 &   0.05 &                            9.0 &              -- &                     0.785 &                      0.023 &                          0.963 \\
	3 &                   0.5 &                 0.8 &   0.20 &                            8.0 &             2.95 &                     0.787 &                      0.160 &                          0.995 \\
	4 &                   0.5 &                 0.8 &   0.20 &                            8.0 &             3.88 &                     0.787 &                      0.160 &                          0.995 \\
	6 &                   0.5 &                 0.8 &   0.20 &                            8.0 &             5.92 &                     0.787 &                      0.160 &                          0.994 \\
	8 &                   0.5 &                 0.8 &   0.20 &                            8.0 &             7.87 &                     0.787 &                      0.161 &                          0.992 \\
	6 &                   0.5 &                 1.2 &   0.20 &                            8.0 &             5.87 &                     0.787 &                      0.159 &                          1.001 \\
	6 &                   0.5 &                 1.5 &   0.20 &                            8.0 &             5.87 &                     0.786 &                      0.153 &                          0.986 \\
	8 &                   0.5 &                 1.5 &   0.20 &                            8.0 &             7.82 &                     0.786 &                      0.153 &                          0.985 \\
	6 &                   0.8 &                 0.8 &   0.20 &                            8.0 &             5.96 &                     0.786 &                      0.159 &                          0.984 \\
	6 &                   1.2 &                 0.8 &   0.20 &                            8.0 &             6.01 &                     0.784 &                      0.154 &                          0.956 \\
	6 &                   1.5 &                 0.8 &   0.20 &                            8.0 &             6.03 &                     0.782 &                      0.149 &                          0.929 \\
	8 &                   1.5 &                 0.8 &   0.20 &                            8.0 &             7.97 &                     0.782 &                      0.150 &                          0.929 \\
	6 &                   1.5 &                 1.5 &   0.20 &                            8.0 &             5.92 &                     0.784 &                      0.140 &                          0.960 \\
	6 &                   0.5 &                 0.8 &   0.05 &                            8.0 &             5.92 &                     0.787 &                      0.040 &                          0.997 \\
	8 &                   0.5 &                 0.8 &   0.05 &                            8.0 &             7.82 &                     0.787 &                      0.040 &                          0.995 \\
	6 &                   0.5 &                 1.5 &   0.05 &                            8.0 &             5.87 &                     0.787 &                      0.038 &                          0.993 \\
	6 &                   1.5 &                 0.8 &   0.05 &                            8.0 &             5.96 &                     0.785 &                      0.035 &                          0.966 \\
	6 &                   1.5 &                 1.5 &   0.05 &                            8.0 &             5.92 &                     0.784 &                      0.023 &                          0.962 \\
 \hline
 \end{tabular}
\end{table}

\begin{table}[htb!]
	\centering
	\caption{Systems properties in non-equilibrium concentration-gradient-driven flow simulations: $d =$ width of transition region, $L_{x/y} =$ system $x$ and $y$ dimensions, $a =$ pore radius, $\epsuw =$ solute--membrane interaction strength, $\siguw =$ solute--membrane interaction range, $\bar{\chi} =$ average solute mole fraction, $\cutarg^+/\cutarg^- =$ target concentration ratio, $\Delta \Ptarg =$ target pressure difference, $\ttotal =$ total simulation duration, $\bar{\rho} =$ average total fluid density in control regions, $\cubar =$ average solute concentration in control regions, $\Delta c = $ average concentration difference between control regions, and $ \Delta \Pi $ and $ \Delta P = $ average osmotic pressure difference and average pressure difference between reservoirs calculated from the applied force balance in the transition region.}
	\label{tbl:neq_Dc_sys}
	\begin{tabular}{cccccccccccccc}
	\hline
%	{\small
	$d$ & $L_{x/y}/\sqrt{2}$ & $a$ & $\epsuw$ & $\siguw$ & $\bar{\chi}$ & $\cutarg^+/\cutarg^-$ & $\Delta \Ptarg$ & $\ttotal$ & $\bar{\rho}$  & $ \cubar $ & $ \Delta \cu $ & $\Delta \Pi $ & $ \Delta P $\\
	($\sigma$) & ($\sigma$) & ($\sigma$) & ($\epsilon$) & ($\sigma$) &  & & ($\epsilon/\sigma^3$) & ($\times 10^4~\tau$) &  ($\sigma^{-3}$) & ($\sigma^{-3}$) & ($\sigma^{-3}$) & ($\epsilon/\sigma^3$) &  ($\epsilon/\sigma^3$) \\
	\hline
	2 &                            50 &              3 &                   0.5 &                 0.8 &   0.20 &            20.0 &                                0.0 &                           16.0 &                       0.787 &                                    0.162 &                                        0.292 &                                              0.324 &                                            0.000 \\
	2 &                            50 &              4 &                   0.5 &                 0.8 &   0.20 &            20.0 &                                0.0 &                           16.0 &                       0.787 &                                    0.163 &                                        0.295 &                                              0.328 &                                           0.000 \\
	2 &                            50 &              6 &                   0.5 &                 0.8 &   0.20 &             2.0 &                                0.0 &                           44.0 &                       0.787 &                                    0.162 &                                        0.108 &                                              0.110 &                                           0.000 \\
	2 &                            50 &              6 &                   0.5 &                 0.8 &   0.20 &             3.0 &                                0.0 &                           50.0 &                       0.787 &                                    0.162 &                                        0.162 &                                              0.169 &                                            0.000 \\
	2 &                            50 &              6 &                   0.5 &                 0.8 &   0.20 &             5.0 &                                0.0 &                           34.0 &                       0.787 &                                    0.163 &                                        0.218 &                                              0.231 &                                            0.000 \\
	2 &                            50 &              6 &                   0.5 &                 0.8 &   0.20 &            20.0 &                                0.0 &                           22.0 &                       0.787 &                                    0.166 &                                        0.300 &                                              0.339 &                                            0.000 \\
	2 &                            50 &              8 &                   0.5 &                 0.8 &   0.20 &            20.0 &                                0.0 &                           24.0 &                       0.787 &                                    0.167 &                                        0.303 &                                              0.344 &                                            0.000 \\
	2 &                            50 &              6 &                   0.5 &                 1.2 &   0.20 &            20.0 &                                0.0 &                           16.0 &                       0.787 &                                    0.165 &                                        0.298 &                                              0.335 &                                            0.000 \\
	2 &                            50 &              6 &                   0.5 &                 1.5 &   0.20 &             3.0 &                                0.0 &                           34.0 &                       0.786 &                                    0.154 &                                        0.154 &                                              0.160 &                                           0.000 \\
	2 &                            50 &              6 &                   0.5 &                 1.5 &   0.20 &            20.0 &                                0.0 &                           16.0 &                       0.786 &                                    0.153 &                                        0.277 &                                              0.310 &                                           0.000 \\
	2 &                            50 &              8 &                   0.5 &                 1.5 &   0.20 &            20.0 &                                0.0 &                           24.0 &                       0.786 &                                    0.152 &                                        0.275 &                                              0.310 &                                            0.000 \\
	2 &                            50 &              6 &                   0.8 &                 0.8 &   0.20 &            20.0 &                                0.0 &                           16.0 &                       0.786 &                                    0.163 &                                        0.295 &                                              0.331 &                                            0.000 \\
	2 &                            50 &              6 &                   1.2 &                 0.8 &   0.20 &            20.0 &                                0.0 &                           16.0 &                       0.785 &                                    0.158 &                                        0.286 &                                              0.321 &                                            0.000 \\
	2 &                            50 &              6 &                   1.5 &                 0.8 &   0.20 &            20.0 &                                0.0 &                           16.0 &                       0.783 &                                    0.152 &                                        0.274 &                                              0.306 &                                           0.000 \\
	2 &                            50 &              8 &                   1.5 &                 0.8 &   0.20 &            20.0 &                                0.0 &                           16.0 &                       0.783 &                                    0.151 &                                        0.274 &                                              0.308 &                                           0.000 \\
	2 &                            50 &              6 &                   1.5 &                 1.5 &   0.20 &             2.0 &                                0.0 &                           35.0 &                       0.784 &                                    0.139 &                                        0.093 &                                              0.097 &                                           0.000 \\
	2 &                            50 &              6 &                   1.5 &                 1.5 &   0.20 &             3.0 &                                0.0 &                           22.0 &                       0.784 &                                    0.138 &                                        0.138 &                                              0.146 &                                           0.000 \\
	2 &                            50 &              6 &                   1.5 &                 1.5 &   0.20 &             5.0 &                                0.0 &                           20.0 &                       0.784 &                                    0.137 &                                        0.182 &                                              0.196 &                                            0.000 \\
	2 &                            50 &              6 &                   1.5 &                 1.5 &   0.20 &            20.0 &                                0.0 &                           22.0 &                       0.785 &                                    0.134 &                                        0.242 &                                              0.274 &                                           0.000 \\
	2 &                            50 &              6 &                   0.5 &                 0.8 &   0.05 &            20.0 &                                0.0 &                           72.0 &                       0.787 &                                    0.040 &                                        0.073 &                                              0.076 &                                            0.000 \\
	2 &                            50 &              8 &                   0.5 &                 0.8 &   0.05 &            20.0 &                                0.0 &                           32.0 &                       0.787 &                                    0.041 &                                        0.074 &                                              0.077 &                                            0.000 \\
	2 &                            50 &              6 &                   0.5 &                 1.5 &   0.05 &            20.0 &                                0.0 &                           64.0 &                       0.787 &                                    0.037 &                                        0.067 &                                              0.070 &                                            0.000 \\
	2 &                            50 &              6 &                   1.5 &                 0.8 &   0.05 &            20.0 &                                0.0 &                           72.0 &                       0.785 &                                    0.036 &                                        0.065 &                                              0.068 &                                            0.000 \\
	2 &                            50 &              6 &                   1.5 &                 1.5 &   0.05 &            20.0 &                                0.0 &                           24.0 &                       0.784 &                                    0.022 &                                        0.040 &                                              0.043 &                                           0.000 \\
	4 &                            50 &              6 &                   0.5 &                 0.8 &   0.20 &            20.0 &                                0.0 &                           16.0 &                       0.787 &                                    0.167 &                                        0.302 &                                              0.341 &                                            0.000 \\
	2 &                            80 &              8 &                   0.5 &                 0.8 &   0.20 &            20.0 &                                0.0 &                           16.0 &                       0.787 &                                    0.163 &                                        0.295 &                                              0.328 &                                           0.000 \\
	2 &                            80 &              8 &                   0.5 &                 1.5 &   0.20 &            20.0 &                                0.0 &                           16.0 &                       0.787 &                                    0.153 &                                        0.277 &                                              0.307 &                                            0.000 \\
	2 &                            50 &              6 &                   0.5 &                 0.8 &   0.20 &             2.0 &                                 -- &                           16.0 &                       0.787 &                                    0.151 &                                        0.101 &                                              0.104 &                                            0.128 \\
	2 &                            50 &              6 &                   0.5 &                 0.8 &   0.20 &             3.0 &                                 -- &                           16.0 &                       0.787 &                                    0.143 &                                        0.143 &                                              0.156 &                                            0.185 \\
	2 &                            50 &              6 &                   0.5 &                 0.8 &   0.20 &             5.0 &                                 -- &                           16.0 &                       0.786 &                                    0.134 &                                        0.179 &                                              0.202 &                                            0.233 \\
	2 &                            50 &              6 &                   0.5 &                 0.8 &   0.05 &            20.0 &                                 -- &                           16.0 &                       0.787 &                                    0.037 &                                        0.066 &                                              0.069 &                                            0.071 \\
\hline
		\end{tabular}
   \end{table}

   \begin{table}[htb!]
	\centering
	\caption{Systems properties in non-equilibrium pressure-driven flow simulations: $d =$ width of transition region,  $L_{x/y} =$ system $x$ and $y$ dimensions, $a =$ pore radius, $\epsuw =$ solute--membrane interaction strength, $\siguw =$ solute--membrane interaction range, $\bar{\chi} =$ solute mole fraction, $\fu =$ force on solute particle, $\fv =$ force on solvent particle, $\ttotal =$ total simulation duration, $\bar{\rho} =$ average total fluid density in control regions, $ \cubar =$ average solute concentration in control regions, $\Delta c = $ average concentration difference between control regions, and $\Delta \Pi $ and $ \Delta P  = $ average osmotic pressure difference and average pressure difference between reservoirs calculated from  the applied force balance in the transition region.}
	\label{tbl:neq_Dp_sys}
	\begin{tabular}{cccccccccccccc}
	\hline
	$d$ & $L_{x/y}/\sqrt{2}$ & $a$ & $\epsuw$ & $\siguw$ & $\bar{\chi}$ &  $\fu = \fv$ & $t_\mathrm{total}$ & $\bar{\rho}$ & $\cubar $ & $\Delta \cu $ & $ \Delta \Pi $ & $ \Delta P$\\
	($\sigma$) & ($\sigma$) & ($\sigma$) & ($\epsilon$) & ($\sigma$) &  & ($\epsilon/\sigma$) & ($\times 10^4~\tau$) & ($\sigma^{-3}$) & ($\sigma^{-3}$) & ($\sigma^{-3}$) & ($\epsilon/\sigma^3$)  & ($\epsilon/\sigma^3$) \\
	\hline
	2 &                            50 &              6 &                   0.5 &                 0.8 &   0.20 &                            0.04 &                          141.8 &                       0.787 &                                    0.161 &                                        0.001 &                                              0.000 &                                            0.039 \\
	2 &                            50 &              6 &                   0.5 &                 0.8 &   0.20 &                            0.08 &                           76.0 &                       0.787 &                                    0.161 &                                        0.001 &                                              0.000 &                                            0.079 \\
	2 &                            50 &              6 &                   0.5 &                 0.8 &   0.20 &                            0.12 &                           47.5 &                       0.787 &                                    0.161 &                                        0.002 &                                              0.000 &                                            0.118 \\
	2 &                            50 &              6 &                   0.5 &                 0.8 &   0.20 &                            0.32 &                           16.0 &                       0.787 &                                    0.161 &                                        0.005 &                                              0.000 &                                            0.315 \\
	2 &                            50 &              6 &                   1.5 &                 1.5 &   0.20 &                            0.04 &                          114.0 &                       0.784 &                                    0.141 &                                       0.000 &                                              0.000 &                                            0.039 \\
	2 &                            50 &              6 &                   1.5 &                 1.5 &   0.20 &                            0.08 &                           53.5 &                       0.784 &                                    0.141 &                                       0.000 &                                              0.000 &                                            0.078 \\
	2 &                            50 &              6 &                   1.5 &                 1.5 &   0.20 &                            0.12 &                           41.5 &                       0.784 &                                    0.141 &                                        0.000 &                                              0.000 &                                            0.118 \\
	2 &                            50 &              6 &                   1.5 &                 1.5 &   0.20 &                            0.16 &                           28.0 &                       0.784 &                                    0.142 &                                        0.000 &                                              0.000 &                                            0.157 \\
	2 &                            50 &              6 &                   1.5 &                 1.5 &   0.20 &                            0.32 &                           18.0 &                       0.784 &                                    0.143 &                                        0.001 &                                              0.000 &                                            0.314 \\
\hline
		\end{tabular}
   \end{table}

   \begin{figure}[htb!]
	\centering
	\begin{subfigure}[b]{0.4\textwidth}
		\centering
		\includegraphics[width=\textwidth]{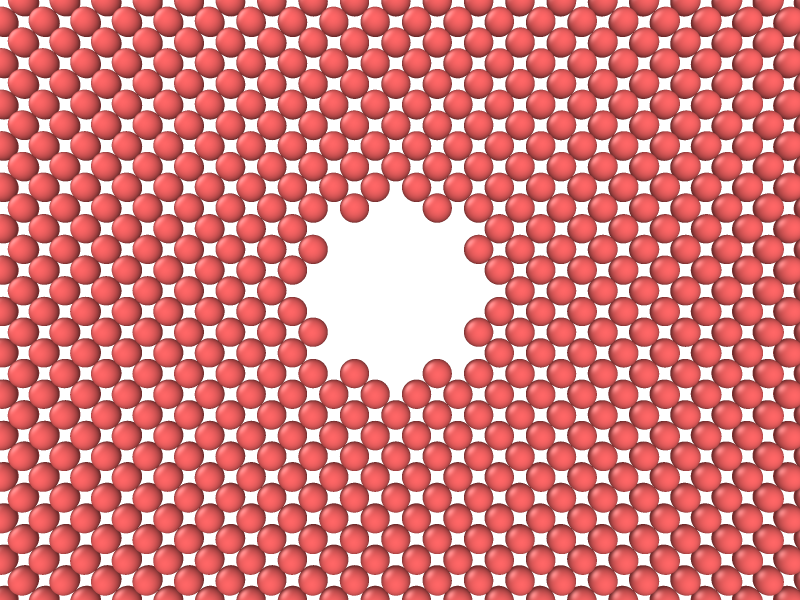}
		\caption{$a = 3\sigma$}
		\label{fig:pore_a3}
	\end{subfigure}
	\hspace*{2em}
	\begin{subfigure}[b]{0.4\textwidth}
		\centering
		\includegraphics[width=\textwidth]{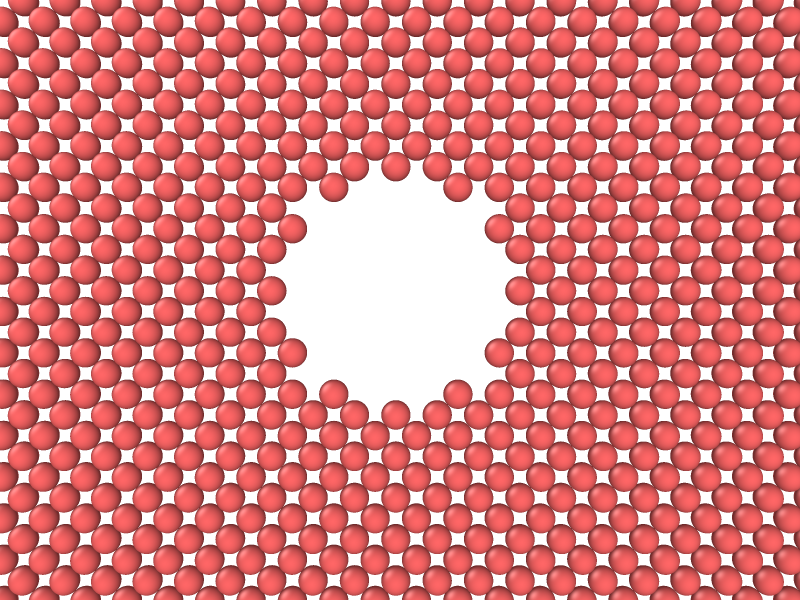}
		\caption{$a = 4\sigma$}
		\label{fig:pore_a4}
	\end{subfigure}
	%\hfill
	\begin{subfigure}[b]{0.4\textwidth}
		\centering
		\includegraphics[width=\textwidth]{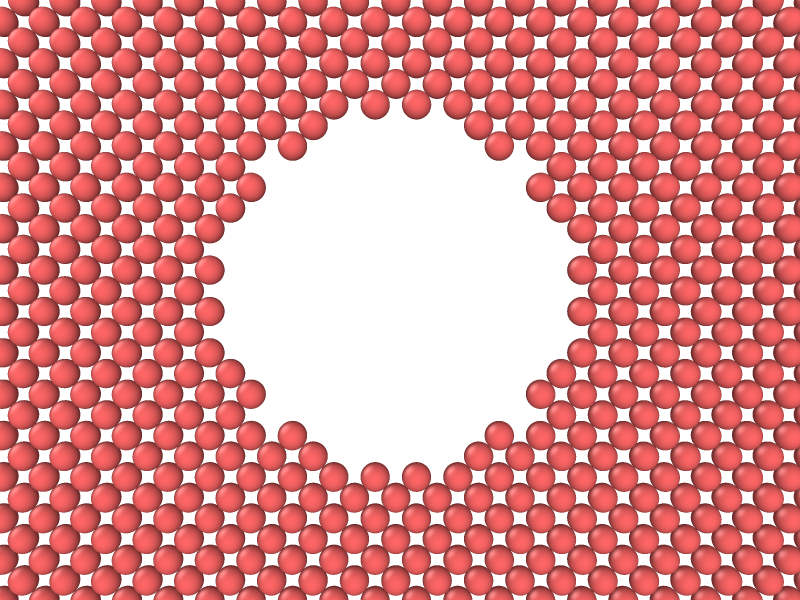}
		\caption{$a = 6\sigma$}
		\label{fig:pore_a6}
	\end{subfigure}
	\hspace*{2em}
	\begin{subfigure}[b]{0.4\textwidth}
		\centering
		\includegraphics[width=\textwidth]{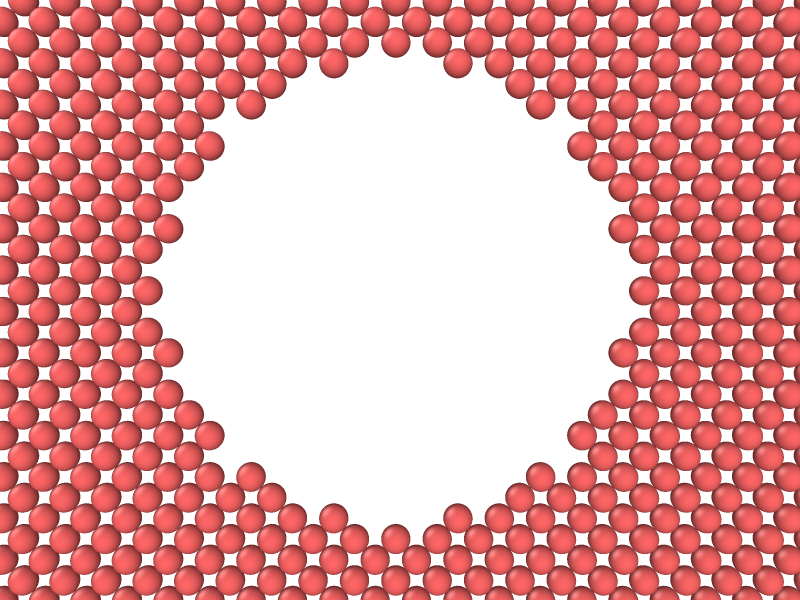}
		\caption{$a = 8\sigma$}
		\label{fig:pore_a8}
	\end{subfigure}
	%\hfill
	\caption{Snapshots of membrane pores of the various radii $a$ that were simulated.}
	\label{fig:pore_image}
\end{figure}

\FloatBarrier

\section{Additional non-equilibrium simulation results}

\begin{figure}[htb!]
	\centering
	\includegraphics{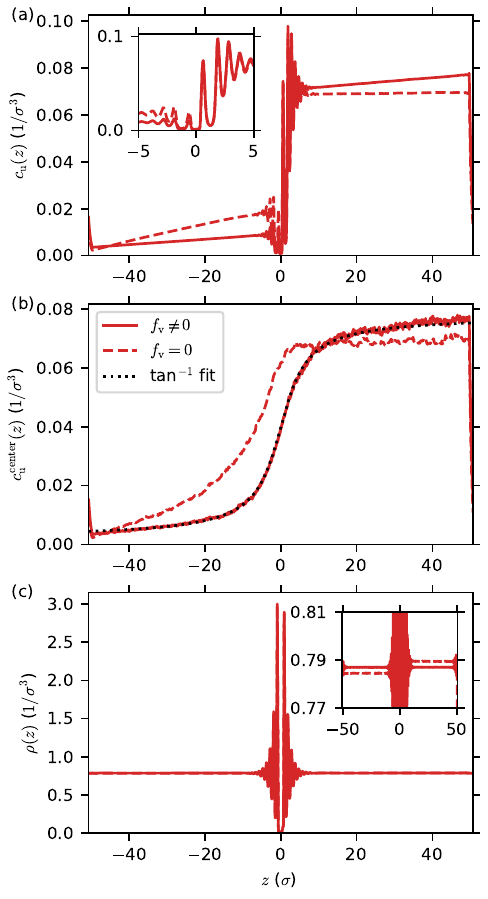}
 \caption{(a) Solute concentration, (b) centerline solute concentration, and (c) total fluid density vs $z$ coordinate in concentration-gradient-driven flow simulations with ($\fv \neq 0$, solid lines) and without ($\fv = 0$, dashed lines) the transmembrane pressure difference constrained to be zero for $a = 6\sigma$ , $\epsuw = 0.5\epsilon$, $\siguw = 0.8\sigma$, $\bar{\chi} = 0.05$, and $\curatio = 20$. The inset in (a) zooms in on $z$ values near the membrane, whereas the inset in (c) zooms in on density values around the bulk density. The dotted black line in (b) is a fit of the solid line to a function of the form $b_0 + b_1\tan^{-1}(z/b_2)$ with fit parameters $b_0$, $b_1$, and $b_2$.}
 \label{fig:conc_profiles_rep_lowc}
\end{figure}

\begin{figure}[htb!]
	\centering
	\includegraphics{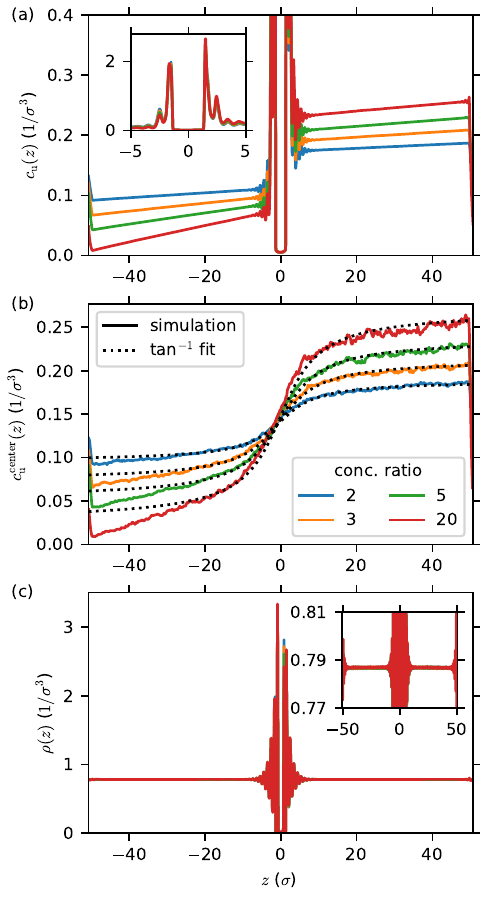}
 \caption{(a) Solute concentration, (b) centerline solute concentration, and (c) total fluid density vs $z$ coordinate in concentration-gradient-driven flow simulations with the transmembrane pressure difference constrained to be zero for $a = 6\sigma$ , $\epsuw = 1.5\epsilon$, $\siguw = 1.5\sigma$, $\bar{\chi} = 0.2$, and various target solute concentration ratios $\curatio$. The inset in (a) zooms in on $z$ values near the membrane, whereas the inset in (c) zooms in on density values around the bulk density. The dotted black lines in (b) are least-squares fit of the solid line to a function of the form $b_0 + b_1\tan^{-1}(z/b_2)$ with fit parameters $b_0$,  $b_1$, and $b_2$.}
 \label{fig:conc_profiles_attr_highc}
\end{figure}

\begin{figure}[htb!]
	\centering
	\includegraphics{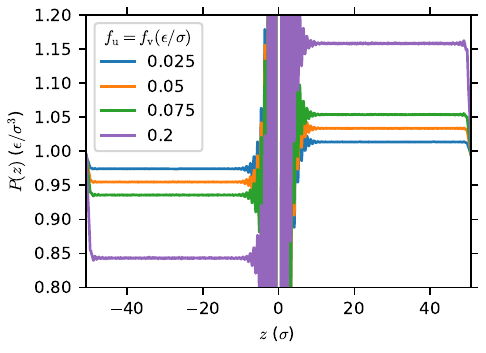}
 \caption{Pressure profile in pressure-driven flow simulations without an imposed concentration difference across the membrane for various values of the applied force $\fu = \fv$ on solute and solvent molecules for the high average solute mole fraction ($\bar{\chi} = 0.2$), repulsive effective solute--membrane interactions ($\epsuw = 0.5\epsilon, \siguw = 0.8\sigma$), and  pore radius $a = 6\sigma$.}
 \label{fig:press_profiles_pdrive_rep_highc}
\end{figure}

\begin{figure}[htb!]
	\centering
	\includegraphics{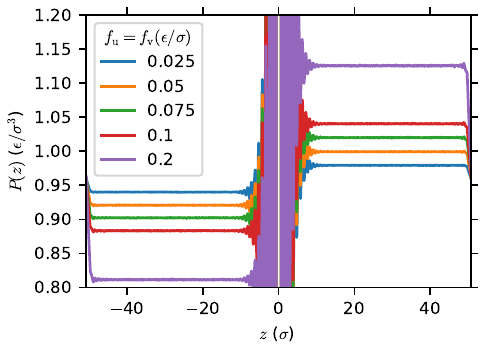}
 \caption{Pressure profile in pressure-driven flow simulations without an imposed concentration difference across the membrane for various values of the applied force $\fu = \fv$ on solute and solvent molecules for the high average solute mole fraction ($\bar{\chi} = 0.2$), attractive effective solute--membrane interactions ($\epsuw = 1.5\epsilon, \siguw = 1.5\sigma$), and  pore radius $a = 6\sigma$.}
 \label{fig:press_profiles_pdrive_attr_highc}
\end{figure}

\begin{figure}[htb!]
	\centering
	\includegraphics{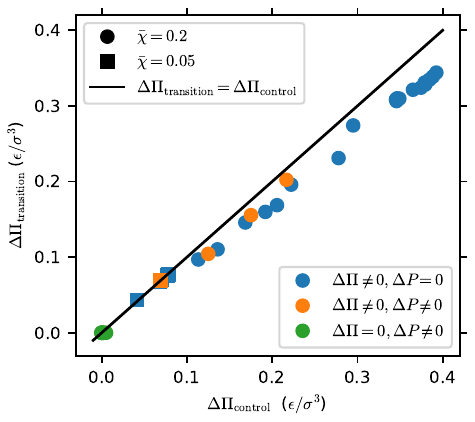}
 \caption{Average osmotic pressure difference calculated from the applied force balance in the transition region vs that calculated from the concentration difference between control regions using Eq.~\eqref{eqn:Pi_id} of the main paper for the high ($\bar{\chi} = 0.2$, circles) and low  ($\bar{\chi} = 0.05$, squares) average solute mole fractions for concentration-gradient-driven flow simulations with (blue symbols) and without (orange symbols) the pressure different across the membrane constrained to be zero and for pressure-driven flow simulations without an imposed concentration difference across the membrane (green symbols). Error bars are smaller than the symbols.}
 \label{fig:DPi_trans_vs_DPi_control}
\end{figure}

\begin{figure}[htb!]
	\centering
	\includegraphics{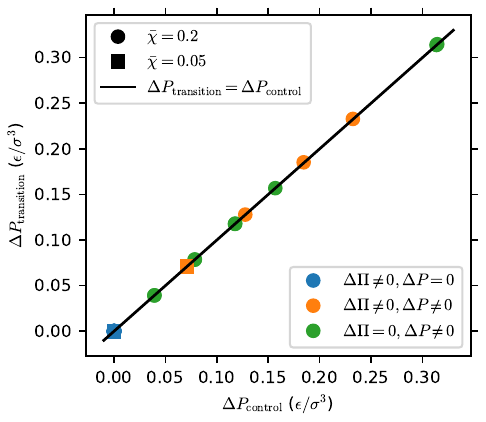}
 \caption{Average pressure difference calculated from the applied force balance in the transition region vs that calculated between control regions for the high ($\bar{\chi} = 0.2$, circles) and low  ($\bar{\chi} = 0.05$, squares) average solute mole fractions for concentration-gradient-driven flow simulations with (blue symbols) and without (orange symbols) the pressure difference across the membrane constrained to be zero and for pressure-driven flow simulations without an imposed concentration difference across the membrane (green symbols). Error bars are smaller than the symbols.}
 \label{fig:Dp_trans_vs_Dp_control}
\end{figure}

\begin{figure}[htb!]
	\centering
	\includegraphics{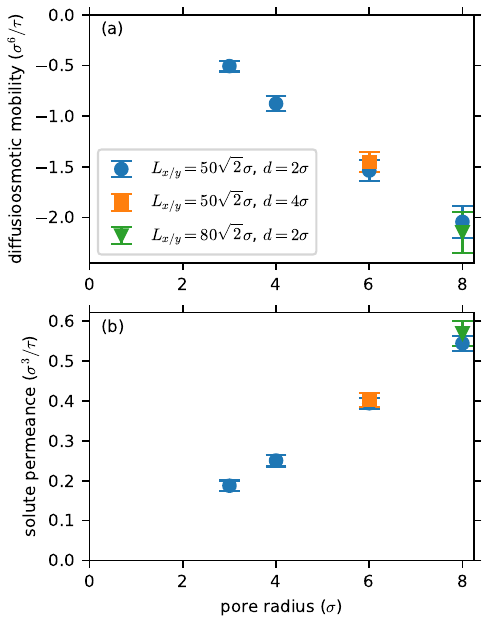}
	\caption{(a) Diffusioosmotic mobility and (b) solute permeance vs pore radius $a$ for fixed average solute mole fraction ($\bar{\chi} = 0.2$) and solute-membrane interaction parameters ($\epsuw = 0.5\epsilon, \siguw = 0.8\sigma$), and different reservoir dimensions $L_{x/y}$ and transition region widths $d$, showing that these transport coefficients are independent of $L_{x/y}$ and $d$.}
	\label{fig:kappaDO_Ps_vs_radius-res_trans_width}
\end{figure}

\begin{figure}[htb!]
	\centering
	\includegraphics{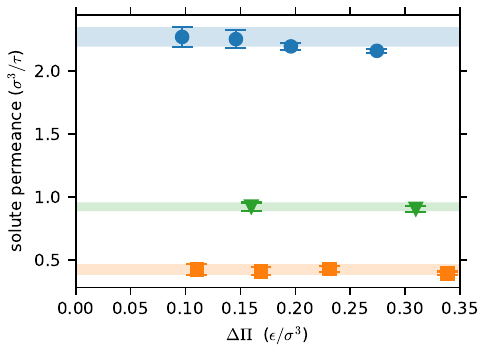}
	\caption{Solute permeance vs osmotic pressure difference $\Delta \Pi$ across the membrane from concentration-gradient-driven flow simulations with target pressure difference $\Delta \Ptarg = 0$ for the high average solute mole fraction ($\bar{\chi} = 0.2$) and a membrane with pore radius $a = 6\sigma$ and various solute--membrane interaction parameters: $\epsuw = 1.5\epsilon, \siguw = 1.5\sigma$ (circles); $\epsuw = 0.5\epsilon, \siguw = 0.8\sigma$ (squares); $\epsuw = 0.5\epsilon, \siguw = 1.5\sigma$ (triangles). The translucent bands are the 95\% confidence intervals for the lowest value of $\Delta \Pi$ for each set of solute--membrane interaction parameters. The fact that all the data points in each set lie within the corresponding band, except for the circle for the highest $\Delta \Pi$, indicates that all the systems except that one are in the linear-response regime.}
	\label{fig:linear_response_solute_permeance}
\end{figure}

\begin{figure}[htb!]
	\centering
	\includegraphics{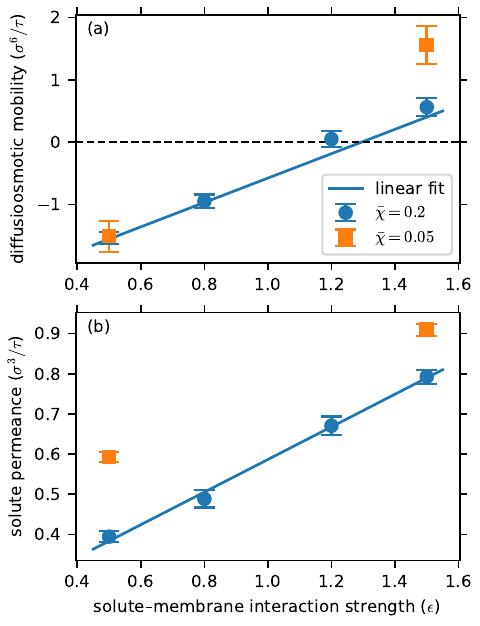}
	\caption{(a) Diffusioosmotic mobility and (b) solute permeance vs solute--membrane interaction strength parameter $\epsuw$ for fixed interaction range parameter ($\siguw = 0.8\sigma$) and pore radius ($a = 6\sigma$) and different average solute mole fractions ($\bar{\chi} = 0.2$ (circles) and  $\bar{\chi} = 0.05$ (squares)). The solid line is a linear fit for the high mole fraction.}
	\label{fig:kappaDO_Ps_vs_eps}
\end{figure}

\begin{figure}[htb!]
	\centering
	\includegraphics{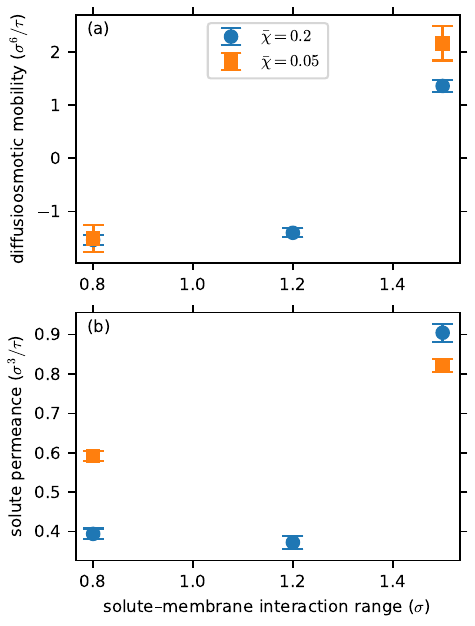}
	\caption{(a) Diffusioosmotic mobility and (b) solute permeance vs solute--membrane interaction range parameter $\siguw$ for fixed interaction strength parameter ($\epsuw = 0.5\epsilon$) and pore radius ($a = 6\sigma$)  and different average solute mole fractions ($\bar{\chi} = 0.2$ (circles) and  $\bar{\chi} = 0.05$ (squares)). }
	\label{fig:kappaDO_Ps_vs_sig}
\end{figure}

\begin{figure}[htb!]
	\centering
	\includegraphics{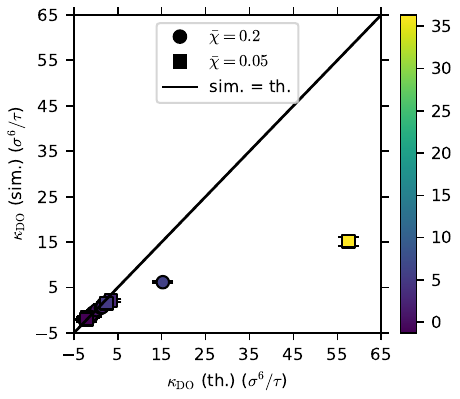}
	\caption{Diffusioosmotic mobility from simulation vs theory for all systems for $\curatio = 20$ except for those with the strongest attractive solute--membrane interactions and high solute mole fraction ($\epsuw = 1.5\epsilon$, $\siguw = 1.5\sigma$, $\bar{\chi} = 0.2$), for which $\curatio = 2$ was used instead. Symbols are colored by the surface solute excess $\Gamma$ (units: $\sigma$) and different symbol shapes distinguish high ($\bar{\chi} = 0.2$, circles) and low ($\bar{\chi} = 0.05$, squares) average solute mole fractions. The simulation and theory values are equal along the solid line.}
	\label{fig:kappaDO_sim_vs_th_Gamma_all}
\end{figure}

\begin{figure}[htb!]
	\centering
	\includegraphics{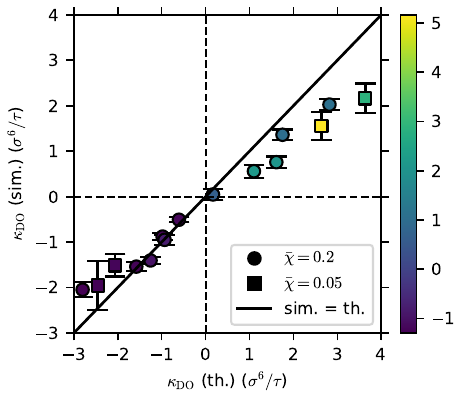}
	\caption{Diffusioosmotic mobility from simulation vs theory, with the actual pore radius $a$ used in the theory instead of the effective pore radius $\ah$,  for all systems except for those with the strongest attractive solute--membrane interactions ($\epsuw = 1.5\epsilon$, $\siguw = 1.5\sigma$) for $\curatio=20$. Symbols are colored by the surface solute excess $\Gamma$  (units: $\sigma$) and different symbol shapes distinguish high ($\bar{\chi} = 0.2$, circles) and low ($\bar{\chi} = 0.05$, squares) average solute mole fractions. The simulation and theory values are equal along the solid line.}
	\label{fig:kappaDO_sim_vs_th_Gamma_a}
\end{figure}

\begin{figure}[htb!]
	\centering
	\includegraphics{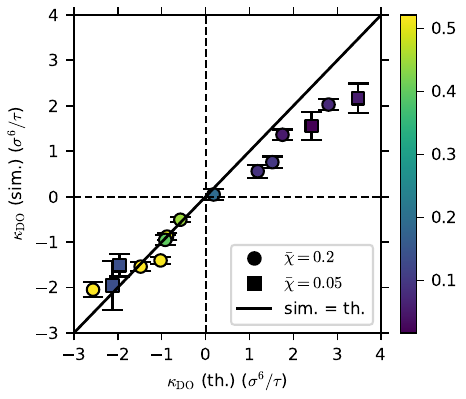}
	\caption{Diffusioosmotic mobility from simulation vs theory for all systems except for those with the strongest attractive solute--membrane interactions ($\epsuw = 1.5\epsilon$, $\siguw = 1.5\sigma$) for $\curatio=20$. Symbols are colored by the P\'eclet number $\mathrm{Pe}$ and different symbol shapes distinguish high ($\bar{\chi} = 0.2$, circles) and low ($\bar{\chi} = 0.05$, squares) average solute mole fractions. The simulation and theory values are equal along the solid line.}
	\label{fig:kappaDO_sim_vs_th_Pe}
\end{figure}

\begin{figure}[htb!]
	\centering
	\includegraphics{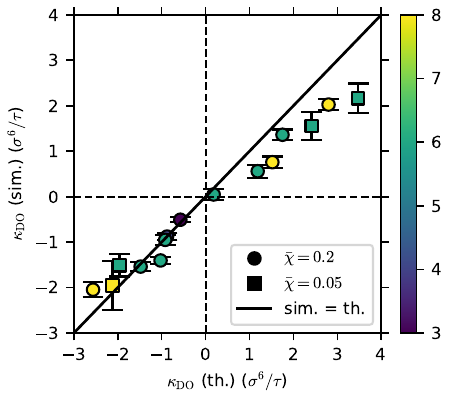}
	\caption{Diffusioosmotic mobility from simulation vs theory for all systems except for those with the strongest attractive solute--membrane interactions ($\epsuw = 1.5\epsilon$, $\siguw = 1.5\sigma$) for $\curatio=20$. Symbols are colored by the pore radius $a$ (units: $\sigma$) and different symbol shapes distinguish high ($\bar{\chi} = 0.2$, circles) and low ($\bar{\chi} = 0.05$, squares) average solute mole fractions. The simulation and theory values are equal along the solid line..}
	\label{fig:kappaDO_sim_vs_th_a}
\end{figure}

\begin{figure}[htb!]
	\centering
	\includegraphics{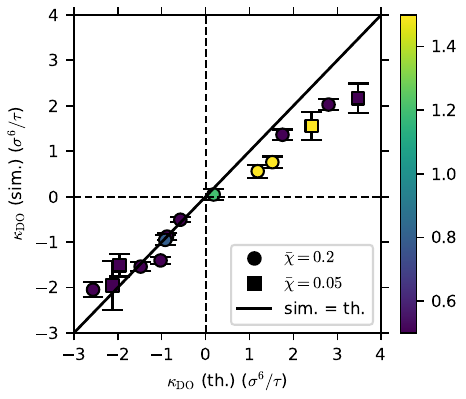}
	\caption{Diffusioosmotic mobility from simulation vs theory for all systems except for those with the strongest attractive solute--membrane interactions ($\epsuw = 1.5\epsilon$, $\siguw = 1.5\sigma$) for $\curatio=20$. Symbols are colored by the solute--membrane interaction strength parameter $\epsuw$ (units: $\epsilon$) and different symbol shapes distinguish high ($\bar{\chi} = 0.2$, circles) and low ($\bar{\chi} = 0.05$, squares) average solute mole fractions. The simulation and theory values are equal along the solid line.}
	\label{fig:kappaDO_sim_vs_th_eps}
\end{figure}

\begin{figure}[htb!]
	\centering
	\includegraphics{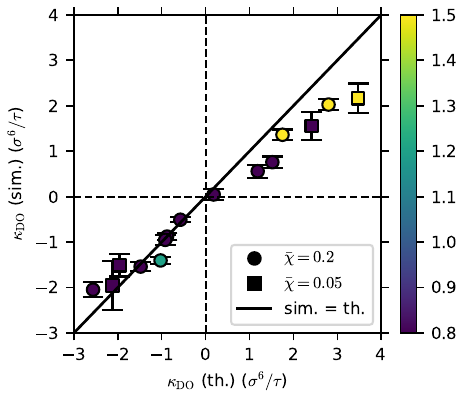}
	\caption{Diffusioosmotic mobility from simulation vs theory for all systems except for those with the strongest attractive solute--membrane interactions ($\epsuw = 1.5\epsilon$, $\siguw = 1.5\sigma$) for $\curatio=20$. Symbols are colored by the solute--membrane interaction range parameter $\siguw$ (units: $\sigma$) and different symbol shapes distinguish high ($\bar{\chi} = 0.2$, circles) and low ($\bar{\chi} = 0.05$, squares) average solute mole fractions. The simulation and theory values are equal along the solid line.}
	\label{fig:kappaDO_sim_vs_th_sig}
\end{figure}

\begin{figure}[htb!]
	\centering
	\includegraphics{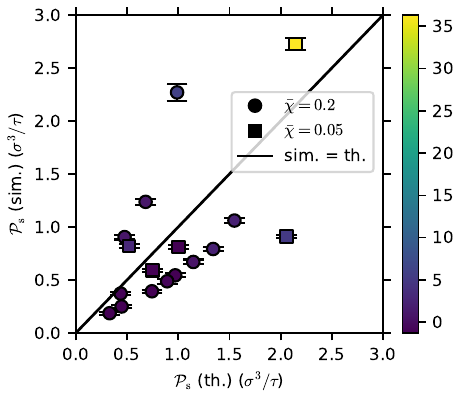}
	\caption{Solute permeance from simulation vs theory for all systems for $\curatio = 20$ except for those with the strongest attractive solute--membrane interactions and high solute mole fraction ($\epsuw = 1.5\epsilon$, $\siguw = 1.5\sigma$, $\bar{\chi} = 0.2$), for which $\curatio = 2$ was used instead. Symbols are colored by the surface solute excess $\Gamma$ (units: $\sigma$) and different symbol shapes distinguish high ($\bar{\chi} = 0.2$, circles) and low ($\bar{\chi} = 0.05$, squares) average solute mole fractions. The simulation and theory values are equal along the solid line.}
	\label{fig:Ps_sim_vs_th_Gamma_all}
\end{figure}

\begin{figure}[htb!]
	\centering
	\includegraphics{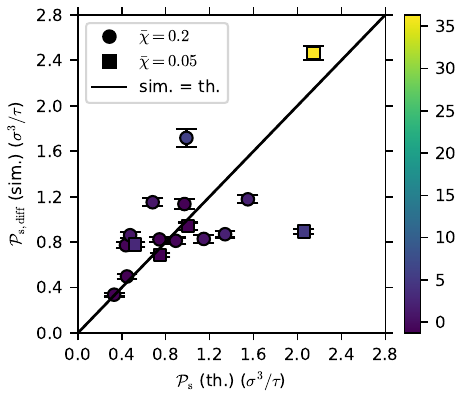}
	\caption{Solute permeance from simulation (calculated from diffusive flux only) vs theory for all systems for $\curatio = 20$ except for those with the strongest attractive solute--membrane interactions and high solute mole fraction ($\epsuw = 1.5\epsilon$, $\siguw = 1.5\sigma$, $\bar{\chi} = 0.2$)), for which $\curatio = 2$ was used instead. Symbols are colored by the surface solute excess $\Gamma$ (units: $\sigma$) and different symbol shapes distinguish high ($\bar{\chi} = 0.2$, circles) and low ($\bar{\chi} = 0.05$, squares) average solute mole fractions. The simulation and theory values are equal along the solid line.}
	\label{fig:Psdiff_sim_vs_th_Gamma_all}
\end{figure}

\begin{figure}[htb!]
	\centering
	\includegraphics{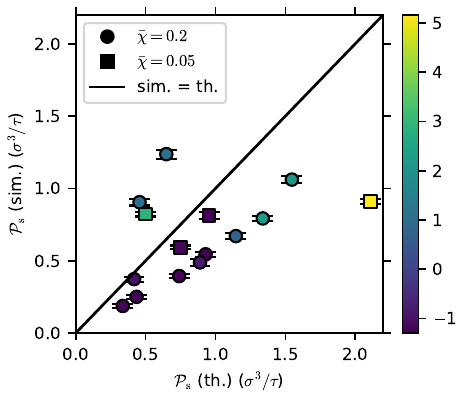}
	\caption{Solute permeance from simulation vs theory, with the actual pore radius $a$ used in the theory instead of the effective pore radius $\ah$, for all systems for $\curatio = 20$ except for those with the strongest attractive solute--membrane interactions ($\epsuw = 1.5\epsilon$, $\siguw = 1.5\sigma$). Symbols are colored by the surface solute excess $\Gamma$ (units: $\sigma$) and different symbol shapes distinguish high ($\bar{\chi} = 0.2$, circles) and low ($\bar{\chi} = 0.05$, squares) average solute mole fractions. The simulation and theory values are equal along the solid line.}
	\label{fig:Ps_sim_vs_th_Gamma_a}
\end{figure}

\clearpage

\begin{figure}[htb!]
	\centering
	\includegraphics{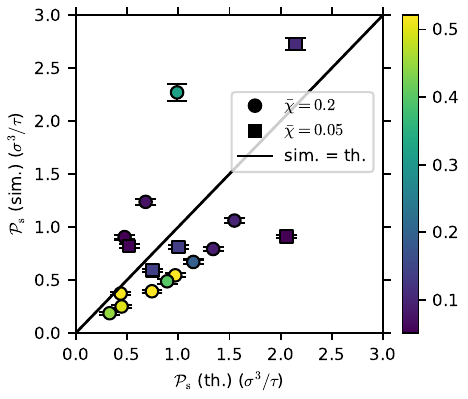}
	\caption{Solute permeance from simulation vs theory for all systems for $\curatio = 20$ except for those with the strongest attractive solute--membrane interactions and high solute mole fraction ($\epsuw = 1.5\epsilon$, $\siguw = 1.5\sigma$, $\bar{\chi} = 0.2$), for which $\curatio = 2$ was used instead. Symbols are colored by the P\'eclet number $\mathrm{Pe}$  and different symbol shapes distinguish high ($\bar{\chi} = 0.2$, circles) and low ($\bar{\chi} = 0.05$, squares) average solute mole fractions. The simulation and theory values are equal along the solid line.}
	\label{fig:Ps_sim_vs_th_Pe_all}
\end{figure}

\begin{figure}[htb!]
	\centering
	\includegraphics{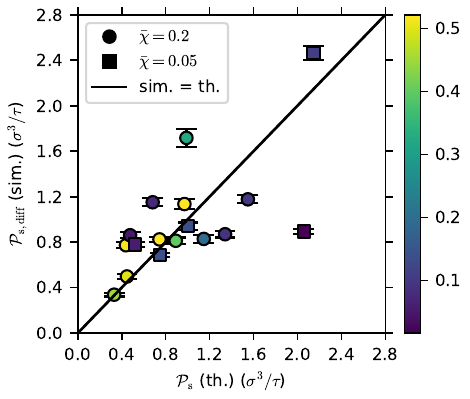}
	\caption{Solute permeance from simulation (calculated from diffusive flux only) vs theory for all systems for $\curatio = 20$ except for those with the strongest attractive solute--membrane interactions and high solute mole fraction ($\epsuw = 1.5\epsilon$, $\siguw = 1.5\sigma$, $\bar{\chi} = 0.2$), for which $\curatio = 2$ was used instead. Symbols are colored by the P\'eclet number $\mathrm{Pe}$  and different symbol shapes distinguish high ($\bar{\chi} = 0.2$, circles) and low ($\bar{\chi} = 0.05$, squares) average solute mole fractions. The simulation and theory values are equal along the solid line.}
	\label{fig:Psdiff_sim_vs_th_Pe_all}
\end{figure}

\begin{figure}[htb!]
	\centering
	\includegraphics{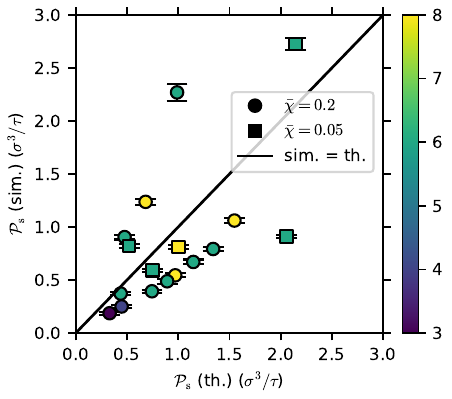}
	\caption{Solute permeance from simulation vs theory for all systems for $\curatio = 20$ except for those with the strongest attractive solute--membrane interactions and high solute mole fraction ($\epsuw = 1.5\epsilon$, $\siguw = 1.5\sigma$, $\bar{\chi} = 0.2$), for which $\curatio = 2$ was used instead. Symbols are colored by the pore radius $a$ (units: $\sigma$) and different symbol shapes distinguish high ($\bar{\chi} = 0.2$, circles) and low ($\bar{\chi} = 0.05$, squares) average solute mole fractions. The simulation and theory values are equal along the solid line.}
	\label{fig:Ps_sim_vs_th_a_all}
\end{figure}

\begin{figure}[htb!]
	\centering
	\includegraphics{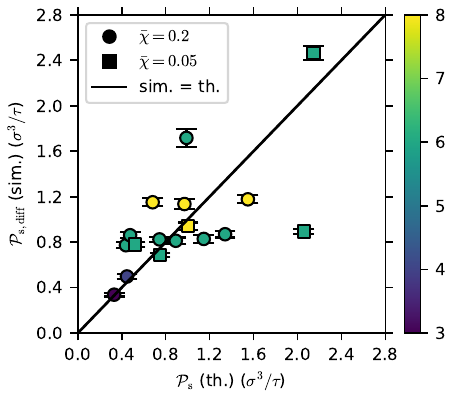}
	\caption{Solute permeance from simulation (calculated from diffusive flux only) vs theory for all systems for $\curatio = 20$ except for those with the strongest attractive solute--membrane interactions and high solute mole fraction ($\epsuw = 1.5\epsilon$, $\siguw = 1.5\sigma$, $\bar{\chi} = 0.2$), for which $\curatio = 2$ was used instead. Symbols are colored by the pore radius $a$ (units: $\sigma$)  and different symbol shapes distinguish high ($\bar{\chi} = 0.2$, circles) and low ($\bar{\chi} = 0.05$, squares) average solute mole fractions. The simulation and theory values are equal along the solid line.}
	\label{fig:Psdiff_sim_vs_th_a_all}
\end{figure}

\begin{figure}[htb!]
	\centering
	\includegraphics{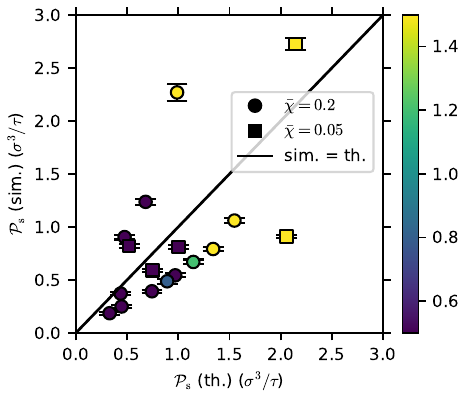}
	\caption{Solute permeance from simulation vs theory for all systems for $\curatio = 20$ except for those with the strongest attractive solute--membrane interactions and high solute mole fraction ($\epsuw = 1.5\epsilon$, $\siguw = 1.5\sigma$, $\bar{\chi} = 0.2$), for which $\curatio = 2$ was used instead. Symbols are colored by the solute--membrane interaction strength parameter $\epsuw$ (units: $\epsilon$) and different symbol shapes distinguish high ($\bar{\chi} = 0.2$, circles) and low ($\bar{\chi} = 0.05$, squares) average solute mole fractions. The simulation and theory values are equal along the solid line.}
	\label{fig:Ps_sim_vs_th_eps_all}
\end{figure}

\begin{figure}[htb!]
	\centering
	\includegraphics{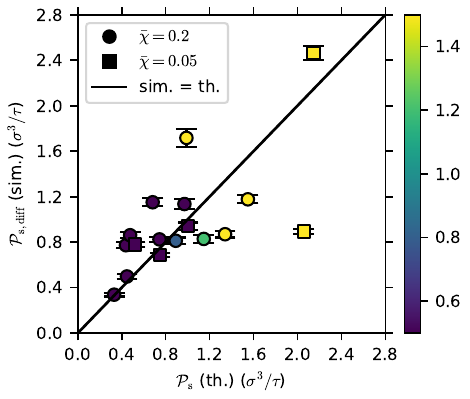}
	\caption{Solute permeance from simulation (calculated from diffusive flux only) vs theory for all systems for $\curatio = 20$ except for those with the strongest attractive solute--membrane interactions and high solute mole fraction ($\epsuw = 1.5\epsilon$, $\siguw = 1.5\sigma$, $\bar{\chi} = 0.2$), for which $\curatio = 2$ was used instead. Symbols are colored by the solute--membrane interaction strength parameter $\epsuw$ (units: $\epsilon$)  and different symbol shapes distinguish high ($\bar{\chi} = 0.2$, circles) and low ($\bar{\chi} = 0.05$, squares) average solute mole fractions. The simulation and theory values are equal along the solid line.}
	\label{fig:Psdiff_sim_vs_th_eps_all}
\end{figure}

\begin{figure}[htb!]
	\centering
	\includegraphics{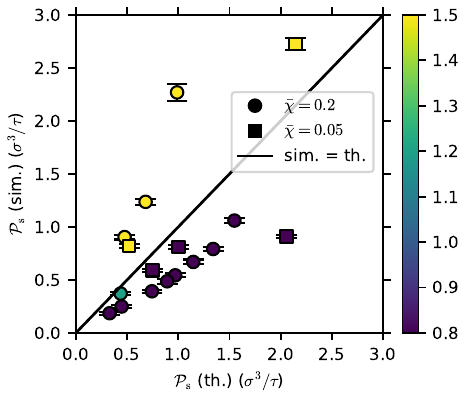}
	\caption{Solute permeance from simulation vs theory for all systems for $\curatio = 20$ except for those with the strongest attractive solute--membrane interactions and high solute mole fraction ($\epsuw = 1.5\epsilon$, $\siguw = 1.5\sigma$, $\bar{\chi} = 0.2$), for which $\curatio = 2$ was used instead. Symbols are colored by the solute--membrane interaction range parameter $\siguw$ (units: $\sigma$) and different symbol shapes distinguish high ($\bar{\chi} = 0.2$, circles) and low ($\bar{\chi} = 0.05$, squares) average solute mole fractions. The simulation and theory values are equal along the solid line.}
	\label{fig:Ps_sim_vs_th_sig_all}
\end{figure}

\begin{figure}[htb!]
	\centering
	\includegraphics{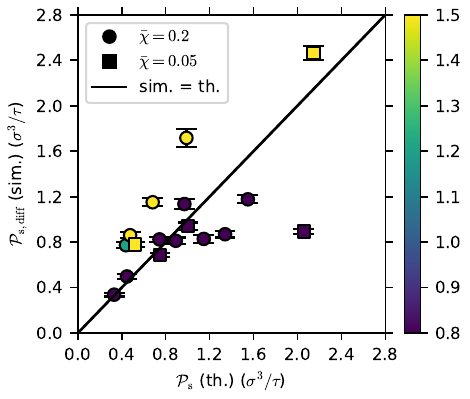}
	\caption{Solute permeance from simulation (calculated from diffusive flux only) vs theory for all systems for $\curatio = 20$ except for those with the strongest attractive solute--membrane interactions and high solute mole fraction ($\epsuw = 1.5\epsilon$, $\siguw = 1.5\sigma$, $\bar{\chi} = 0.2$), for which $\curatio = 2$ was used instead. Symbols are colored by the solute--membrane interaction range parameter $\siguw$ (units: $\sigma$)   and different symbol shapes distinguish high ($\bar{\chi} = 0.2$, circles) and low ($\bar{\chi} = 0.05$, squares) average solute mole fractions. The simulation and theory values are equal along the solid line.}
	\label{fig:Psdiff_sim_vs_th_sig_all}
\end{figure}

\FloatBarrier

\section{Equilibrium simulation results}

\begin{figure}[ht!]
	\centering
	\begin{subfigure}[b]{0.32\textwidth}
		\centering
		\includegraphics[width=\textwidth]{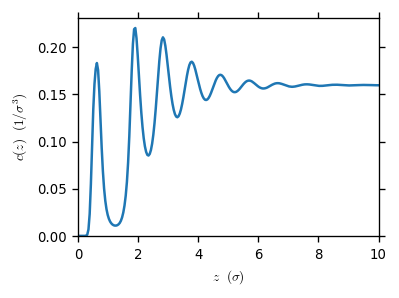}
		\caption{$\epsuw = 0.5\epsilon$, $\siguw = 0.8\sigma$, $\bar{\chi} = 0.2$}
		\label{fig:cuz_eq_0.2_0.5_0.8}
	\end{subfigure}
	\hfill
	\begin{subfigure}[b]{0.32\textwidth}
		\centering
		\includegraphics[width=\textwidth]{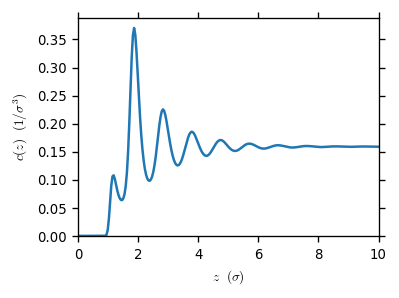}
		\caption{$\epsuw = 0.5\epsilon$, $\siguw = 1.2\sigma$, $\bar{\chi} = 0.2$}
		\label{fig:cuz_eq_0.2_0.5_1.2}
	\end{subfigure}
	\hfill
	\begin{subfigure}[b]{0.32\textwidth}
		\centering
		\includegraphics[width=\textwidth]{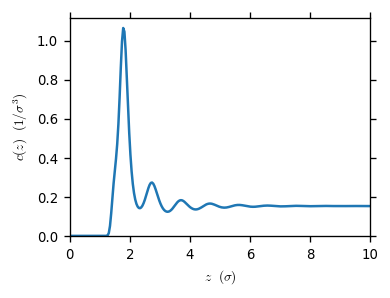}
		\caption{$\epsuw = 0.5\epsilon$, $\siguw = 1.5\sigma$, $\bar{\chi} = 0.2$}
		\label{fig:cuz_eq_0.2_0.5_1.5}
	\end{subfigure}	
	\hfill
	\begin{subfigure}[b]{0.32\textwidth}
		\centering
		\includegraphics[width=\textwidth]{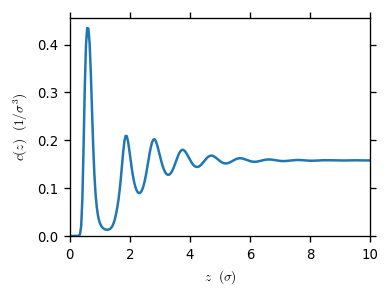}
		\caption{$\epsuw = 0.8\epsilon$, $\siguw = 0.8\sigma$, $\bar{\chi} = 0.2$}
		\label{fig:cuz_eq_0.2_0.8_0.8}
	\end{subfigure}
		\hfill
	\begin{subfigure}[b]{0.32\textwidth}
		\centering
		\includegraphics[width=\textwidth]{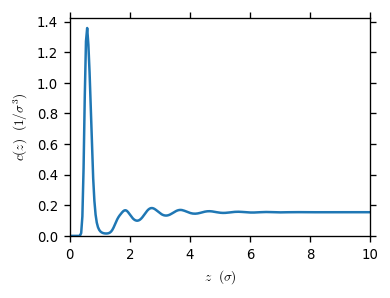}
		\caption{$\epsuw = 1.2\epsilon$, $\siguw = 0.8\sigma$, $\bar{\chi} = 0.2$}
		\label{fig:cuz_eq_0.2_1.2_0.8}
	\end{subfigure}
		\hfill
	\begin{subfigure}[b]{0.32\textwidth}
		\centering
		\includegraphics[width=\textwidth]{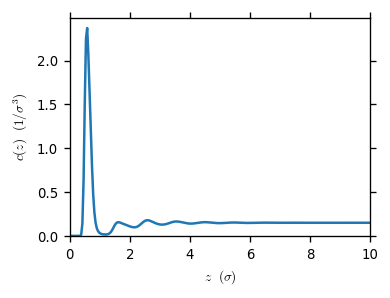}
		\caption{$\epsuw = 1.5\epsilon$, $\siguw = 0.8\sigma$, $\bar{\chi} = 0.2$}
		\label{fig:cuz_eq_0.2_1.5_0.8}
	\end{subfigure}
	\hfill
	\begin{subfigure}[b]{0.32\textwidth}
	\centering
	\includegraphics[width=\textwidth]{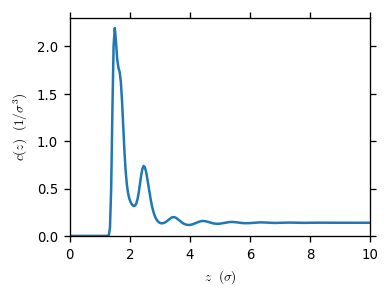}
	\caption{$\epsuw = 1.5\epsilon$, $\siguw = 1.5\sigma$, $\bar{\chi} = 0.2$}
	\label{fig:cuz_eq_0.2_1.5_1.5}
	\end{subfigure}
	\hfill
	\begin{subfigure}[b]{0.32\textwidth}
		\centering
		\includegraphics[width=\textwidth]{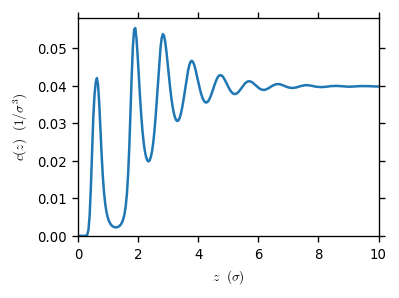}
		\caption{$\epsuw = 0.5\epsilon$, $\siguw = 0.8\sigma$, $\bar{\chi} = 0.05$}
		\label{fig:cuz_eq_0.05_0.5_0.8}
	\end{subfigure}
		\hfill
	\begin{subfigure}[b]{0.32\textwidth}
		\centering
		\includegraphics[width=\textwidth]{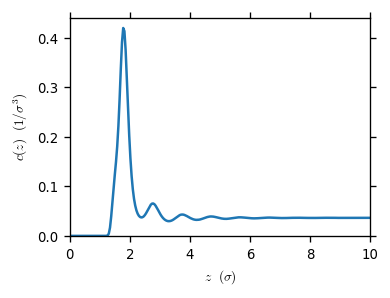}
		\caption{$\epsuw = 0.5\epsilon$, $\siguw = 1.5\sigma$, $\bar{\chi} = 0.05$}
		\label{fig:cuz_eq_0.05_0.5_1.5}
	\end{subfigure}
	\hfill
	\begin{subfigure}[b]{0.32\textwidth}
		\centering
		\includegraphics[width=\textwidth]{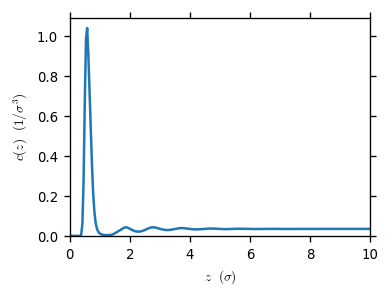}
		\caption{$\epsuw = 1.5\epsilon$, $\siguw = 0.8\sigma$, $\bar{\chi} = 0.05$}
		\label{fig:cuz_eq_0.05_1.5_0.8}
	\end{subfigure}
	\hfill
	\begin{subfigure}[b]{0.32\textwidth}
	\centering
	\includegraphics[width=\textwidth]{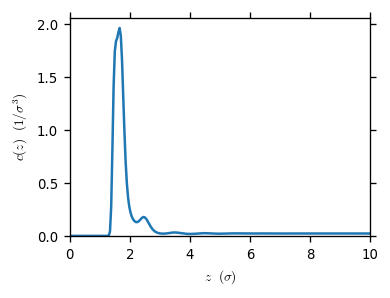}
	\caption{$\epsuw = 1.5\epsilon$, $\siguw = 1.5\sigma$, $\bar{\chi} = 0.05$}
	\label{fig:cuz_eq_0.05_1.5_1.5}
	\end{subfigure}
	\hfill
	\caption{Solute concentration vs distance perpendicular to a poreless membrane for various solute--membrane interaction parameters, $\epsuw$ and $\siguw$, and solute mole fractions, $\bar{\chi}$. Due to symmetry, the distribution is only shown for $z\geq 0$.}
	\label{fig:cuz_eq}
\end{figure}

\begin{figure}[htb!]
	\centering
	\begin{subfigure}[b]{0.32\textwidth}
		\centering
		\includegraphics[width=\textwidth]{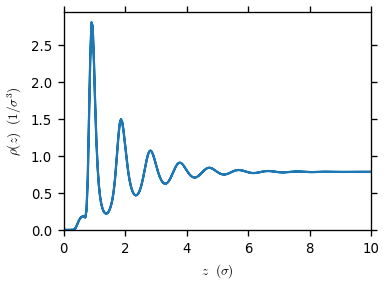}
		\caption{$\epsuw = 0.5\epsilon$, $\siguw = 0.8\sigma$, $\bar{\chi} = 0.2$}
		\label{fig:cz_eq_0.2_0.5_0.8}
	\end{subfigure}
	\hfill
	\begin{subfigure}[b]{0.32\textwidth}
		\centering
		\includegraphics[width=\textwidth]{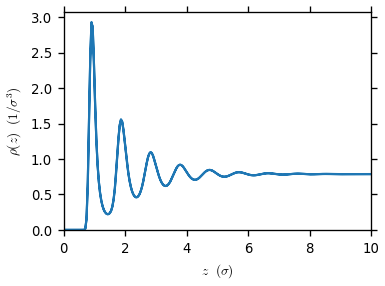}
		\caption{$\epsuw = 0.5\epsilon$, $\siguw = 1.2\sigma$, $\bar{\chi} = 0.2$}
		\label{fig:cz_eq_0.2_0.5_1.2}
	\end{subfigure}
	\hfill
	\begin{subfigure}[b]{0.32\textwidth}
		\centering
		\includegraphics[width=\textwidth]{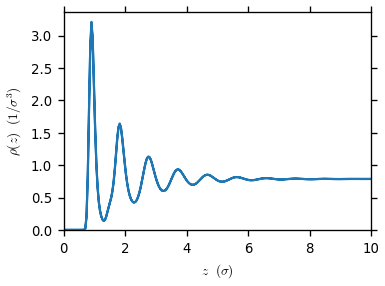}
		\caption{$\epsuw = 0.5\epsilon$, $\siguw = 1.5\sigma$, $\bar{\chi} = 0.2$}
		\label{fig:cz_eq_0.2_0.5_1.5}
	\end{subfigure}	
	\hfill
	\begin{subfigure}[b]{0.32\textwidth}
		\centering
		\includegraphics[width=\textwidth]{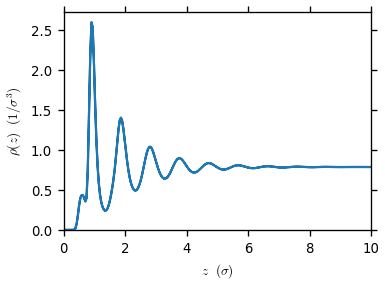}
		\caption{$\epsuw = 0.8\epsilon$, $\siguw = 0.8\sigma$, $\bar{\chi} = 0.2$}
		\label{fig:cz_eq_0.2_0.8_0.8}
	\end{subfigure}
		\hfill
	\begin{subfigure}[b]{0.32\textwidth}
		\centering
		\includegraphics[width=\textwidth]{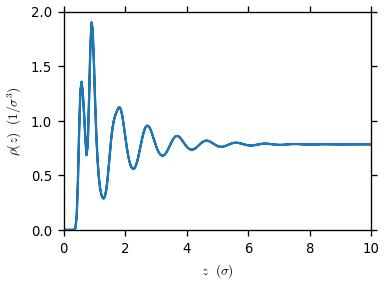}
		\caption{$\epsuw = 1.2\epsilon$, $\siguw = 0.8\sigma$, $\bar{\chi} = 0.2$}
		\label{fig:cz_eq_0.2_1.2_0.8}
	\end{subfigure}
		\hfill
	\begin{subfigure}[b]{0.32\textwidth}
		\centering
		\includegraphics[width=\textwidth]{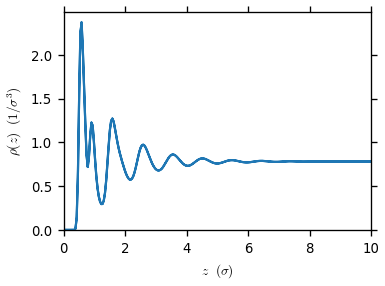}
		\caption{$\epsuw = 1.5\epsilon$, $\siguw = 0.8\sigma$, $\bar{\chi} = 0.2$}
		\label{fig:cz_eq_0.2_1.5_0.8}
	\end{subfigure}
	\hfill
	\begin{subfigure}[b]{0.32\textwidth}
	\centering
	\includegraphics[width=\textwidth]{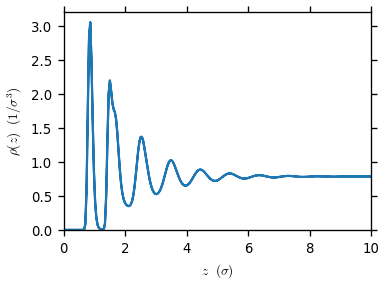}
	\caption{$\epsuw = 1.5\epsilon$, $\siguw = 1.5\sigma$, $\bar{\chi} = 0.2$}
	\label{fig:cz_eq_0.2_1.5_1.5}
	\end{subfigure}
	\hfill
	\begin{subfigure}[b]{0.32\textwidth}
		\centering
		\includegraphics[width=\textwidth]{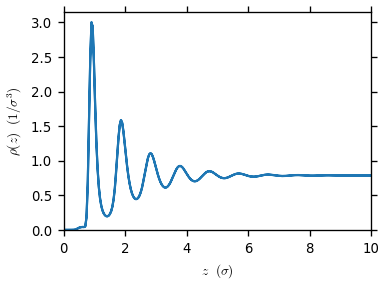}
		\caption{$\epsuw = 0.5\epsilon$, $\siguw = 0.8\sigma$, $\bar{\chi} = 0.05$}
		\label{fig:cz_eq_0.05_0.5_0.8}
	\end{subfigure}
		\hfill
	\begin{subfigure}[b]{0.32\textwidth}
		\centering
		\includegraphics[width=\textwidth]{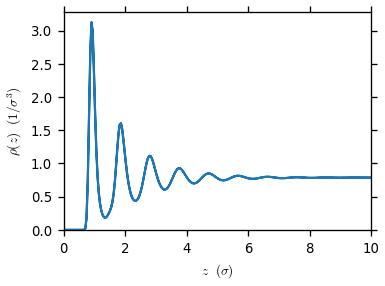}
		\caption{$\epsuw = 0.5\epsilon$, $\siguw = 1.5\sigma$, $\bar{\chi} = 0.05$}
		\label{fig:cz_eq_0.05_0.5_1.5}
	\end{subfigure}
	\hfill
	\begin{subfigure}[b]{0.32\textwidth}
		\centering
		\includegraphics[width=\textwidth]{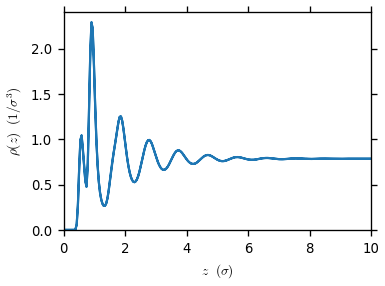}
		\caption{$\epsuw = 1.5\epsilon$, $\siguw = 0.8\sigma$, $\bar{\chi} = 0.05$}
		\label{fig:cz_eq_0.05_1.5_0.8}
	\end{subfigure}
	\hfill
	\begin{subfigure}[b]{0.32\textwidth}
	\centering
	\includegraphics[width=\textwidth]{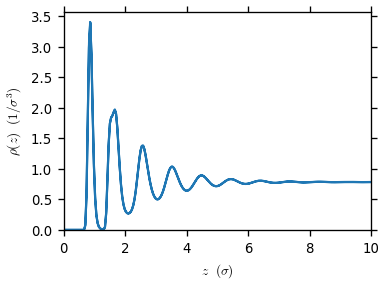}
	\caption{$\epsuw = 1.5\epsilon$, $\siguw = 1.5\sigma$, $\bar{\chi} = 0.05$}
	\label{fig:cz_eq_0.05_1.5_1.5}
	\end{subfigure}
	\hfill
	\caption{Total fluid density vs distance perpendicular to a poreless membrane for  various solute--membrane interaction parameters, $\epsuw$ and $\siguw$, and solute mole fractions, $\bar{\chi}$. Due to symmetry, the distribution is only shown for $z\geq 0$.}
	\label{fig:cz_eq}
\end{figure}

\begin{figure}[htb!]
	\centering
	\begin{subfigure}[b]{0.32\textwidth}
		\centering
		\includegraphics[width=\textwidth]{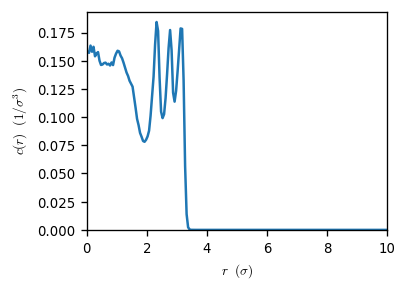}
		\caption{$\epsuw = 0.5\epsilon$, $\siguw = 0.8\sigma$, $a = 3\sigma$}
		\label{fig:cur_eq_0.2_3_0.5_0.8}
	\end{subfigure}
	\hfill
	\begin{subfigure}[b]{0.32\textwidth}
		\centering
		\includegraphics[width=\textwidth]{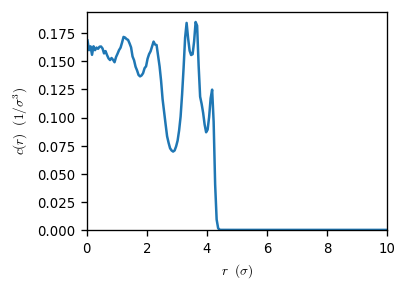}
		\caption{$\epsuw = 0.5\epsilon$, $\siguw = 0.8\sigma$, $a = 4\sigma$}
		\label{fig:cur_eq_0.2_4_0.5_0.8}
	\end{subfigure}
	\hfill	
	\begin{subfigure}[b]{0.32\textwidth}
		\centering
		\includegraphics[width=\textwidth]{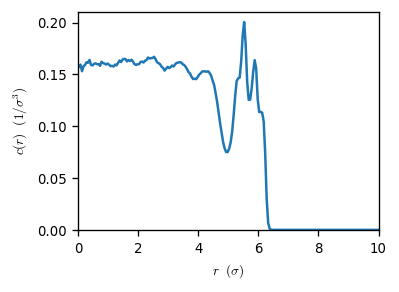}
		\caption{$\epsuw = 0.5\epsilon$, $\siguw = 0.8\sigma$, $a = 6\sigma$}
		\label{fig:cur_eq_0.2_6_0.5_0.8}
	\end{subfigure}
	\hfill
	\begin{subfigure}[b]{0.32\textwidth}
		\centering
		\includegraphics[width=\textwidth]{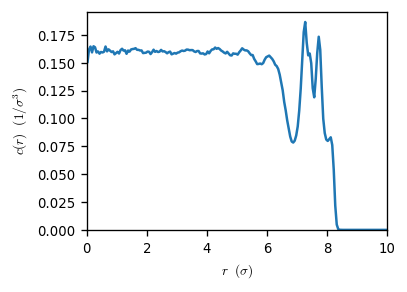}
		\caption{$\epsuw = 0.5\epsilon$, $\siguw = 0.8\sigma$, $a = 8\sigma$}
		\label{fig:cur_eq_0.2_8_0.5_0.8}
	\end{subfigure}
	\hfill
	\begin{subfigure}[b]{0.32\textwidth}
		\centering
		\includegraphics[width=\textwidth]{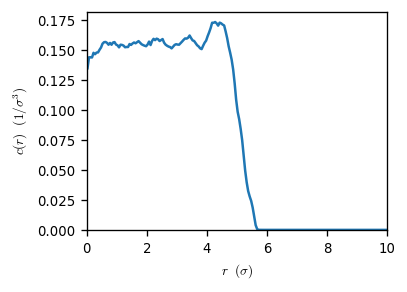}
		\caption{$\epsuw = 0.5\epsilon$, $\siguw = 1.2\sigma$, $a = 6\sigma$}
		\label{fig:cur_eq_0.2_6_0.5_1.2}
	\end{subfigure}
	\hfill
	\begin{subfigure}[b]{0.32\textwidth}
		\centering
		\includegraphics[width=\textwidth]{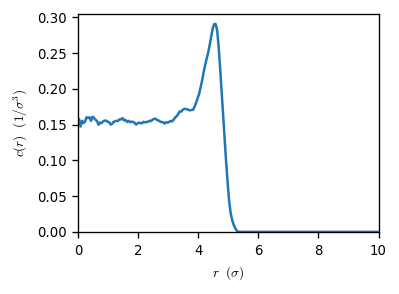}
		\caption{$\epsuw = 0.5\epsilon$, $\siguw = 1.5\sigma$, $a = 6\sigma$}
		\label{fig:cur_eq_0.2_6_0.5_1.5}
	\end{subfigure}
	\hfill
	\begin{subfigure}[b]{0.32\textwidth}
		\centering
		\includegraphics[width=\textwidth]{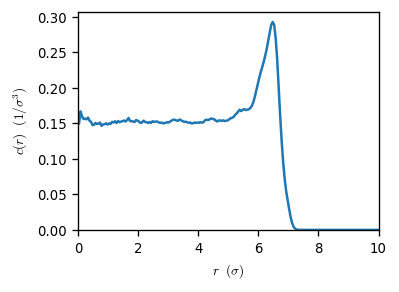}
		\caption{$\epsuw = 0.5\epsilon$, $\siguw = 1.5\sigma$, $a = 8\sigma$}
		\label{fig:cur_eq_0.2_8_0.5_1.5}
	\end{subfigure}
	\hfill
	\begin{subfigure}[b]{0.32\textwidth}
		\centering
		\includegraphics[width=\textwidth]{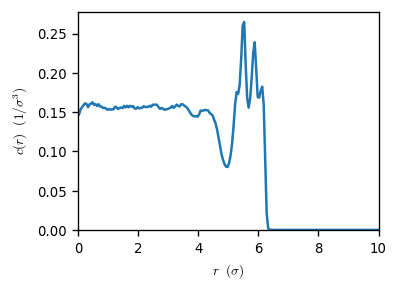}
		\caption{$\epsuw = 0.8\epsilon$, $\siguw = 0.8\sigma$, $a = 6\sigma$}
		\label{fig:cur_eq_0.2_6_0.8_0.8}
	\end{subfigure}
	\hfill
	\begin{subfigure}[b]{0.32\textwidth}
		\centering
		\includegraphics[width=\textwidth]{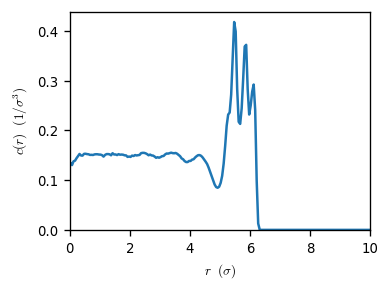}
		\caption{$\epsuw = 1.2\epsilon$, $\siguw = 0.8\sigma$, $a = 6\sigma$}
		\label{fig:cur_eq_0.2_6_1.2_0.8}
	\end{subfigure}
	\hfill
	\begin{subfigure}[b]{0.32\textwidth}
		\centering
		\includegraphics[width=\textwidth]{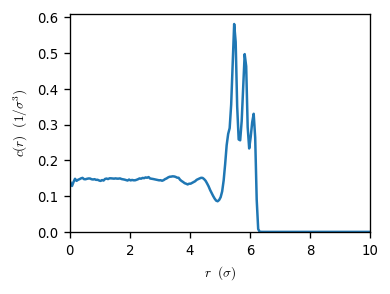}
		\caption{$\epsuw = 1.5\epsilon$, $\siguw = 0.8\sigma$, $a = 6\sigma$}
		\label{fig:cur_eq_0.2_6_1.5_0.8}
	\end{subfigure}
	\hfill
	\begin{subfigure}[b]{0.32\textwidth}
		\centering
		\includegraphics[width=\textwidth]{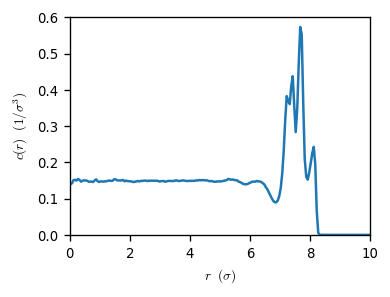}
		\caption{$\epsuw = 1.5\epsilon$, $\siguw = 0.8\sigma$, $a = 8\sigma$}
		\label{fig:cur_eq_0.2_8_1.5_0.8}
	\end{subfigure}
	\hfill
	\begin{subfigure}[b]{0.32\textwidth}
		\centering
		\includegraphics[width=\textwidth]{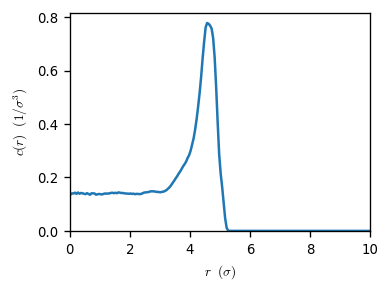}
		\caption{$\epsuw = 1.5\epsilon$, $\siguw = 1.5\sigma$, $a = 6\sigma$}
		\label{fig:cur_eq_0.2_6_1.5_1.5}
	\end{subfigure}
	\hfill
	\caption{Solute concentration vs radial coordinate inside the membrane pore (for axial coordinate $-\sigma/10 \leq z \leq \sigma/10$) for solute mole fraction $\bar{\chi} = 0.2$ and various pore radii, $a$, and solute--membrane interaction parameters, $\epsuw$ and $\siguw$.}
	\label{fig:cur_eq_chi0.2}
\end{figure}

\begin{figure}
	\centering
	\begin{subfigure}[b]{0.32\textwidth}
		\centering
		\includegraphics[width=\textwidth]{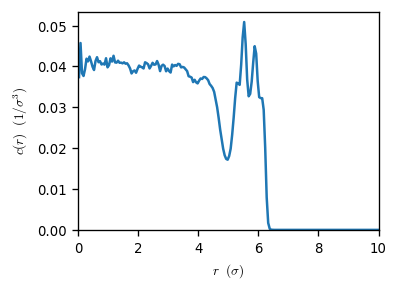}
		\caption{$\epsuw = 0.5\epsilon$, $\siguw = 0.8\sigma$, $a = 6\sigma$}
		\label{fig:cur_eq_0.05_6_0.5_0.8}
	\end{subfigure}
	\hfill
	\begin{subfigure}[b]{0.32\textwidth}
		\centering
		\includegraphics[width=\textwidth]{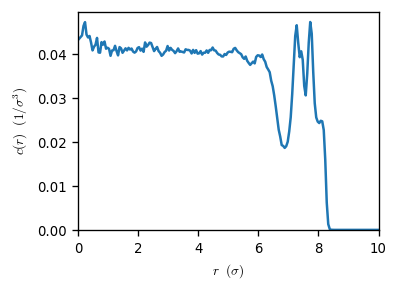}
		\caption{$\epsuw = 0.5\epsilon$, $\siguw = 0.8\sigma$, $a = 8\sigma$}
		\label{fig:cur_eq_0.05_8_0.5_0.8}
	\end{subfigure}
	\hfill
	\begin{subfigure}[b]{0.32\textwidth}
		\centering
		\includegraphics[width=\textwidth]{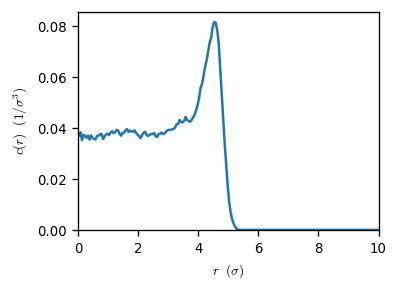}
		\caption{$\epsuw = 0.5\epsilon$, $\siguw = 1.5\sigma$, $a = 6\sigma$}
		\label{fig:cur_eq_0.05_6_0.5_1.5}
	\end{subfigure}
	\hfill
	\begin{subfigure}[b]{0.32\textwidth}
		\centering
		\includegraphics[width=\textwidth]{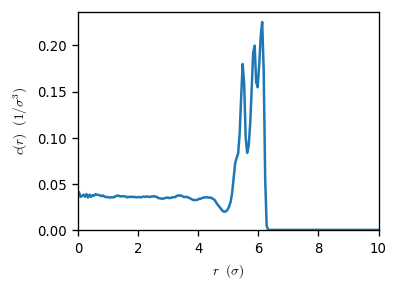}
		\caption{$\epsuw = 1.5\epsilon$, $\siguw = 0.8\sigma$, $a = 6\sigma$}
		\label{fig:cur_eq_0.05_6_1.5_0.8}
	\end{subfigure}
	\hfill
	\begin{subfigure}[b]{0.32\textwidth}
		\centering
		\includegraphics[width=\textwidth]{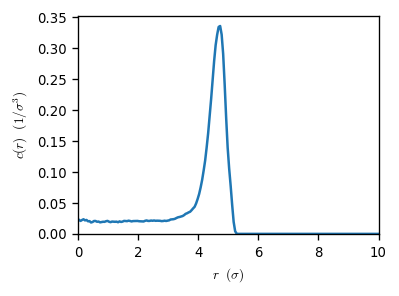}
		\caption{$\epsuw = 1.5\epsilon$, $\siguw = 1.5\sigma$, $a = 6\sigma$}
		\label{fig:cur_eq_0.05_6_1.5_1.5}
	\end{subfigure}
	\hfill
	\caption{Solute concentration versus radial coordinate inside the membrane pore (for axial coordinate $-\sigma/10 \leq z \leq \sigma/10$) for solute mole fraction $\bar{\chi} = 0.05$ and various pore radii, $a$, and solute--membrane interaction parameters, $\epsuw$ and $\siguw$. }
	\label{fig:cur_eq_chi0.05}
\end{figure}

\begin{figure}
	\centering
	\begin{subfigure}[b]{0.32\textwidth}
		\centering
		\includegraphics[width=\textwidth]{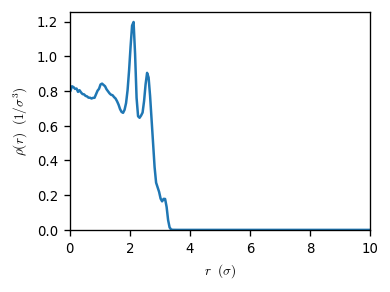}
		\caption{$\epsuw = 0.5\epsilon$, $\siguw = 0.8\sigma$, $a = 3\sigma$}
		\label{fig:cr_eq_0.2_3_0.5_0.8}
	\end{subfigure}
	\hfill
	\begin{subfigure}[b]{0.32\textwidth}
		\centering
		\includegraphics[width=\textwidth]{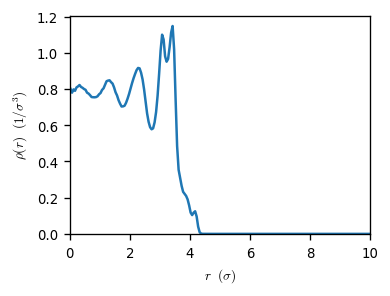}
		\caption{$\epsuw = 0.5\epsilon$, $\siguw = 0.8\sigma$, $a = 4\sigma$}
		\label{fig:cr_eq_0.2_4_0.5_0.8}
	\end{subfigure}
	\hfill	
	\begin{subfigure}[b]{0.32\textwidth}
		\centering
		\includegraphics[width=\textwidth]{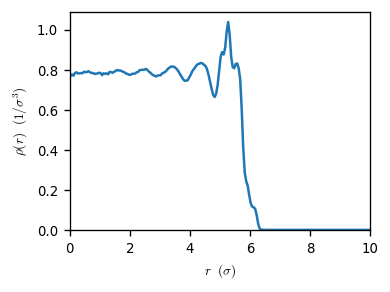}
		\caption{$\epsuw = 0.5\epsilon$, $\siguw = 0.8\sigma$, $a = 6\sigma$}
		\label{fig:cr_eq_0.2_6_0.5_0.8}
	\end{subfigure}
	\hfill
	\begin{subfigure}[b]{0.32\textwidth}
		\centering
		\includegraphics[width=\textwidth]{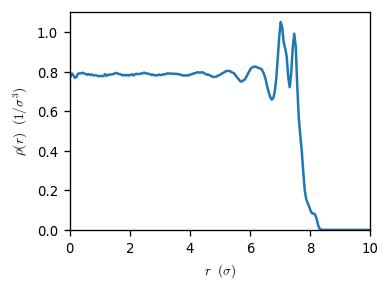}
		\caption{$\epsuw = 0.5\epsilon$, $\siguw = 0.8\sigma$, $a = 8\sigma$}
		\label{fig:cr_eq_0.2_8_0.5_0.8}
	\end{subfigure}
	\hfill
	\begin{subfigure}[b]{0.32\textwidth}
		\centering
		\includegraphics[width=\textwidth]{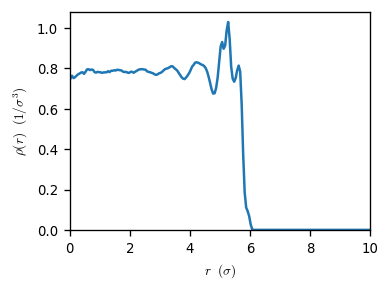}
		\caption{$\epsuw = 0.5\epsilon$, $\siguw = 1.2\sigma$, $a = 6\sigma$}
		\label{fig:cr_eq_0.2_6_0.5_1.2}
	\end{subfigure}
	\hfill
	\begin{subfigure}[b]{0.32\textwidth}
		\centering
		\includegraphics[width=\textwidth]{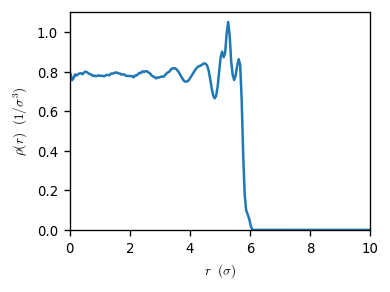}
		\caption{$\epsuw = 0.5\epsilon$, $\siguw = 1.5\sigma$, $a = 6\sigma$}
		\label{fig:cr_eq_0.2_6_0.5_1.5}
	\end{subfigure}
	\hfill
	\begin{subfigure}[b]{0.32\textwidth}
		\centering
		\includegraphics[width=\textwidth]{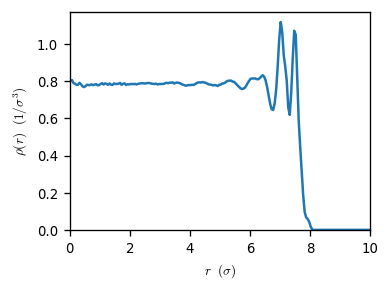}
		\caption{$\epsuw = 0.5\epsilon$, $\siguw = 1.5\sigma$, $a = 8\sigma$}
		\label{fig:cr_eq_0.2_8_0.5_1.5}
	\end{subfigure}
	\hfill
	\begin{subfigure}[b]{0.32\textwidth}
		\centering
		\includegraphics[width=\textwidth]{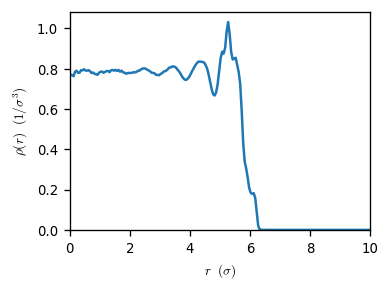}
		\caption{$\epsuw = 0.8\epsilon$, $\siguw = 0.8\sigma$, $a = 6\sigma$}
		\label{fig:cr_eq_0.2_6_0.8_0.8}
	\end{subfigure}
	\hfill
	\begin{subfigure}[b]{0.32\textwidth}
		\centering
		\includegraphics[width=\textwidth]{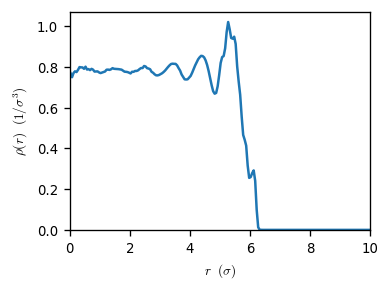}
		\caption{$\epsuw = 1.2\epsilon$, $\siguw = 0.8\sigma$, $a = 6\sigma$}
		\label{fig:cr_eq_0.2_6_1.2_0.8}
	\end{subfigure}
	\hfill
	\begin{subfigure}[b]{0.32\textwidth}
		\centering
		\includegraphics[width=\textwidth]{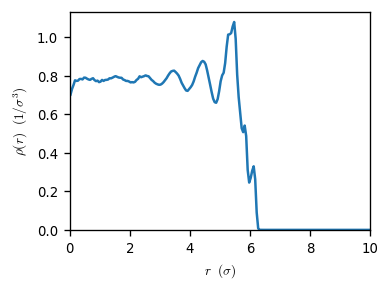}
		\caption{$\epsuw = 1.5\epsilon$, $\siguw = 0.8\sigma$, $a = 6\sigma$}
		\label{fig:cr_eq_0.2_6_1.5_0.8}
	\end{subfigure}
	\hfill
	\begin{subfigure}[b]{0.32\textwidth}
		\centering
		\includegraphics[width=\textwidth]{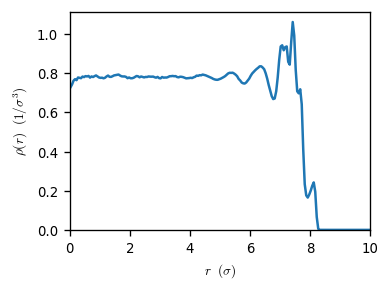}
		\caption{$\epsuw = 1.5\epsilon$, $\siguw = 0.8\sigma$, $a = 8\sigma$}
		\label{fig:cr_eq_0.2_8_1.5_0.8}
	\end{subfigure}
	\hfill
	\begin{subfigure}[b]{0.32\textwidth}
		\centering
		\includegraphics[width=\textwidth]{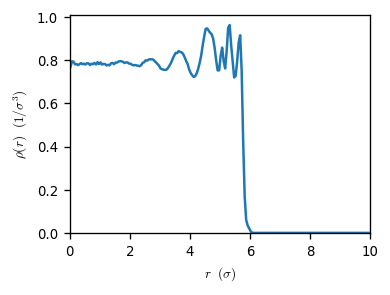}
		\caption{$\epsuw = 1.5\epsilon$, $\siguw = 1.5\sigma$, $a = 6\sigma$}
		\label{fig:cr_eq_0.2_6_1.5_1.5}
	\end{subfigure}
	\hfill
	\caption{Total fluid density vs radial coordinate inside the membrane pore (for axial coordinate $-\sigma/10 \leq z \leq \sigma/10$) for solute mole fraction $\bar{\chi} = 0.2$ and various pore radii, $a$, and solute--membrane interaction parameters, $\epsuw$ and $\siguw$.}
	\label{fig:cr_eq_chi0.2}
\end{figure}

\begin{figure}
	\centering
	\begin{subfigure}[b]{0.32\textwidth}
		\centering
		\includegraphics[width=\textwidth]{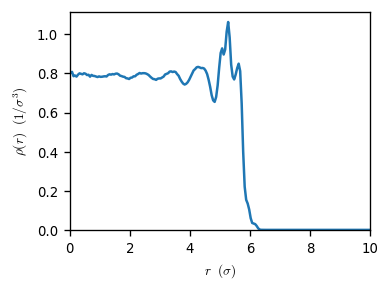}
		\caption{$\epsuw = 0.5\epsilon$, $\siguw = 0.8\sigma$, $a = 6\sigma$}
		\label{fig:cr_eq_0.05_6_0.5_0.8}
	\end{subfigure}
	\hfill
	\begin{subfigure}[b]{0.32\textwidth}
		\centering
		\includegraphics[width=\textwidth]{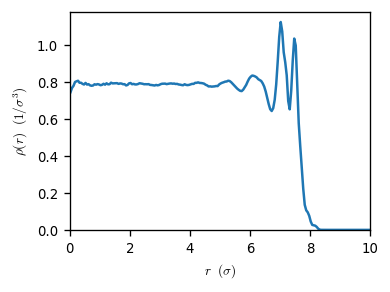}
		\caption{$\epsuw = 0.5\epsilon$, $\siguw = 0.8\sigma$, $a = 8\sigma$}
		\label{fig:cr_eq_0.05_8_0.5_0.8}
	\end{subfigure}
	\hfill
	\begin{subfigure}[b]{0.32\textwidth}
		\centering
		\includegraphics[width=\textwidth]{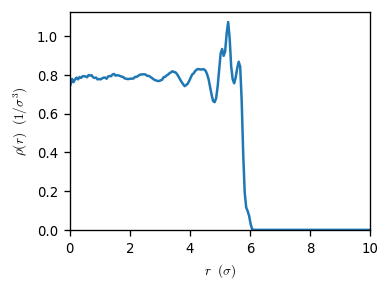}
		\caption{$\epsuw = 0.5\epsilon$, $\siguw = 1.5\sigma$, $a = 6\sigma$}
		\label{fig:cr_eq_0.05_6_0.5_1.5}
	\end{subfigure}
	\hfill
	\begin{subfigure}[b]{0.32\textwidth}
		\centering
		\includegraphics[width=\textwidth]{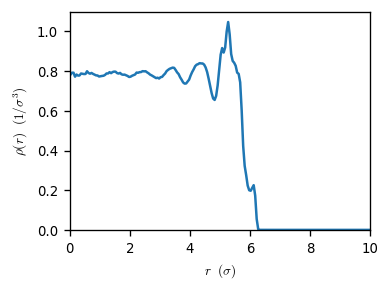}
		\caption{$\epsuw = 1.5\epsilon$, $\siguw = 0.8\sigma$, $a = 6\sigma$}
		\label{fig:cr_eq_0.05_6_1.5_0.8}
	\end{subfigure}
	\hfill
	\begin{subfigure}[b]{0.32\textwidth}
		\centering
		\includegraphics[width=\textwidth]{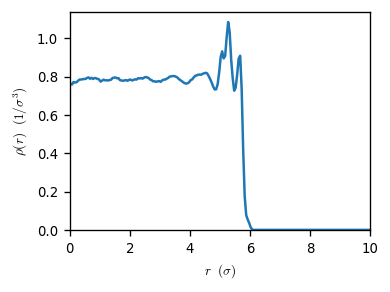}
		\caption{$\epsuw = 1.5\epsilon$, $\siguw = 1.5\sigma$, $a = 6\sigma$}
		\label{fig:cr_eq_0.05_6_1.5_1.5}
	\end{subfigure}
	\hfill
	\caption{Total fluid density vs radial coordinate inside the membrane pore (for axial coordinate $-\sigma/10 \leq z \leq \sigma/10$) for solute mole fraction $\bar{\chi} = 0.05$ and various pore radii, $a$, and solute--membrane interaction parameters, $\epsuw$ and $\siguw$.}
	\label{fig:cr_eq_chi0.05}
\end{figure}

\begin{figure}
	\centering
	\begin{subfigure}[b]{0.32\textwidth}
		\centering
		\includegraphics[width=\textwidth]{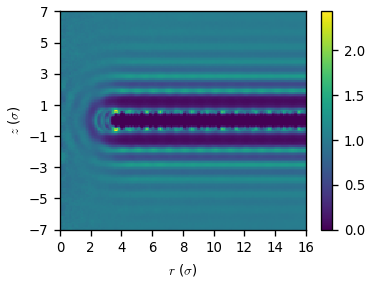}
		\caption{$\epsuw = 0.5\epsilon$, $\siguw = 0.8\sigma$, $a = 3\sigma$}
		\label{fig:curz_eq_0.2_3_0.5_0.8}
	\end{subfigure}
	\hfill
	\begin{subfigure}[b]{0.32\textwidth}
		\centering
		\includegraphics[width=\textwidth]{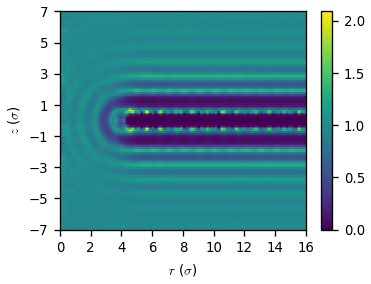}
		\caption{$\epsuw = 0.5\epsilon$, $\siguw = 0.8\sigma$, $a = 4\sigma$}
		\label{fig:curz_eq_0.2_4_0.5_0.8}
	\end{subfigure}
	\hfill	
	\begin{subfigure}[b]{0.32\textwidth}
		\centering
		\includegraphics[width=\textwidth]{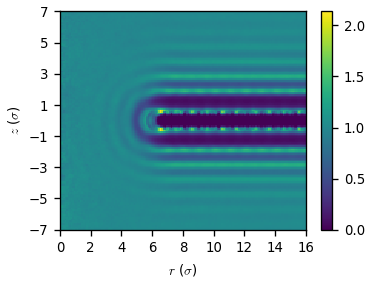}
		\caption{$\epsuw = 0.5\epsilon$, $\siguw = 0.8\sigma$, $a = 6\sigma$}
		\label{fig:curz_eq_0.2_6_0.5_0.8}
	\end{subfigure}
	\hfill
	\begin{subfigure}[b]{0.32\textwidth}
		\centering
		\includegraphics[width=\textwidth]{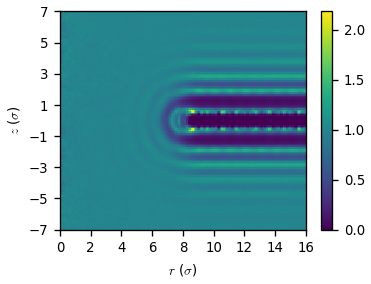}
		\caption{$\epsuw = 0.5\epsilon$, $\siguw = 0.8\sigma$, $a = 8\sigma$}
		\label{fig:curz_eq_0.2_8_0.5_0.8}
	\end{subfigure}
	\hfill
	\begin{subfigure}[b]{0.32\textwidth}
		\centering
		\includegraphics[width=\textwidth]{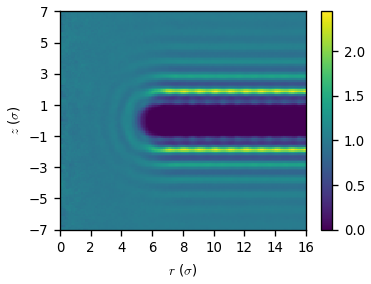}
		\caption{$\epsuw = 0.5\epsilon$, $\siguw = 1.2\sigma$, $a = 6\sigma$}
		\label{fig:curz_eq_0.2_6_0.5_1.2}
	\end{subfigure}
	\hfill
	\begin{subfigure}[b]{0.32\textwidth}
		\centering
		\includegraphics[width=\textwidth]{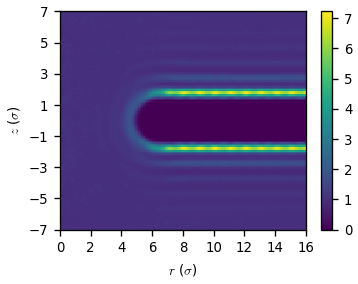}
		\caption{$\epsuw = 0.5\epsilon$, $\siguw = 1.5\sigma$, $a = 6\sigma$}
		\label{fig:curz_eq_0.2_6_0.5_1.5}
	\end{subfigure}
	\hfill
	\begin{subfigure}[b]{0.32\textwidth}
		\centering
		\includegraphics[width=\textwidth]{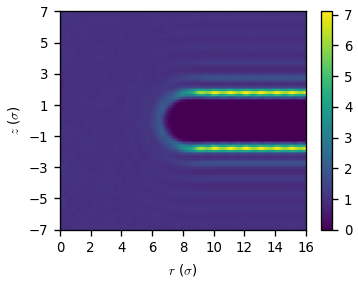}
		\caption{$\epsuw = 0.5\epsilon$, $\siguw = 1.5\sigma$, $a = 8\sigma$}
		\label{fig:curz_eq_0.2_8_0.5_1.5}
	\end{subfigure}
	\hfill
	\begin{subfigure}[b]{0.32\textwidth}
		\centering
		\includegraphics[width=\textwidth]{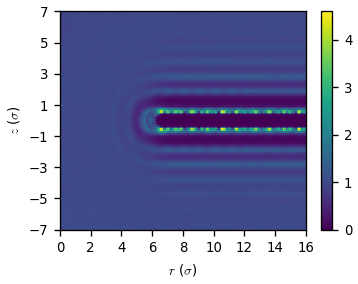}
		\caption{$\epsuw = 0.8\epsilon$, $\siguw = 0.8\sigma$, $a = 6\sigma$}
		\label{fig:curz_eq_0.2_6_0.8_0.8}
	\end{subfigure}
	\hfill
	\begin{subfigure}[b]{0.32\textwidth}
		\centering
		\includegraphics[width=\textwidth]{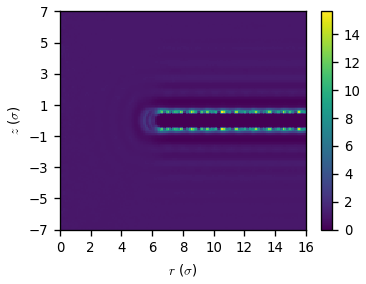}
		\caption{$\epsuw = 1.2\epsilon$, $\siguw = 0.8\sigma$, $a = 6\sigma$}
		\label{fig:curz_eq_0.2_6_1.2_0.8}
	\end{subfigure}
	\hfill
	\begin{subfigure}[b]{0.32\textwidth}
		\centering
		\includegraphics[width=\textwidth]{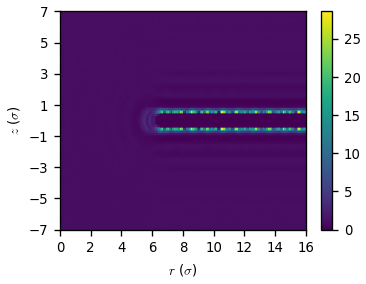}
		\caption{$\epsuw = 1.5\epsilon$, $\siguw = 0.8\sigma$, $a = 6\sigma$}
		\label{fig:curz_eq_0.2_6_1.5_0.8}
	\end{subfigure}
	\hfill
	\begin{subfigure}[b]{0.32\textwidth}
		\centering
		\includegraphics[width=\textwidth]{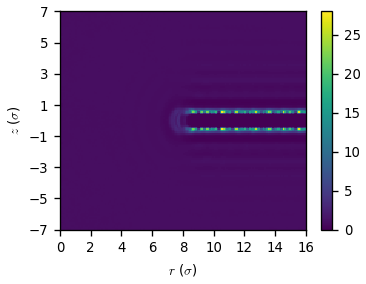}
		\caption{$\epsuw = 1.5\epsilon$, $\siguw = 0.8\sigma$, $a = 8\sigma$}
		\label{fig:curz_eq_0.2_8_1.5_0.8}
	\end{subfigure}
	\hfill
	\begin{subfigure}[b]{0.32\textwidth}
		\centering
		\includegraphics[width=\textwidth]{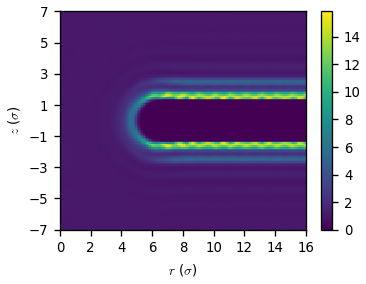}
		\caption{$\epsuw = 1.5\epsilon$, $\siguw = 1.5\sigma$, $a = 6\sigma$}
		\label{fig:curz_eq_0.2_6_1.5_1.5}
	\end{subfigure}
	\hfill
	\caption{Contour plots of solute concentration relative to the bulk, $\cu(r,z)/\cubulk$, near the pore vs radial and axial coordinates for solute mole fraction $\bar{\chi} = 0.2$ and various pore radii, $a$, and solute--membrane interaction parameters, $\epsuw$ and $\siguw$.}
	\label{fig:curz_eq_chi0.2}
\end{figure}

\begin{figure}
	\centering
	\begin{subfigure}[b]{0.32\textwidth}
		\centering
		\includegraphics[width=\textwidth]{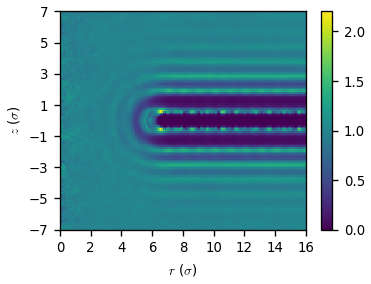}
		\caption{$\epsuw = 0.5\epsilon$, $\siguw = 0.8\sigma$, $a = 6\sigma$}
		\label{fig:curz_eq_0.05_6_0.5_0.8}
	\end{subfigure}
	\hfill
	\begin{subfigure}[b]{0.32\textwidth}
		\centering
		\includegraphics[width=\textwidth]{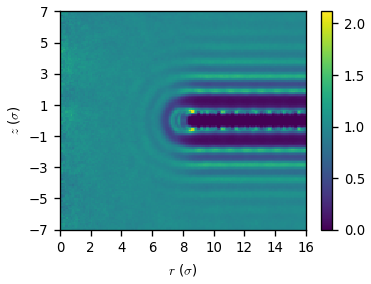}
		\caption{$\epsuw = 0.5\epsilon$, $\siguw = 0.8\sigma$, $a = 8\sigma$}
		\label{fig:curz_eq_0.05_8_0.5_0.8}
	\end{subfigure}
	\hfill
	\begin{subfigure}[b]{0.32\textwidth}
		\centering
		\includegraphics[width=\textwidth]{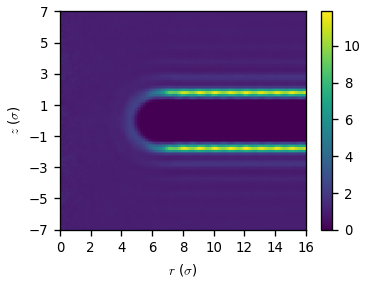}
		\caption{$\epsuw = 0.5\epsilon$, $\siguw = 1.5\sigma$, $a = 6\sigma$}
		\label{fig:curz_eq_0.05_6_0.5_1.5}
	\end{subfigure}
	\hfill
	\begin{subfigure}[b]{0.32\textwidth}
		\centering
		\includegraphics[width=\textwidth]{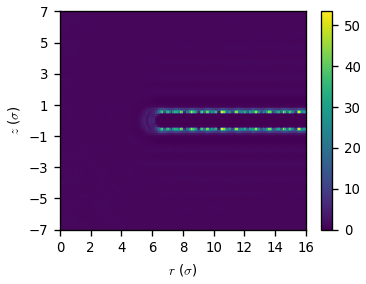}
		\caption{$\epsuw = 1.5\epsilon$, $\siguw = 0.8\sigma$, $a = 6\sigma$}
		\label{fig:curz_eq_0.05_6_1.5_0.8}
	\end{subfigure}
	\hfill
	\begin{subfigure}[b]{0.32\textwidth}
		\centering
		\includegraphics[width=\textwidth]{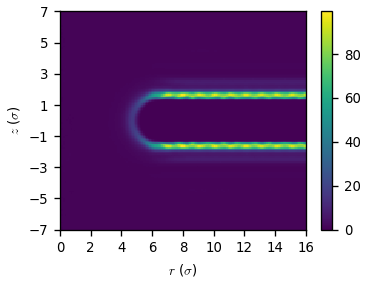}
		\caption{$\epsuw = 1.5\epsilon$, $\siguw = 1.5\sigma$, $a = 6\sigma$}
		\label{fig:curz_eq_0.05_6_1.5_1.5}
	\end{subfigure}
	\hfill
	\caption{Contour plots of solute concentration relative to the bulk, $\cu(r,z)/\cubulk$, near the pore vs radial and axial coordinates for solute mole fraction $\bar{\chi} = 0.05$ and various pore radii, $a$, and solute--membrane interaction parameters, $\epsuw$ and $\siguw$. }
	\label{fig:curz_eq_chi0.05}
\end{figure}

\begin{figure}
	\centering
	\begin{subfigure}[b]{0.32\textwidth}
		\centering
		\includegraphics[width=\textwidth]{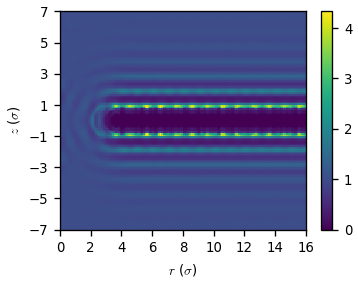}
		\caption{$\epsuw = 0.5\epsilon$, $\siguw = 0.8\sigma$, $a = 3\sigma$}
		\label{fig:crz_eq_0.2_3_0.5_0.8}
	\end{subfigure}
	\hfill
	\begin{subfigure}[b]{0.32\textwidth}
		\centering
		\includegraphics[width=\textwidth]{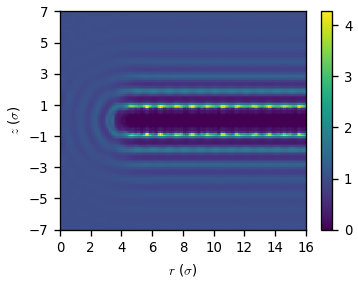}
		\caption{$\epsuw = 0.5\epsilon$, $\siguw = 0.8\sigma$, $a = 4\sigma$}
		\label{fig:crz_eq_0.2_4_0.5_0.8}
	\end{subfigure}
	\hfill	
	\begin{subfigure}[b]{0.32\textwidth}
		\centering
		\includegraphics[width=\textwidth]{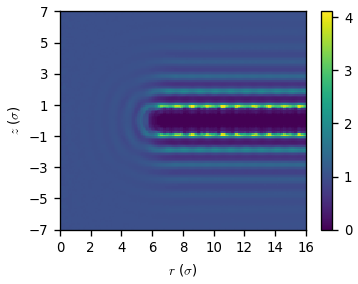}
		\caption{$\epsuw = 0.5\epsilon$, $\siguw = 0.8\sigma$, $a = 6\sigma$}
		\label{fig:crz_eq_0.2_6_0.5_0.8}
	\end{subfigure}
	\hfill
	\begin{subfigure}[b]{0.32\textwidth}
		\centering
		\includegraphics[width=\textwidth]{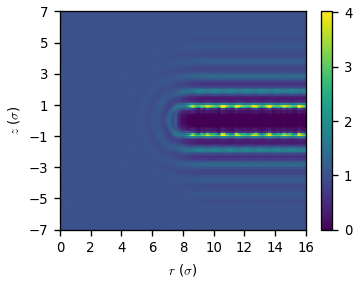}
		\caption{$\epsuw = 0.5\epsilon$, $\siguw = 0.8\sigma$, $a = 8\sigma$}
		\label{fig:crz_eq_0.2_8_0.5_0.8}
	\end{subfigure}
	\hfill
	\begin{subfigure}[b]{0.32\textwidth}
		\centering
		\includegraphics[width=\textwidth]{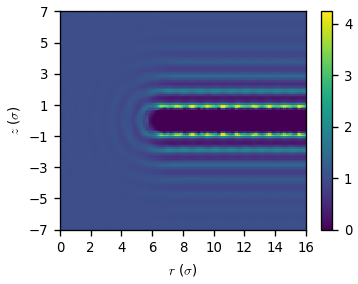}
		\caption{$\epsuw = 0.5\epsilon$, $\siguw = 1.2\sigma$, $a = 6\sigma$}
		\label{fig:crz_eq_0.2_6_0.5_1.2}
	\end{subfigure}
	\hfill
	\begin{subfigure}[b]{0.32\textwidth}
		\centering
		\includegraphics[width=\textwidth]{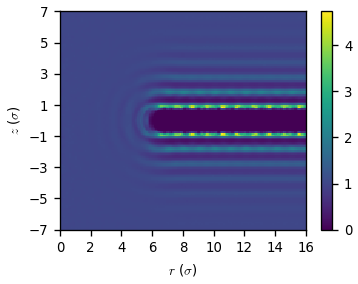}
		\caption{$\epsuw = 0.5\epsilon$, $\siguw = 1.5\sigma$, $a = 6\sigma$}
		\label{fig:crz_eq_0.2_6_0.5_1.5}
	\end{subfigure}
	\hfill
	\begin{subfigure}[b]{0.32\textwidth}
		\centering
		\includegraphics[width=\textwidth]{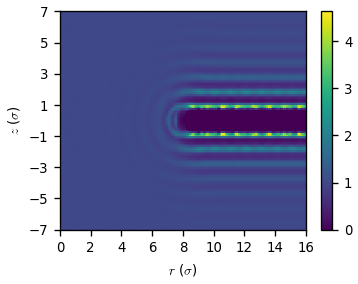}
		\caption{$\epsuw = 0.5\epsilon$, $\siguw = 1.5\sigma$, $a = 8\sigma$}
		\label{fig:crz_eq_0.2_8_0.5_1.5}
	\end{subfigure}
	\hfill
	\begin{subfigure}[b]{0.32\textwidth}
		\centering
		\includegraphics[width=\textwidth]{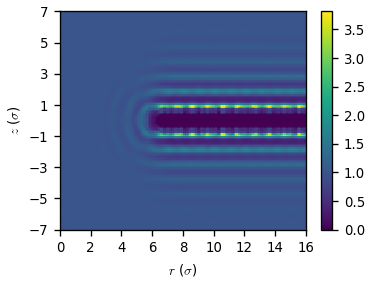}
		\caption{$\epsuw = 0.8\epsilon$, $\siguw = 0.8\sigma$, $a = 6\sigma$}
		\label{fig:crz_eq_0.2_6_0.8_0.8}
	\end{subfigure}
	\hfill
	\begin{subfigure}[b]{0.32\textwidth}
		\centering
		\includegraphics[width=\textwidth]{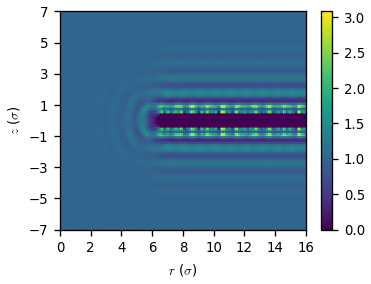}
		\caption{$\epsuw = 1.2\epsilon$, $\siguw = 0.8\sigma$, $a = 6\sigma$}
		\label{fig:crz_eq_0.2_6_1.2_0.8}
	\end{subfigure}
	\hfill
	\begin{subfigure}[b]{0.32\textwidth}
		\centering
		\includegraphics[width=\textwidth]{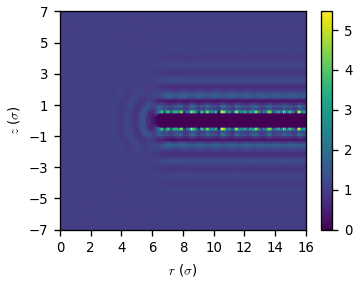}
		\caption{$\epsuw = 1.5\epsilon$, $\siguw = 0.8\sigma$, $a = 6\sigma$}
		\label{fig:crz_eq_0.2_6_1.5_0.8}
	\end{subfigure}
	\hfill
	\begin{subfigure}[b]{0.32\textwidth}
		\centering
		\includegraphics[width=\textwidth]{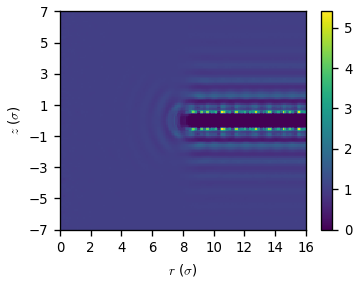}
		\caption{$\epsuw = 1.5\epsilon$, $\siguw = 0.8\sigma$, $a = 8\sigma$}
		\label{fig:crz_eq_0.2_8_1.5_0.8}
	\end{subfigure}
	\hfill
	\begin{subfigure}[b]{0.32\textwidth}
		\centering
		\includegraphics[width=\textwidth]{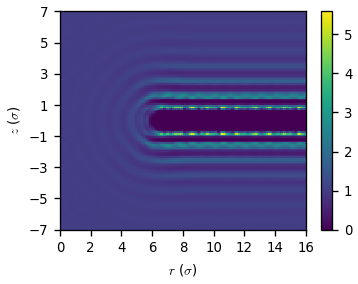}
		\caption{$\epsuw = 1.5\epsilon$, $\siguw = 1.5\sigma$, $a = 6\sigma$}
		\label{fig:crz_eq_0.2_6_1.5_1.5}
	\end{subfigure}
	\hfill
	\caption{Contour plots of total fluid density relative to the bulk, $\rho(r,z)/\rhobulk$, near the pore vs radial and axial coordinates for solute mole fraction $\bar{\chi} = 0.2$ and various pore radii, $a$, and solute--membrane interaction parameters, $\epsuw$ and $\siguw$.}
	\label{fig:crz_eq_chi0.2}
\end{figure}

\begin{figure}
	\centering
	\begin{subfigure}[b]{0.32\textwidth}
		\centering
		\includegraphics[width=\textwidth]{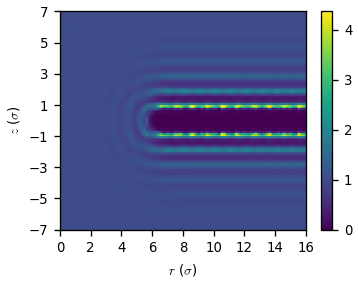}
		\caption{$\epsuw = 0.5\epsilon$, $\siguw = 0.8\sigma$, $a = 6\sigma$}
		\label{fig:crz_eq_0.05_6_0.5_0.8}
	\end{subfigure}
	\hfill
	\begin{subfigure}[b]{0.32\textwidth}
		\centering
		\includegraphics[width=\textwidth]{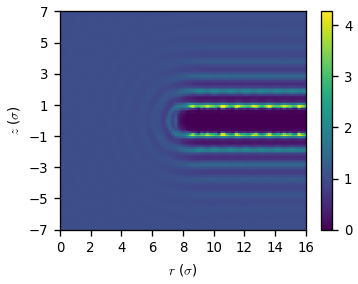}
		\caption{$\epsuw = 0.5\epsilon$, $\siguw = 0.8\sigma$, $a = 8\sigma$}
		\label{fig:crz_eq_0.05_8_0.5_0.8}
	\end{subfigure}
	\hfill
	\begin{subfigure}[b]{0.32\textwidth}
		\centering
		\includegraphics[width=\textwidth]{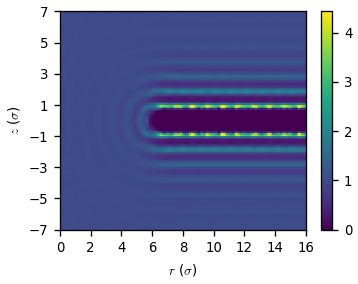}
		\caption{$\epsuw = 0.5\epsilon$, $\siguw = 1.5\sigma$, $a = 6\sigma$}
		\label{fig:crz_eq_0.05_6_0.5_1.5}
	\end{subfigure}
	\hfill
	\begin{subfigure}[b]{0.32\textwidth}
		\centering
		\includegraphics[width=\textwidth]{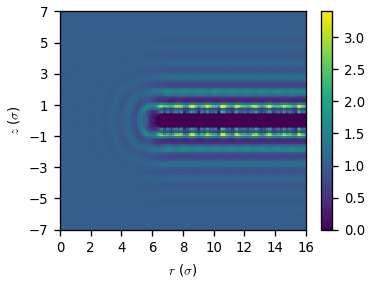}
		\caption{$\epsuw = 1.5\epsilon$, $\siguw = 0.8\sigma$, $a = 6\sigma$}
		\label{fig:crz_eq_0.05_6_1.5_0.8}
	\end{subfigure}
	\hfill
	\begin{subfigure}[b]{0.32\textwidth}
		\centering
		\includegraphics[width=\textwidth]{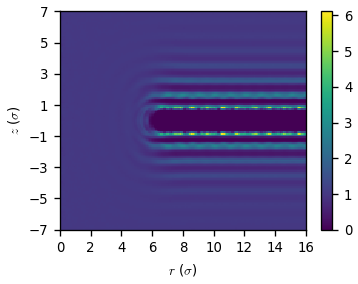}
		\caption{$\epsuw = 1.5\epsilon$, $\siguw = 1.5\sigma$, $a = 6\sigma$}
		\label{fig:crz_eq_0.05_6_1.5_1.5}
	\end{subfigure}
	\hfill
	\caption{Contour plots of total fluid density relative to the bulk, $\rho(r,z)/\rhobulk$, near the pore vs radial and axial coordinates for solute mole fraction $\bar{\chi} = 0.05$ and various pore radii, $a$, and solute--membrane interaction parameters, $\epsuw$ and $\siguw$.}
	\label{fig:crz_eq_chi0.05}
\end{figure}

\begin{figure}[htb!]
	\centering
	\includegraphics{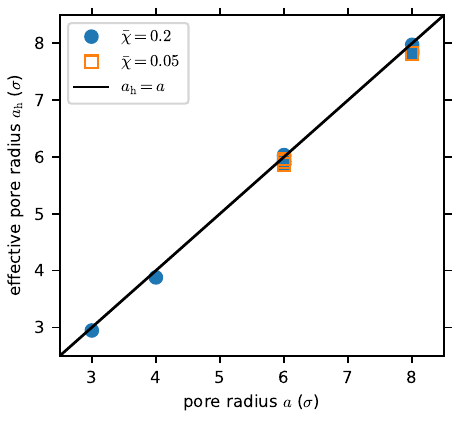}
	\caption{Effective pore radius (calculated from the Gibbs dividing surface in equilibrium simulations) vs pore radius (radius within which there were no membrane atom centers) for different average solute mole fractions ($\bar{\chi} = 0.2$ (circles) and  $\bar{\chi} = 0.05$ (squares)).}
	\label{fig:a_h_vs_a}
\end{figure}

\begin{figure}[htb!]
	\centering
	\includegraphics{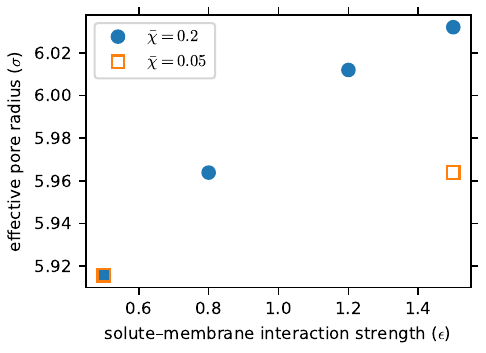}
	\caption{Effective pore radius vs solute--membrane interaction strength parameter $\epsuw$ for fixed interaction range parameter ($\siguw =  0.8\sigma$) and pore radius ($a = 6\sigma$) and different average solute mole fractions ($\bar{\chi} = 0.2$ (circles) and  $\bar{\chi} = 0.05$ (squares)).}
	\label{fig:a_h_vs_eps}
\end{figure}

\begin{figure}[htb!]
	\centering
	\includegraphics{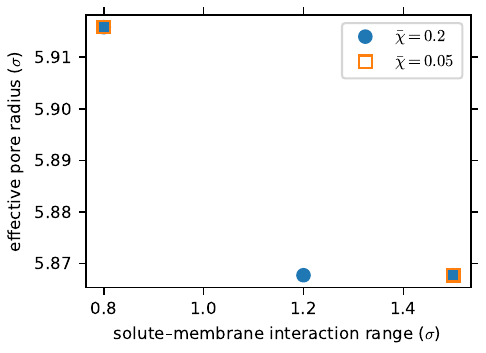}
	\caption{Effective pore radius vs solute--membrane interaction range parameter $\siguw$ for fixed interaction strength parameter ($\epsuw = 0.5\epsilon$) and pore radius ($a = 6\sigma$) and different average solute mole fractions ($\bar{\chi} = 0.2$ (circles) and  $\bar{\chi} = 0.05$ (squares))..}
	\label{fig:a_h_vs_sig}
\end{figure}

% \cite{rankin2019}

% \clearpage
% \bibliographystyle{aipnum4-1}
% \bibliography{lj-2d-cgrad}

% add cross references to external document
\makeatletter\@input{lj-2d-cgrad-aux.tex}\makeatother